\documentclass[12pt]{article}
\usepackage{amsmath}
\usepackage{multirow}
\usepackage{amsfonts}
\usepackage{amssymb,graphics,psfrag}
\usepackage{array,epsfig,multirow,graphicx}
\usepackage{comment}
\usepackage{feynmp}

\numberwithin{equation}{section}
\numberwithin{table}{section}

\usepackage{hyperref}

\def\hybrid{\topmargin -20pt    \oddsidemargin 0pt
        \headheight 0pt \headsep 0pt
        \textwidth 6.25in      
        \textheight 9 in      
        \marginparwidth .875in
        \parskip 5pt plus 1pt
          \jot = 1.5ex
  }
\hybrid

\newcommand{\be}{\begin{equation}}
\newcommand{\ee}{\end{equation}} 
\newcommand{\bea}{\begin{eqnarray}}
\newcommand{\eea}{\end{eqnarray}}

\newcommand{\nn}{\nonumber}

\newcommand{\beq}{\begin{equation}}
\newcommand{\eeq}{\end{equation}}

\newcommand{\cref}{{\bf [check ref]}}

\def\blfootnote{\xdef\@thefnmark{}\@footnotetext}
\long\def\symbolfootnote[#1]#2{\begingroup%
\def\thefootnote{\fnsymbol{footnote}}\footnote[#1]{#2}\endgroup}

\begin{document}

\baselineskip=15pt

\begin{titlepage}
\begin{flushright}
\parbox[t]{1.08in}{UPR-1249-T}
\end{flushright}

\begin{center}

\vspace*{ 1.2cm}

{\Large \bf \phantom{.}\text{\hspace{-0.5cm}F-Theory Compactifications with Multiple U(1)-Factors:} Constructing  Elliptic Fibrations with\\ Rational Sections}

\vskip 1.2cm

\renewcommand{\thefootnote}{}
\begin{center}
 {Mirjam Cveti\v{c}$^{1,2}$,  Denis Klevers$^1$ Hernan Piragua$^1$}
\end{center}
\vskip .2cm
\renewcommand{\thefootnote}{\arabic{footnote}} 

$\,^1$ {Department of Physics and Astronomy,\\
University of Pennsylvania, Philadelphia, PA 19104-6396, USA} \\[.3cm]

$\,^2$ {Center for Applied Mathematics and Theoretical Physics,\\ University of Maribor, Maribor, Slovenia}\\[.3cm]

{cvetic\ \textsf{at}\ cvetic.hep.upenn.edu, klevers\ \textsf{at}\ sas.upenn.edu, hpiragua\ \textsf{at}\ sas.upenn.edu, } 

 \vspace*{0.8cm}

\end{center}

\vskip 0.2cm
 
\begin{center} {\bf ABSTRACT } \end{center}

We study F-theory compactifications with U$(1)\times$U(1) 
gauge symmetry on elliptically fibered Calabi-Yau manifolds with a rank
two Mordell-Weil group. We find that the natural presentation of an elliptic curve 
$\mathcal{E}$ with two rational points and a zero point is the generic Calabi-Yau onefold in $dP_2$.
We determine the birational map to its Tate and Weierstrass form and the coordinates
of the two rational points in Weierstrass form.  We discuss its resolved elliptic fibrations over a 
general base $B$ and classify them in the case of $B=\mathbb{P}^2$. A thorough  analysis
of the generic codimension two singularities of these elliptic Calabi-Yau manifolds is presented. 
This determines  the general U$(1)\times$U(1)-charges of matter in corresponding
F-theory compactifications.  The matter multiplicities for the fibration over $\mathbb{P}^2$ are determined explicitly and shown 
to be consistent with anomaly cancellation. Explicit toric examples are constructed, both with U$(1)\times$U(1)
and SU$(5)\times$U$(1)\times$U(1) gauge symmetry. As a by-product, we  prove the birational equivalence 
of the two elliptic fibrations with elliptic fibers in the two blow-ups $Bl_{(1,0,0)}\mathbb{P}^2(1,2,3)$ and  
$Bl_{(0,1,0)}\mathbb{P}^{2}(1,1,2)$ employing birational maps and extremal transitions.

\hfill {March, 2013}
\end{titlepage}

\tableofcontents

\addtolength\topmargin{50pt}
\addtolength\textheight{-105pt}

%
\section{Introduction and Summary of Results}

In recent years significant progress has been made on the construction of phenomenologically 
appealing models of particle physics from F-theory compactifications, initiated by the local model approach of 
\cite{Donagi:2008ca,Beasley:2008dc,Beasley:2008kw,Donagi:2008kj}. 
The road to the construction of semi-realistic compact F-theory GUTs models with SU$(5)$ or SO$(10)$
gauge group has been paved in \cite{Blumenhagen:2009yv,Marsano:2009wr,Chen:2010ts,Grimm:2009yu,Knapp:2011ip}, 
although a model passing most phenomenological tests has yet to be constructed.
Our understanding of both the local approach as well its embedding into global compactifications has 
improved vastly over the last years \cite{Heckman:2010bq,Weigand:2010wm,Maharana:2012tu}.
Since the development of F-theory \cite{Vafa:1996xn,Morrison:1996na,Morrison:1996pp} 
the construction of non-Abelian gauge symmetry in F-theory compactifcations  is well understood due to the full classification 
of codimension one singularities of elliptically fibered Calabi-Yau manifolds \cite{kodaira1963compact,tate1975algorithm,Bershadsky:1996nh}.  
A recent reconsideration and careful analysis has closed remaining gaps in the understanding of singularities at higher codimension  
\cite{Esole:2011sm,Marsano:2011hv,Lawrie:2012gg}. Constructive algorithms for the  explicit construction of non-Abelian gauge 
symmetries in toric Calabi-Yau manifolds are developed  \cite{Candelas:1996su,Candelas:1997eh} and have recently been employed for an 
exhaustive exploration of 6d gauge symmetries from F-theory on toric elliptic fibrations over 
$\mathbb{P}^2$ \cite{Braun:2011ux}.

In contrast to this, the construction of Abelian gauge symmetries in F-theory is much less understood. One reason 
is the dependence of this question on the global geometry of the elliptic fibration.  However, aspects of the physics 
of  U(1)-symmetries can already be studied in local F-theory models employing spectral cover methods 
\cite{Donagi:2009ra,Marsano:2009gv,Marsano:2009wr,Dudas:2009hu,Cvetic:2010rq,Dudas:2010zb,Dolan:2011iu,Marsano:2012yc}. 
F-theory compactifications with an Abelian gauge theory sector arise from
compactifications on elliptic Calabi-Yau manifolds $X$ with general fiber being an elliptic curve 
with rational points. Elliptic curves $\mathcal{E}$ with a so-called non-trivial Mordell-Weil group of rational points 
are a classical subject in mathematics \cite{neron1964modeles,shioda1989,shioda1990mordell,Wazir:2001,silverman2009arithmetic}.
These rational points lift to rational sections of the fibration $X$, that contribute 
new harmonic two-forms to the cohomology that support Abelian gauge fields in the F-theory 
effective action. The number of Abelian gauge fields is set by the rank of the Mordell-Weil 
group of the elliptic curve and its torsion subgroup gives rise to non-simply laced groups 
\cite{Morrison:1996pp,Aspinwall:1998xj,Aspinwall:2000kf}. 
Rank one Mordell-Weil groups in compact elliptic fibrations have been
studied recently in the F-theory literature in varying contexts 
\cite{Grimm:2010ez,Braun:2011zm,Krause:2011xj,Grimm:2011fx,Morrison:2012ei,Cvetic:2012xn,Mayrhofer:2012zy,Braun:2013yti}.

The gauge theory arising from an F-theory compactification changes significantly in 
the presence of  U(1)-gauge symmetries. Additional codimension two singularities
support singlet fields only charged under the U(1)-gauge fields
and the charged matter fields of the non-Abelian sector  split 
into multiple fields differing by their U(1)-charges. 
Furthermore, the structure of Yukawa-couplings  are dominated by U(1)-selection rules.
This additional structure has rich phenomenological implications in Beyond the Standard Model (BSM)
model building, where it can be used to avoid dangerous proton decay operators, engineer realistic Yukawa 
textures or provide a solution
to the $\mu$-problem \cite{Marsano:2009wr,Dudas:2009hu,Hayashi:2010zp,Grimm:2010ez,Marsano:2010sq,Dolan:2011iu}.

To broaden the applicability of F-theory to BSM physics it is desirable to
construct Abelian sectors of higher rank in F-theory. For example, conservative  SM-extensions
by only a single U(1),  e.g.~in models $G_{SM}\times$U(1) with a $Z'$-boson,  
have an Abelian sector U$(1)\times $U$(1)_Y$ of rank two and would require the understanding
of rank two Mordell-Weil groups in F-theory. In addition, a classification of the possible
Abelian gauge sector in F-theory analogous to the well-studied non-Abelian 
sector is lacking. However, see \cite{Grassi:2012qw} for a systematic study of rational sections on toric  
K3-surfaces. We make some progress in this direction by constructing F-theory compactifications 
with a U$(1)\times$U(1)-sector\footnote{For a study of Tate models with multiple rational 
points at $z=0$ see \cite{Mayrhofer:2012zy}.}. 

\subsection{Summary of Results}

In this work we construct for the first time an elliptic curve $\mathcal{E}$ and a resolved elliptic 
Calabi-Yau fibration $\hat{X}$ with Mordell-Weil (MW) group of rational sections of rank two.
We  systematically  derive by purely algebraic methods the concrete presentation of the elliptic curve $\mathcal{E}$ as the Calabi-Yau 
onefold in the toric
del Pezzo surface $dP_2$\footnote{We note that a model with elliptic fiber in $dP_1$ has been
studied in  \cite{Braun:2013yti}.}.  An F-theory compactification on $\hat{X}$ will have two U$(1)\times$U(1)-gauge symmetry and charged matter.
We determine the full matter spectrum including matter multiplicities for the first time and show consistency with six-dimensional anomaly 
cancellation. The general procedure employed here can readily be generalized to  higher rank MW-groups, and might be 
relevant for the classification of the possible Abelian sectors in F-theory. 

It is the pivotal result of our analysis that it is necessary   to depart from the Weierstrass and Tate form to study
the curve $\mathcal{E}$ and its corresponding elliptic fibrations.
The reasons for this can be summarized by the following key findings of our analysis: \vspace{-0.4cm}
\begin{itemize}
	 \item[i)]  \textbf{Abandoning the paradigm of a holomorphic zero section:} General  models with two rational sections require that the zero 
	 section is only a rational, not a holomorphic section. \vspace{-0.2cm}
	\item[ii)] \textbf{Departure from Tate/Weierstrass form:}  The zero section in the Tate or Weierstrass model is always holomorphic. Thus
	we have to work in a realization of elliptic fibrations that naturally incorporate a rational zero section such as toric geometry.\vspace{-0.2cm}
	\item[iii)] \textbf{Toric elliptic curves:} Toric geometry greatly facilitates the construction of some higher rank MW-groups. 
	Rational sections are directly realized as toric divisors\footnote{This fact has also been  exploited in  \cite{Grassi:2012qw,Morrison:2012ei,Mayrhofer:2012zy,Braun:2013yti}.}.\vspace{-0.2cm}
	\item[iv)] \textbf{Birational maps:}  Due to i) the map of the toric elliptic curve iii) with rational zero section to the Weierstrass
	model ii) with holomorphic zero section is rational. \vspace{-0.6cm}
\end{itemize}
We emphasize that it is also imperative to present $\mathcal{E}$  as the hypersurface in $dP_2$ because the Weierstrass model of
the elliptic fibration $\hat{X}$ will be generically singular. The birational map from the Weierstrass form to $dP_2$ will automatically 
resolve all singularities in any codimension, which is in accord with the smoothness of generic toric Calabi-Yau 
hypersurfaces \cite{Batyrev:1994hm}. Thus, the understanding of the rational map involved is also the key to analyze 
higher codimension singularities which determine the F-theory matter spectrum.

 In the following we  summarize the key points of  our program leading to the construction of the curve $\mathcal{E}$ in $dP_2$ and the analysis  
 of the associated elliptic fibrations $\hat{X}$, its rational sections, its Tate\footnote{As will be evident from the concrete exposition in section 
 \ref{sec:construct3points}, a global Tate model might not exist for all elliptic fibrations.} and Weierstrass form and codimension two 
 singularities.
\begin{itemize}
	\item We first derive the representation of an elliptic curve $\mathcal{E}$ over a field $K$ with two rational points
	$Q$, $R$ and a zero point $P$. To this end we consider a degree three line bundle $M=\mathcal{O}(P+Q+R)$ with holomorphic
	sections $u$, $v$ and $w$ and find a cubic relation among these holomorphic sections.  This naturally leads to an embedding of the elliptic 
	curve $\mathcal{E}$ as the non-generic cubic curve in 
	$\mathbb{P}^2$, which is resolved into the Calabi-Yau onefold in the del Pezzo surface 
	$dP_2$. We determine the birational map of this elliptic curve in $dP_2$ to its Tate form and Weierstrass form with respect
	to the zero point $P$. This birational map allows us to determine 
	the coordinates of the rational points $Q$ and $R$ in Weierstrass form.
	Then we construct an elliptically fibered Calabi-Yau threefold $\mathcal{E}\rightarrow \hat{X}\rightarrow B$
	over a base $B$ with general fiber being the elliptic curve $\mathcal{E}$ in $dP_2$. The point $P$ becomes the
	zero section $\hat{s}_P$ of the fibration and the rational points $Q$, $R$ lift to rational sections $\hat{s}_R$, $\hat{s}_Q$.
	\item We find the matter content of an F-theory compactification on $\hat{X}$ by analyzing the codimension two 
	singularities of the elliptic fibration, by deriving the Weierstrass representation of the fibration and the rational sections. 
	The resolved fiber in $dP_2$ over all codimension two loci is a reducible $I_2$ curve, i.e.~the elliptic 
	curve $\mathcal{E}$ splits into two rational curves $\mathcal{E}=c_1+c_2$ intersecting in two points. This can  
	be viewed as the Dynkin diagram of an extended SU$(2)$ group arising in
	codimension two. The behavior of the rational sections is crucial for the analysis of matter. 
	In the generic elliptic fibration with
	our $dP_2$-fiber we identify six different codimension two loci. Along three loci one of the sections $\hat{s}_P$, $\hat{s}_Q$
	and $\hat{s}_R$ are ill-behaved and we have to resolve $\hat{X}$ in the base $B$ along the
    corresponding loci. In the resolved space the section wraps an entire fiber component. Physically, these
	loci are distinguished as the loci of matter of more exotic U$(1)\times$U(1)-charges with charges $(-1,1)$, $(-1,-2)$
	and $(0,2)$. Further matter is located whenever one of our two rational sections $\hat{s}_Q$, $\hat{s}_R$ intersect the zero 
	section $\hat{s}_P$. These degenerations contribute matter of charges $(1,0)$ and $(0,1)$ depending on which section touches 
	the zero section. Finally, a third type of singularity arises when  $\hat{s}_Q$ and $\hat{s}_R$ coincide and touch 
	the zero section $\hat{s}_R$. This is still a codimension two phenomenon leading to matter of charge $(1,1)$.
	\item Finally, we determine the multiplicities of matter fields. We demonstrate that this counting is comparably involved
	since the various codimension two loci corresponding to different matter fields intersect. Concretely, assuming matter
	of charges $(-1,1)$, $(-1,-2)$ and $(0,2)$ is located at a collection of points $loc_1$ and matter with charges $(1,0)$, $(0,1)$ and 
	$(1,1)$ is located at points $loc_2$, we find that $loc_2$ is automatically obeyed at $loc_1$, i.e.~$loc_1\cap loc_2\neq 0$.
	The appropriate multiplicity of the latter matter fields at $loc_2$ is then determined by subtracting the multiplicities of the former matter fields
	at $loc_1$, taking into account the order of vanishing of $loc_2$ along $loc_1$. This complication also arose in \cite{Morrison:2012ei}.
	We show that the appropriate treatment involves the application of the resultant of $loc_2$ with $loc_1$ as its root.
\end{itemize}

We would also like to emphasize that we find all six-dimensional anomalies to be canceled for the matter charges
and multiplicities found. This gives us some evidence for the completeness of our
analysis of codimension two singularities of the fibration $\hat{X}$. 
Since the matter content is a codimension two phenomenon, we expect
our results for the matter charges to hold in general, in particular for Calabi-Yau fourfolds. 

We construct three global models as toric Calabi-Yau hypersurfaces with base $\mathbb{P}^2$.
We present two different models with U$(1)\times$U(1) gauge group, that differ in the complexity of the behavior
of the rational sections $\hat{s}_Q$, $\hat{s}_R$. In the first model both sections never intersect the zero sections and
only charge $(1,0)$, $(0,1)$, $(1,1)$ matter is realized. In the second model we allow for a non-trivial behavior of the 
section $\hat{s}_R$ and confirm the existence of all six different matter multiplets. In both cases we confirm by anomaly cancellation
the consistency and completeness of our analysis. We explicitly write down
the 4d polytope realizing a model with SU$(5)\times$U$(1)^2$ gauge group. We observe that 
matter in the $\mathbf{5}$-representation splits into five matter
representation $\mathbf{5}_{(q_1,q_2)}$ differing by their U$(1)^2$-charges. The $\mathbf{10}$ matter
curves does not split. 

Finally we exploit the use of birational geometry to study extremal transitions between elliptic fibrations 
of two different fiber types \cite{Klemm:1996hh}  yielding two equivalent, but different descriptions of the
same F-theory vacua with one U(1). 
We perform the extremal transition from the elliptic curve in $\mathbb{P}^{2}(1,2,3)$, which is the standard Tate form, 
to the curve in $Bl_{(1,0,0)}\mathbb{P}^{2}(1,2,3)$, which is still in Tate form, but with Tate coefficient $a_6\equiv 0$. 
Ultimately, simple toric techniques  allow us to readily obtain the birational map from the latter model to the resolved quartic curve in 
$Bl_{(0,1,0)}\mathbb{P}^{2}(1,1,2)$, which immediately reproduces 
the map found in \cite{Morrison:2012ei}. Extending this birational map to elliptic fibrations, this thereby establishes the birational 
equivalence of the U(1)-restricted Tate-model \cite{Grimm:2010ez}  
and the fibration with elliptic curve in $Bl_{(0,1,0)}\mathbb{P}^{2}(1,1,2)$. 
We emphasize, however, that this equivalence only holds for elliptic fibrations of a certain class.  

This paper is organized as follows. In section \ref{sec:FTheoryWithMW} we introduce the basic concepts and constructions for elliptic fibrations with
a Mordell-Weil group for F-theory compactifications with an Abelian gauge theory sector. In section \ref{sec:Fibrations_with_2secs}
we derive the form of the general elliptic Calabi-Yau manifold $\hat{X}$ with two rational sections. For this purpose we
first derive the presentation of an elliptic curve $\mathcal{E}$ over a field $K$ with two rational points as the hypersurface in $dP_2$ 
and then readily generalize to Calabi-Yau elliptic fibrations, that we classify for base $B=\mathbb{P}^2$. We elucidate the structure of the 
resolved geometry of the elliptic curve $\mathcal{E}$ in $dP_2$. In section 
\ref{sec:Matter} we study in detail the general matter spectrum of the F-theory compactification on the resolved threefold $\hat{X}$. We show that 
both its singular Weierstrass model as well as the resolution to $\hat{X}$ with $dP_2$-fiber are crucial to 
understand the structure of all codimension two singularities. We determine both U$(1)\times$U(1)-charges and matter
multiplicities. In section \ref{sec:anomalies} we check that our spectrum is in general consistent with anomaly cancellation in 6d.
We present three concrete toric realizations of elliptic threefolds $\hat{X}$ over $B=\mathbb{P}^2$  in section \ref{sec:examples}. 
Two examples have U$(1)\times$U(1) gauge symmetry whereas we add an SU(5)-GUT sector in a third example.
We conclude by an application of birational maps in section \ref{sec:birationalmodels}  to show the equivalence of elliptic
fibrations with fibers of two different types, both of which with rank one MW-group. Section \ref{sec:conclusions} contains our
outlook. We have two appendices \ref{app:detailsOf2U1model} and \ref{app:NagellsAlgorithm} with
details on the Weierstrass model of the elliptic curve $\mathcal{E}$ and Nagell's algorithm of the cubic.

While we were preparing this manuscript the paper \cite{Borchmann:2013jwa} appeared which has some overlap with
our discussion.

 \section{Basics of Abelian Gauge Sectors in F-theory}
 \label{sec:FTheoryWithMW}

In this section we review the notion of the Mordell-Weil group as the
group of rational points on an elliptic curve in section \ref{sec:ellipticCurvesWithRP}. Then we 
discuss the geometry of Calabi-Yau manifolds formed as elliptic fibrations 
with general fiber given by such an elliptic curve  in section \ref{sec:CYwithMW}. The rational
points are lifted to rational sections of the elliptic fibration and contribute 
additional cycles to the homology group of these manifolds. The Abelian sector of an
F-theory compactification on these elliptically fibered Calabi-Yau manifolds, as explained in section 
\ref{sec:AbelianSector}, is determined by the structure of these rational sections. 
The reader interested in the results of our analysis can skip this section
and directly proceed with section \ref{sec:construct3points}.
 
 \subsection{Elliptic Curves with Multiple Rational Points}
 \label{sec:ellipticCurvesWithRP}

To set the stage for our later discussion, we begin with the most concrete definition of an elliptic 
curve $\mathcal{E}$ over a field $K$ in terms of a Weierstrass model. 
The Tate form of the Weierstrass model reads
\beq \label{eq:Tateform}
	y^2+a_1 x y z +a_3 y z^3=x^3+a_2 x^2 z^2+a_4xz^4+a_6z^6
\eeq	
with $a_i$ denoting numbers in a field $K$. In elliptic fibrations of $\mathcal{E}$ over a base $B$, $K$ will be 
identified with the function field of $B$.  In the chosen projectivization \eqref{eq:Tateform} is a one-dimensional 
Calabi-Yau hypersurface in the weighted projective space $\mathbb{P}^2(1,2,3)$ with homogeneous coordinates $[z:x:y]$.
The Tate form can be brought into the reduced Weierstrass form  
\beq \label{eq:WSF}
		\tilde{y}^2=\tilde{x}^3+f\tilde{x}z^4+gz^6
\eeq
by the variable transformation
\beq \label{eq:varTrafoToWSF}
	\tilde{x}=  x+\tfrac{1}{12}b_2 z^2\,,\qquad \tilde{y}= y+\tfrac12 a_1 x z+\tfrac12 a_3z^3
\eeq
with the following definitions
\bea\label{eq:defsWSF}
	f&=&-\tfrac{1}{48}(b_2^2-24 b_4)\,,\qquad g=-\tfrac{1}{864}(-b_2^3 + 36 b_2 b_4 -216 b_6)\,,
	\nn\\
	b_2&=&a_1^2+4a_2\,,\qquad b_4=a_1 a_3+2 a_4\,,
	\qquad b_6=a_3^2+4 a_6\,,\nn\\
	\Delta&=&-16(4f^3+27 g^2)= -8b_4^3+\tfrac{1}{4}b_2^2b_4^2+9b_2 b_4 b_6-
	\tfrac{1}{4}b_6 b_2^3 - 27 b_6 \,.
\eea
Here the quantity $\Delta$ defined in the last line of 
\eqref{eq:defsWSF} is the discriminant of the Weierstrass equation
\eqref{eq:WSF}. If $\Delta=0$ then the elliptic curve defined by 
\eqref{eq:WSF} is singular.

Both the Tate form 
\eqref{eq:Tateform} as well as the reduced Weierstrass form \eqref{eq:WSF} have 
one distinguished point $P$ at $[z:x:y]=[0:1:1]$, referred to as the zero point.
In general, a rational point on the elliptic curve $\mathcal{E}$ is defined as a point
with coordinates in the designated field $K$.  
The set of rational points on $\mathcal{E}$ form an Abelian group under addition that is
naturally defined in the Weierstrass form \eqref{eq:WSF}. The group of rational
points, with $P$ as the zero, is the \textit{Mordell-Weil group} of the curve $\mathcal{E}$. 
The Mordell-Weil group is finitely generated, i.e.~it is the sum of a torsion subgroup
and $\mathbb{Z}^r$,  and $r$ is the Mordell-Weil rank. We introduce a basis of the 
torsionless subgroup of the Mordell-Weil group as $Q_m$, $m=1,\ldots, r$.

We note that the presence of a rational point can imply a factorization of the Weierstrass
form. A case of particular interest later in this work is a point $Q$ of the form $[z:\tilde{x}:\tilde{y}]=[1:A:B]$
for given numbers $A$, $B$ in $K$. In this case, the Weierstrass equation \eqref{eq:WSF}
factorizes as
\beq \label{eq:factorz=1}
	(\tilde{y}-Bz^3)(\tilde{y}+Bz^3)=(\tilde{x}-Az^2)(\tilde{x}^2+A\tilde{x}z^2+Cz^4)\,,
\eeq
as is easily checked by plugging in the point $Q$. This implies that the coefficients $f$, $g$
in \eqref{eq:WSF} as well as the discriminant in \eqref{eq:defsWSF} can be parametrized as
\beq \label{eq:factorz=1_fg}
	f=C-A^2\,,\quad g=B^2-A C\,,\quad \Delta=16 (27B^2(2A C-B^2) + (A^2-4 C)(2A^2+C)^2)
\eeq
This parametrization will prove useful in the discussion of matter with only charge 1
under the Abelian gauge field corresponding to the rational point $Q$ \cite{Morrison:2012ei}, 
i.e.~codimension two singularities only due to the presence of the point $Q$ in the fiber $\mathcal{E}$ alone. 
We add that factorization properties of the Tate form \eqref{eq:Tateform} due to rational points 
of the form $z=0$ have been discussed in \cite{Mayrhofer:2012zy}.

 \subsection{Elliptic Calabi-Yau Manifolds with Rational Sections}
 \label{sec:CYwithMW}

In this section we briefly discuss the geometry of an elliptically fibered Calabi-Yau threefold $X$,
although we note that the following holds for general complex dimension of $X$.

By definition an elliptic fibration over a base $B$ is defined  by a holomorphic projection $\pi:\,X\rightarrow B$ 
to the base $B$.  We will be interested in fibrations with general fiber $\mathcal{E}=\pi^{-1}(pt)$ over a generic 
point $pt$ in $B$ given by elliptic curve with a zero point and a number of rational points.
As mentioned before we can always describe an elliptic fibration over $B$ by its Weierstrass model
\eqref{eq:WSF}, where the field $K$ is replaced by the function field of $B$. 
Concretely, by the Calabi-Yau condition $f$, $g$ are sections of the line bundles $K_{B}^{-4}$ respectively  
$K_{B}^{-6}$, where $K_B$ denotes the canonical bundle of the base. If it exists globally we can 
also construct the Tate form\footnote{See \cite{Katz:2011qp} for a reconsideration of the validity of the Tate model
in global F-theory compactifications.} \eqref{eq:Tateform} where the coefficients $a_i$ take values in $K_B^{-i}$.
The holomorphic zero section in the Tate and Weierstrass form is given by $z=0$.

However, when having global questions in mind, such as the global resolution of singularities of the elliptic 
fibration $X$ or the construction of rational sections, it is of advantage to consider elliptic fibrations $X$ with 
the general elliptic fiber $\mathcal{E}$ given as the Calabi-Yau hypersurface in one of the 16 two-dimensional 
toric varieties. It is always possible, as discussed at the end of section \ref{sec:ellipticCurvesWithRP}, to 
obtain the Weierstrass form of these fibrations by a birational map. In this note we will focus on the 
elliptic fibration with general fiber in $dP_2$, cf.~section \ref{sec:construct3points}.

When we fiber an elliptic curve $\mathcal{E}$ over a base $B$ its zero point $P$ becomes the zero section 
$\hat{s}_{P}$ and its rational points $Q_m$ lift to rational sections $\hat{s}_m\equiv \hat{s}_{Q_m}$ of the 
elliptic fibration $\pi:\,X\rightarrow B$. 
All of these sections define injective maps $\hat{s}_P,\hat{s}_m:\,B\hookrightarrow X$ and the group generated by the $\hat{s}_m$ is
the Mordell-Weil group of the elliptic fibration $X$. 
A holomorphic section, that is in the literature typically denoted by $\sigma$, 
defines a holomorphic injection $\sigma:\,B\hookrightarrow X$ on all of $B$.
However, a rational section does in general not vary holomorphically over the base $B$. 
Indeed, over codimension two or higher the rational section can be ill-defined and wrap components of
the reducible fiber over the singular loci \cite{Morrison:2012ei,Braun:2013yti}. Thus, rational sections 
$\hat{s}_m$ can only be defined on the  blow-up $\pi_B:\,\hat{B}\rightarrow B$ of $B$ along the singular 
loci of $\hat{s}_m$ with $\pi_B$ denoting the blow-down map. Consequently, a rational section defines only a birational map 
$\hat{s}_m:\,\hat{B}\rightarrow B\hookrightarrow X$ of the base $B$ into $X$.

In sections \ref{sec:construct3points} and \ref{sec:Matter}, we study
elliptic fibrations with general fiber given by an elliptic curve $\mathcal{E}$ with two rational points $Q$,
and $R$ and a zero section $P$. As we see there, in these cases even the zero section $\hat{s}_P$ is not 
defined over codimension two and wraps fiber components of the elliptic fiber. A Calabi-Yau 
elliptic fibration without a holomorphic zero section still defines a valid F-theory background, although
these most general fibrations have only recently drawn attention in the F-theory literature. As we discuss in 
section \ref{sec:Matter}, the behavior of the three rational sections $\hat{s}_P$ and $\hat{s}_Q$, $\hat{s}_R$ 
leads to a rich structure of charged matter.

The group of divisors, or its dual group $H^{(1,1)}(\hat{X})$, on the smooth elliptic Calabi-Yau manifold 
$\hat{X}\rightarrow X$ arising from $X$ by resolving all singularities, is generated by divisors $D_A$ 
that fall into four different classes of divisors:\vspace{-0.3cm}
\begin{itemize}
\item the zero section $\hat{s}_P$ of the fibration with homology class $S_P$, which in the case of a holomorphic section $\sigma$ agrees with the 
class of the base $B$,\vspace{-0.2cm}
\item the rational sections $\hat{s}_m$, $m=1,\ldots, r$, with divisor classes $S_m$  generating the 
Mordell-Weil group of rational sections of the elliptic fibration of $X$,\vspace{-0.2cm}
\item the vertical divisors $D_\alpha=\pi^*(D_\alpha^{\rm b})$, $\alpha=1,\dots, h^{(1,1)}(B)$, of the fibration 
that are inherited from divisors $D_\alpha^b$ in the base $B$,\vspace{-0.2cm}
\item exceptional divisors $D_{i_I}$ resolving singularities of $X$ from singularities in its elliptic fibration at
irreducible components $\Delta_{I}=0$ of the discriminant locus $\Delta=0$ in $B$.\vspace{-0.1cm}
\end{itemize}
We summarize this basis of divisors on $\hat{X}$ as 
\beq \label{eq:basisH^11}
	D_A=(B,S_m,D_\alpha,D_i)\,, \quad A=0,1,\ldots,h^{(1,1)}(\hat{X})\,.
\eeq  

We conclude with some key intersection properties of the divisors $D_A$.
The exceptional divisors $D_{i_I}$ over one common discriminant locus intersect as
\beq \label{eq:CartanMatrix}
	D_{i_I}\cdot D_{i_J}\cdot D_{\alpha}=-C^{(I)}_{i_Ij_I} S_{(I)}\cdot B\cdot D_{\alpha}
\eeq
where $C^{(I)}_{i_Ij_I}$ denotes the Cartan matrix of an ADE-group $G_{(I)}$ in case of an ADE-singularity
in the fibration. In this case we denote the $D_{i_I}$ as the Cartan divisors of the corresponding ADE group.
Here the divisors $S_{(I)}=\pi^*(S_{(I)}^{\rm b})$ are related to the components of the discriminant
and the $S_{(I)}^{\rm b}$ are the loci in the base $B$ wrapped by 7-branes supporting a gauge group $G_{(I)}$ 
in F-theory. Upon intersecting the Cartan 
divisors $D_{i_I}$ with a divisor $\tilde{D}$ in $B$ that intersects $S_{(I)}$ in a point,
we obtain a rational curve that is localized over $S_{(I)}$ . 
The relation \eqref{eq:CartanMatrix} teaches us that this curve is to be identified with
minus the root $\alpha_{i_I}$ of the Lie-algebra of the ADE-group under consideration, i.e.
\beq \label{eq:C_alpha}
	\mathcal{C}_{-\alpha_{i_I}}:= D_{i_I}\cdot \tilde{D}\,\qquad \text{for   }\quad\, 
	\tilde{D}\cdot S_{(I)}\cdot B=1\,.
\eeq
Next we turn to the divisors $S_P$ and $S_m$ of the sections. By construction
the divisors corresponding to a section will intersect the general fiber $F\cong \mathcal{E}$ as
\beq \label{eq:intsSectionsF}
	S_P\cdot F=S_m \cdot F=1\,.
\eeq 
In addition we note that any section of $X$ also has to obey \cite{Morrison:2012ei}
\beq \label{eq:intsSections}
	B^2=-[c_1(B)]\cdot B\,\,,\qquad S_m^2\cdot D_\alpha=-[c_1(B)]\cdot S_m\cdot D_\alpha\,,
\eeq 
where the first relation holds in homology and $c_1(B)$ is the first Chern class of the 
base.
 
 \subsection{Abelian Gauge Sectors in F-Theory}
 \label{sec:AbelianSector}

In an F-theory compactification on $X$ gauge fields $A^a$ arise by expanding the 
M-theory three-form $C_3$ in the dual M-theory compactification\footnote{See \cite{Grimm:2010ks}
for a general discussion of this and a derivation of the full F-theory effective action.} along appropriate 
$(1,1)$-forms $\omega_a$,
\beq \label{eq:expC3}
	C_3=\sum_a A^a\omega_a\,.
\eeq
These are gauge fields along the Cartan generators of all $G_{(I)}$ and along U(1) groups.
The U(1)-gauge fields correspond to rational sections $\hat{s}_m$ 
and are constructed from the cohomology classes dual to the divisors $S_m$ of the 
rational sections by the Shioda map. 

The Shioda map roughly speaking describes an orthogonalization procedure
in the (co)homology group of $\hat{X}$. It is defined as the map from the Mordell-Weil
group to $H^{(1,1)}(\hat{X})$ given by \cite{Morrison:2012ei,Park:2011ji,Cvetic:2012xn}
\beq \label{eq:ShiodaMap}	
	\sigma(\hat{s}_m) := S_m-\tilde{S}_P-(S_m\cdot\tilde{S}_P\cdot D_\alpha)
	\eta^{\alpha\beta}	D_\beta+
	\sum_{I}(S_m\cdot \mathcal{C}_{-\alpha_{i_I}})
	(C_{(I)}^{-1})^{i_Ij_I}D_{j_I}\ , 
\eeq 
where the curves $\mathcal{C}_{-\alpha}$ have been defined in \eqref{eq:C_alpha}
and $C_{(I)}^{-1}$ is the inverse of the Cartan matrix determined in \eqref{eq:CartanMatrix}.
The divisor $\tilde{S}_P=S_P+\tfrac{1}{2}\pi^*c_1(B)$ has been introduced for convenience,
and is of relevance for the match of the F- and M-theory dual effective actions 
\cite{Grimm:2011sk,Bonetti:2011mw}. The matrix $\eta^{\alpha\beta}$ is the inverse of the intersection
form of the divisors on the base,
\beq
	\eta_{\alpha\beta}=D^{\rm b}_\alpha\cdot D^{\rm b}_\beta\,.
\eeq
The Shioda map \eqref{eq:ShiodaMap} maps into the orthogonal complement\footnote{The inner product is the
N\'{e}ron-Tate height pairing \cite{Wazir:2001}.} in $H^{(1,1)}(\hat{X})$ generated by
the zero section $S_P$, the vertical divisors $D_\alpha$ and the Cartan divisors $D_{i_I}$.
Thus it ensures that the gauge field associated to $\sigma(\hat{s}_m)$ by the reduction
\eqref{eq:expC3} along the $(1,1)$-form $\omega_m$ defined as the Poincare dual of 
$\sigma(\hat{s}_m)$ is a U(1)-gauge field.

Having defined the Shioda map \eqref{eq:ShiodaMap} it is straightforward to calculate 
U(1)-charges of matter fields in F-theory. The matter fields arise from M2-branes wrapping rational curves $c$
in the fiber of the elliptic fibration that are localized in codimension two in the base $B$. 
Under the assumption of a holomorpic zero section with $S_P=B$, the charge of the M2-brane state under the 
U(1)-gauge field corresponding to the rational section $\sigma_m$ is then calculated as 
\cite{Morrison:2012ei}
\beq \label{eq:U1charge}
	\sigma(\hat{s}_m)\cdot c=(S_m\cdot c)+\sum_I(S_m\cdot \mathcal{C}_{-\alpha_{i_I}})
	(C_{(I)}^{-1})^{i_Ij_I}(D_{j_I}\cdot c)\,.
\eeq
This follows readily using the definition of the Shioda map \eqref{eq:ShiodaMap} and
the fact, that the isolated curves $c$ neither intersect 
vertical divisors nor the holomorphic zero section. 

We note that the intersection matrix between
the $\sigma(\hat{s}_m)$ projected into the homology of the base by $\pi:\,\hat{X}\rightarrow B$ is
\beq \label{eq:anomalycoeff}
	\pi(\sigma(\hat{s}_m)\cdot\sigma(\hat{s}_n))=\pi(S_m\cdot S_n)+[K_B]-\pi(S_m\cdot B)-\pi(S_n\cdot B)
	+(\mathcal{C}_{(I)}^{-1})^{i_I j_I}(S_m\cdot \mathcal{C}_{-\alpha_{i_I}})(S_n\cdot \mathcal{C}_{-\alpha_{j_I}})S_{(I)}^{\rm b}\,.
\eeq
Here, we introduced the definition
\beq \label{eq:height_pairing}
	\pi(\mathcal{C})=(\mathcal{C}\cdot D_\alpha)\eta^{\alpha\beta}D^{\rm b}_\alpha
\eeq
of the N\'{e}ron-Tate height pairing for a curve $\mathcal{C}$ and employed the normalized coroot matrix
of the $I$-th 7-brane gauge group
\beq
	\mathcal{C}^{(I)}_{i_{I}j_I}=\frac{2}{\lambda_I\langle\alpha_{i_I},\alpha_{j_I}\rangle}C^{(I)}_{i_Ij_I}\,.
\eeq
In this expression we denote the inner product on the Lie-algebra by $\langle\cdot,\cdot\rangle$ and introduced
the length squared $\lambda_I=\frac{2}{\langle \alpha_0,\alpha_0\rangle}$ of the maximal root $\alpha_0$.
The intersections \eqref{eq:anomalycoeff} will be relevant for the discussion of anomaly
cancellation in section \ref{sec:anomalies}.

We conclude by noting that if the zero section $\hat{s}_P$ is only rational, the formula for the
U(1)-charge has to be modified. The reason for this is that we have to relax the condition $B\cdot c=0$
and in general assume a non-trivial intersection
\beq
	S_P\cdot c\neq 0\,,
\eeq
where $S_P$ denotes the homology class of the rational zero section. Then \eqref{eq:U1charge} has to 
be replaced by
\beq \label{eq:U1charge1}
	\sigma(\hat{s}_m)\cdot c=(S_m\cdot c)-(S_P\cdot c)+\sum_I(S_m\cdot \mathcal{C}_{-\alpha_{i_I}})
	(C_{(I)}^{-1})^{i_Ij_I}(D_{j_I}\cdot c)\,.
\eeq
In contrast, the formula \eqref{eq:anomalycoeff} for the intersections of sections does not have 
to be modified. 
We note that \eqref{eq:U1charge1} has also been used in \cite{Braun:2013yti}. 

\section{Elliptic Fibrations with Two Rational Sections}
\label{sec:Fibrations_with_2secs}

In this section we determine an elliptic fibration with two rational 
sections and a zero section, i.e.~an elliptic curve with Mordell-Weil 
group of rank two. This curve serves as the model for the general
elliptic fiber in an elliptically fibered Calabi-Yau manifold. As discussed in section \ref{sec:FTheoryWithMW} 
the  F-theory compactification on such a Calabi-Yau admits two U(1)-gauge groups.

We first find in section \ref{sec:construct3points} that any elliptic curve $\mathcal{E}$ with three 
marked points (two rational and the zero point) has  a representation as a non-generic cubic 
in $\mathbb{P}^2$. We argue further that this elliptic curve should be properly 
viewed as the generic Calabi-Yau onefold in the del Pezzo surface $dP_2$, which is the blow 
up of $\mathbb{P}^2$ at two generic points. However, we first focus on the 
singular model of $\mathcal{E}$ by neglecting the blow-ups in $\mathbb{P}^2$. 
This allows us to compute the Weierstrass model for $\mathcal{E}$
with respect to one of the three rational points on it
and derive the location of the two other rational points in Weierstrass 
coordinates. 

Then, in section \ref{sec:blowupgeo}, we discuss the resolution 
geometry of $\mathcal{E}$ in $dP_2$ that we obtain by performing the two 
blow-ups in $\mathbb{P}^2$. The understanding of the resolved geometry is relevant
to resolve elliptic fibrations with general fiber being the curve $\mathcal{E}$.
We conclude by constructing elliptic fibrations with $\mathcal{E}$ and find
that all elliptic fibrations with base $B=\mathbb{P}^2$ are classified by two
integers $n_2$ and $n_{12}$. We will consequently denote the families of 
$dP_2$-fibrations over $\mathbb{P}^2$ we have found as $dP_2(n_2,n_{12})$.

Our discussion partly follows and extends the techniques of 
appendix B of \cite{Morrison:2012ei}.

\subsection{Constructing an Elliptic Curve with two Rational Points}
\label{sec:construct3points}

Our discussion in this section is based on the following basic 
mathematical fact:  Given an algebraic variety X with a very 
ample line bundle\footnote{A very ample line bundle is a line bundle 
that has ``enough'' global sections so that its base variety is 
embedable into projective space.} $M$ defined over it, we can find  an embedding of $X$ into projective 
space. If this is the case, we will have enough independent global sections 
$a_0, \cdots, a_n$ such that for every point $x\in X$ 
there exists at least one section not vanishing at this point. Thus, 
there is an immersion
\beq
 f:\,\,\,\, X \rightarrow \mathbb{P}^n\,\qquad x \mapsto [a_0(x): \cdots : a_n(x)]\,
\eeq
with $M\cong f^*(\mathcal{O}(1))$. 

The algebraic variety we are 
interested in is an elliptic curve $\mathcal{E}$ over a field $K$. We find 
the embedding of the curve $\mathcal{E}$ into projective space, or more generally 
a toric variety, explicitly as a hypersurface. For this purpose we 
consider the global sections of powers of $M^k$. For sufficiently 
large $k$, in our case $k=3$, not all sections are independent and the relation between them yield the 
desired hypersurface equation. 

As a warm-up we use this logic to derive the Tate form \eqref{eq:Tateform} of an 
elliptic curve $\mathcal{E}$ with only the zero point $P$ as follows. 
We introduce the degree one line bundle $M=\mathcal{O}(P)$ over $\mathcal{E}$
and consider the homogeneous  coordinates $[z:x:y]$ on $\mathbb{P}^2(1,2,3)$ as 
sections of the line bundles $M$, $M^2$ and $M^3$, respectively.
We recall that in the case of an elliptic curve $\mathcal{E}$ the Riemann-Roch 
theorem tells us that the number of independent global holomorphic sections of a 
line bundle $M$ of degree $d$ is $h^0(X,M)=d$. Then the bundle
$M^6$ has six independent holomorphic sections, however, we can construct seven sections from $z$, $x$,
$y$. Thus, there has to be a relation between these  sections which
yields precisely the Tate form \eqref{eq:Tateform} in $\mathbb{P}^{2} (1,2,3)$.

\begin{figure}[ht!]
\centering
\begin{tabular}{cc}
 \includegraphics[scale=0.4]{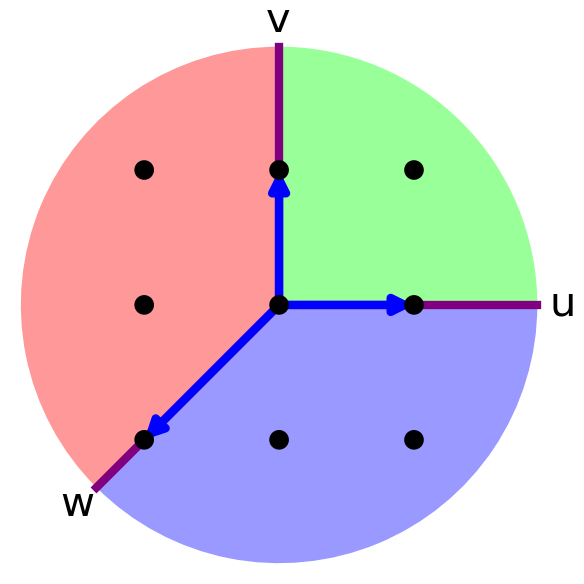} &\hspace{2cm}
 \includegraphics[scale=0.3]{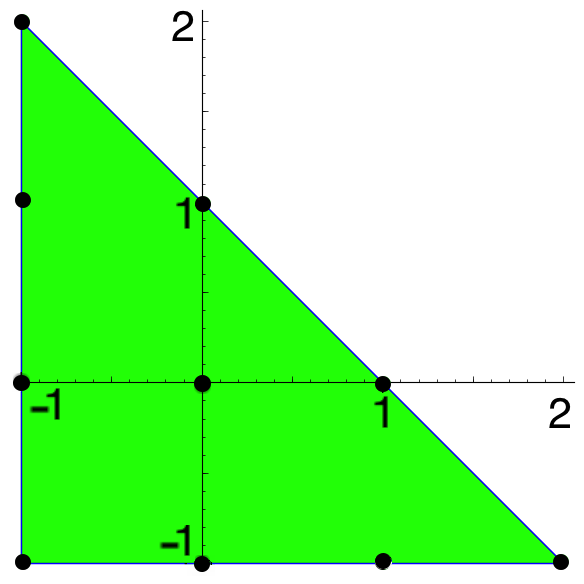} 
 \end{tabular}
 \caption{Fan of $\mathbb{P}^2$ on the left and its dual polytope on the right.}
 \label{fig:p2poly}
\end{figure}

The same  strategy applies to finding the equation that describes an elliptic curve
with rational points $Q_m$. Concretely, we first find the sections $[z:x:y]$ of 
$\mathcal{O}(kP)$, $k=1,2,3$, and then determine the relation between sections of higher
degree generated from  $[z:x:y]$.  We perform this
procedure in the following for an elliptic curve $\mathcal{E}$ with a zero point $P$ and two
rational points denoted $Q$ and $R$.
In the same spirit as before we first start with a general line 
bundle $M$ of degree three that we then specialize to 
$M=\mathcal{O}(P+Q+R)$. The group of holomorphic section $H^0(M)$ is 
generated by three sections denoted $u,v,w$. The space $H^0(2M)$ has 
dimension six with sections $u^2$, $v^2$, $w^2$, $uv$, $vw$ and $wu$. 
The space $H^0(3M)$ must have nine independent sections, however we know ten of them. 
Consequently, there has to be a linear relation of the form
\begin{equation}
 s_1 u^3 + s_2 u^2 v + s_3 u v^2 + s_4 v^3 + s_5 u^2 w + s_6 uvw 
 + s_7 v^2 w + s_8 u w^2 
 + s_9 v w^2 + s_{10}w^3=0\,,
 \label{eq:uvwp2}
\end{equation}
with coefficients $s_i$ in the field $K$. This equation can be 
interpreted as the cubic hypersurface in $\mathbb{P}^2$. Even more it is 
a section of its anti-canonical bundle $\mathcal{O}(3H)$, see figure 
\ref{fig:p2poly}, where $H$ denotes the hyperplane class in 
$\mathbb{P}^2$. Thus the zero of this section defines the 
one-dimensional Calabi-Yau manifold, i.e.~the torus\footnote{There is a
one-to-one correspondence between elliptic curves and two-tori.}.

\begin{figure}[ht!]
\centering
\begin{tabular}{cc}
 \includegraphics[scale=0.4]{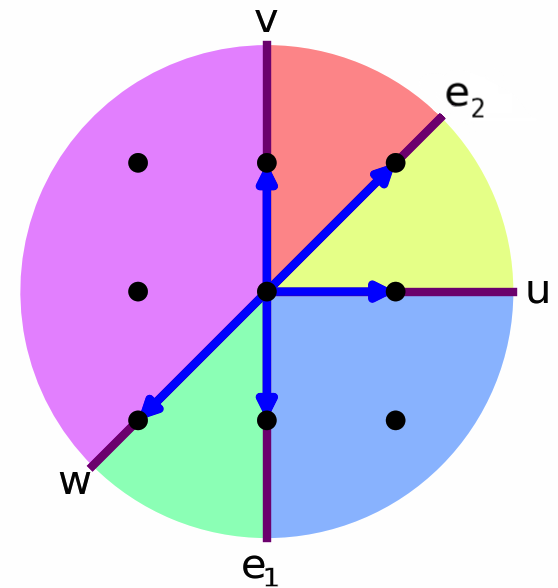} &\hspace{2cm}
 \includegraphics[scale=0.3]{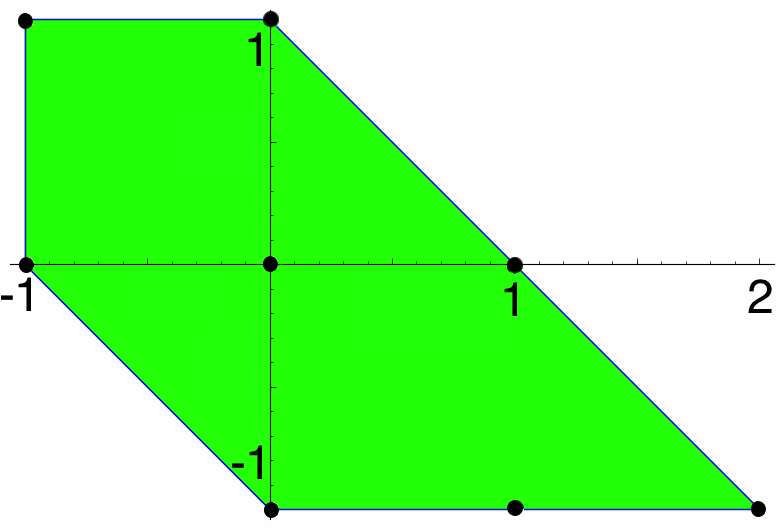}
 \end{tabular}
 \caption{Fan  of  $dP_2$ on the left and its dual polytope on the right.}
 \label{fig:dp2poly}
\end{figure}

Now we specialize to $M=\mathcal{O}(P+Q+R)$. Let us assume that the 
section $u$ vanishes at the three points $P$, $Q$ and $R$, then equation 
\eqref{eq:uvwp2} simplifies to
\beq \label{eq:eq@u=0}
  s_4 v^3 + s_7 v^2 w +  s_9 v w^2 + s_{10}w^3=0\,.
\eeq
Performing appropriate shifts of the coordinates $v$ and $w$ we can 
always get rid of the coefficients $s_4$ and $s_{10}$. However, these 
variable transformations involve square roots
of the coefficients $s_i$ that are generically not defined over the 
field $K$. Thus, we specialize the constraint \eqref{eq:uvwp2}
by setting the coefficients $s_4$ and $s_{10}$ to zero. This specialization
can also be viewed as changing the toric ambient variety such that the 
coefficients $s_4$ and $s_{10}$ are automatically absent by means of the
toric construction. This is achieved by going from $\mathbb{P}^2$ to $dP_2$, 
which is the blow-up of $\mathbb{P}^2$ at two generic points. In figure \ref{fig:dp2poly} 
we have depicted the polytope of $dP_2$ for the  blow-up at $[u:v:w]=[0:0:1]$ 
and $[u:v:w]=[0:1:0]$. We note that the blow-ups introduces new exceptional divisors $E_i$ 
with coordinates $e_i$, $i=1,2$. 

This latter perspective on the specialization of the constraint \eqref{eq:uvwp2}
as the Calabi-Yau hypersurface in $dP_2$ is of particular relevance for the construction 
of smooth elliptic fibrations in F-theory. However, the blown-up geometry $dP_2$
is of no importance for the determination of the Weierstrass model of the curve
elliptic curve $\mathcal{E}$ with the rational points $Q$ and $R$, with which we proceed 
in the remainder of this section. Thus, we discuss the resolved geometry of $dP_2$
with all divisor classes including the exceptional divisors $E_i$ separately in section \ref{sec:blowupgeo} 
and work for the following in the patch
$e_1=1$, $e_2=1$.

The restricted hypersurface following from \eqref{eq:uvwp2} with $s_4=s_{10}=0$ takes now the form
\begin{equation}
 p\equiv s_1 u^3 + s_2 u^2 v + s_3 u v^2  + s_5 u^2 w + s_6 uvw 
 + s_7 v^2 w + s_8 u w^2 
 + s_9 v w^2 =0\,.
 \label{eq:uvwdp2}
\end{equation}
Then, this hypersurface constraint specializes at $u=0$ to
\beq
  vw ( s_7 v  +  s_9 w )=0\,,
\eeq
which vanishes at the three different points with coordinates 
\begin{equation}
P=[0:0:1], \,\,\,\,\, Q=[0:1:0],  \,\,\,\,\, R=[0:s_9:-s_7]\,.
\label{eq:points-uvw}
\end{equation}
We emphasize that the points $P$ and $Q$ coincide with precisely those points at which we blow up $\mathbb{P}^2$
into $dP_2$, see section \ref{sec:blowupgeo} for a more detailed discussion.

We note that we could set even more coefficients $s_i$ to zero or to one by using the automorphisms
of the ambient space, e.g.~by shifting or rescaling variables and the equation, if we were interested in the 
properties of the elliptic curve alone. For example the shift symmetry of $w$ by an appropriate factor of $u$
is still unbroken in the presence of the rational points $Q$ and $R$ since these are left invariant. 
For example, we can perform the transformation $w\rightarrow w^{\prime} - \frac{s_5}{2s_8} u$ to eliminate the term $u^2 w$ in \eqref{eq:uvwdp2}\footnote{The coefficients will take the form
\begin{equation}
 s_1^\prime= s_1 - \frac{s_5^2}{4 s_8}, \,\,\,\,\
 s_2^\prime= s_2 - \frac{s_5 s_6}{2s_8} + \frac{s_5^2 s_9}{4 s_8^2}, \,\,\,\,\
 s_3^\prime= s_3 - \frac{s_5 s_7}{2s_8}, \,\,\,\,\
 s_6^\prime= s_6 - \frac{s_5 s_9}{s_8}, \\ 
\end{equation}
with the other coefficients unchanged.}. 
However, we use later in section \ref{sec:examples} the elliptic curve 
defined by \eqref{eq:uvwdp2} as the general elliptic fiber in an elliptically fibered Calabi-Yau manifold. 
Then the coefficients $s_i$ are lifted to sections over the base of the fibration. In this case the 
embedding of the fiber into the Calabi-Yau manifold will in general forbid eliminating more $s_i$ than those
in \eqref{eq:uvwdp2}. This can be seen explicitly in the toric example constructed in section \ref{sec:examples}.
Therefore, we work with the most general form \eqref{eq:uvwdp2} of an elliptic curve with two rational points.

To obtain the Weierstrass equation with respect to $P$, we need to find the three sections 
of $H^0(kP)=H^0(kM-kQ-kR)$ for $k=1,2,3$. These sections are precisely the projective coordinates 
$z$, $\tilde{x}$, $\tilde{y}$ of the Weierstrass model in $\mathbb{P}^{2}(1,2,3)$. For the purpose of constructing
these sections we determine sections of $H^0(kM)$ on the ambient space $dP_2$ that simultaneously 
vanish $k$ times both at $Q$ and $R$\footnote{We note that this somewhat counter-intuitive fact follows from the fact that 
we implicitly divide by a function $g$ vanishing at $P+Q+R$. Indeed, we recall the definition of the line bundle associated to a 
Cartier divisor $D$, $\mathcal{O}(D):=\{\frac{f}{g}\vert \text{div}(\frac{f}{g})+D\geq 0\}$ for local 
functions $f$, $g$, cf.~\cite{griffiths2011principles}.  Applying this for $\mathcal{O}(P+Q+R)$ we obtain that
$g$ has three zeros precisely at $P$, $Q$, $R$, with $f$ of degree three vanishing elsewhere. For $\mathcal{O}(P)$ we obtain 
$g$ has to vanish at $P$ and $f$ is of degree one. Then, sections of $\mathcal{O}(P)$ are a subset 
of sections of $\mathcal{O}(P+Q+R)$ by demanding that 
$f$ in the latter vanishes precisely at $Q$ and $R$.} when restricted to $\mathcal{E}$. 

The general derivation of this birational transformation from the cubic \eqref{eq:uvwdp2} to the
Tate model \eqref{eq:Tateform} and then the Weierstrass form \eqref{eq:WSF} is quite lengthy.
Therefore we first summarize the results and present the proof in the remainder of the section
for the interested reader. The birational map from the projective coordinates $[u:v:w]$ of $dP_2$
to $[z:\tilde{x}:\tilde{y}]$
in the Weierstrass model \eqref{eq:WSF} reads
\bea \label{eq:xyzWSFdP2}
   z&=&  u\,,\nn\\
   \tilde{x}&=&s_3 s_9 u v +( s_6 s_9- 
 s_7 s_8 )u w  + s_7 s_9 v w + s_9^2 w^2+(\tfrac{1}{12}s_6^2 - \tfrac{1}{3} s_3 s_8-\tfrac{1}{3}s_5 s_7+ \tfrac{2}{3} s_2 s_9 )z^2  \nn\\
 \tilde{y}&=& \tfrac{1}{2} \left[ s_3 (s_6 s_9-2 s_7 s_8  ) u^2v + 
    (s_6^2 s_9 -s_6 s_7 s_8 + 2 s_9 (-s_5 s_7 - s_3 s_8 + s_2 s_9))u^2 w\right. \nn\\ &&  + ( s_6 s_7 s_9-2s_7^2 s_8 + 2 s_3 s_9^2 ) u v w+( 3 s_6 s_9^2-4 s_7 s_8 s_9) u w^2+ s_7 s_9^2 v w^2 + 2 s_9^3 w^3
 \nn \\ 
   && \left.   +(s_2 s_6 s_9-s_2 s_7 s_8 - s_3 s_5 s_9  - s_1 s_7 s_9) z^3\right] \,.
\eea
The sections defined like this obey the Weierstrass constraint \eqref{eq:WSF},
\beq \label{eq:WSFdP2curve}
	\tilde{y}^2=\tilde{x}^3+f \tilde{x}z^4+g z^6
\eeq
 with the parameters $f$ and $g$ determined 
via \eqref{eq:defsWSF} in terms of the three polynomials
\bea \label{eq:dP2WSF}
 	b_2&=&s_6^2 - 4 s_5 s_7 - 4 s_3 s_8 + 8 s_2 s_9\,,\nn\\ 
b_4&=&2 s_2^2 s_9^2 + s_1 s_7 (2 s_7 s_8 - s_6 s_9) + 
 s_2 ( s_6^2 s_9-s_6 s_7 s_8 - 2 s_5 s_7 s_9)  \nn \\
& &+s_3 (2 s_5 s_7 s_8 - s_5 s_6 s_9 - 2 s_2 s_8 s_9 + 2 s_1 s_9^2)\,,\nn\\ 
b_6&=&(s_2 s_7 s_8 + s_3 s_5 s_9 - s_2 s_6 s_9 + s_1 s_7 s_9)^2\nn\\ 
&&-4 s_1 s_3 (s_7^2 s_8^2 + s_9^2 (s_3 s_8 - s_2 s_9) +  s_7 s_9 (s_5 s_9-s_6 s_8 ))\,.  
\eea
We have summarized the somewhat lengthy expressions for $f$, $g$ along with the Tate coefficient $a_i$ and the
discriminant in the appendix \ref{app:detailsOf2U1model}.

Along similar lines outlined below, we obtain the coordinates for the generators of the Mordell-Weil group of $\mathcal{E}$ 
given by the rational points $Q$ and $R$ in Weierstrass form. The result reads
\beq
 Q=[\tilde{x}_Q :\tilde{y}_Q:z_Q]=   [\tfrac{1}{12} (s_6^2 - 4 s_5 s_7 + 8 s_3 s_8 - 4 s_2 s_9),\tfrac{1}{2} (s_3 s_6 s_8 - s_2 s_7 s_8 - s_3 s_5 s_9 + s_1 s_7 s_9):1] 
 \label{eq:MWGQ}
\eeq
for $Q$ and analogously for $R=[\tilde{x}_R:\tilde{y}_R:z_R]$,
\bea
  \tilde{x}_R &=&  \tfrac{1}{12} (12 s_7^2 s_8^2 + s_9^2 (s_6^2 + 8 s_3 s_8 - 4 s_2 s_9) + 
 4 s_7 s_9 (-3 s_6 s_8 + 2 s_5 s_9))\,, \nn \\ 
     \tilde{y}_R&=& \tfrac{1}{2} (2 s_7^3 s_8^3 + s_3 s_9^3 (-s_6 s_8 + s_5 s_9) + 
 s_7^2 s_8 s_9 (-3 s_6 s_8 + 2 s_5 s_9)  \nn \\ 
 & &+s_7 s_9^2 (s_6^2 s_8 + 2 s_3 s_8^2 - s_5 s_6 s_9 - s_2 s_8 s_9 + s_1 s_9^2)\,, \nn \\ 
    z_R&=& s_9  \,.
    \label{eq:MWGR}
\eea

In the remainder of the section we derive the final results \eqref{eq:xyzWSFdP2} and \eqref{eq:dP2WSF}. 
The reader less interest in these details can safely skip to section \ref{sec:blowupgeo}.
Beginning with $H^0(M)$, we already know that the section $u=0$, by  assumption, vanishes 
at the three points $P$, $Q$ and $R$.  Thus we set
\begin{equation}
 z:= u\,.
\end{equation}
Next we need two find a section of $H^0(2M)$ with double zeros at $Q$ and $R$ on $\mathcal{E}$. There are 
two such sections, one of which is given by the section  $u^2=z^2$. The other one is constructed 
from an appropriate linear combination of the other elements of $H^0(2M)$. We make the ansatz
\begin{equation} \label{eq:ansatzx}
 x:= a v^2 + b uv + c w^2 + d v w + e uw\,.
\end{equation}
The coefficients are fixed requiring that the section $x$ vanishes at degree two on both points
$Q$ and $R$. This gives us four equations. The other coefficient can be rescaled away using the scaling of $x$. 

A nice pictorial way to think about these conditions on the coefficients in $x$ is to realize 
that $x=0$ and $p=0$ in \eqref{eq:uvwdp2} are two curves in the ambient space  $dP_2$. Then the first 
two of the four conditions on the coefficients enforce that $x$ vanishes at the points $P$, $Q$, 
which are also automatically the intersections of $p=0$ and $x=0$. The other two conditions on the 
coefficients then make $Q$, $R$ a double zero or equivalently let the curves intersect tangentially.

A systematic way of finding the coefficients is to solve for one of the variables $u$, $v$, $w$
in one of the equations $p=0$, $x=0$ and then to use the other one to determine the unknown coefficients 
in \eqref{eq:ansatzx} in order to obtain the right order of vanishing. First we note that both $Q$, and $R$
can be described in the affine patch $v=1$. To solve for one of the remaining variables $u$, $w$ 
instead of dealing with radicals we can approximate the curve $\mathcal{E}$ by a Taylor expansion. 
We solve for $w\equiv w(u)$ order by order in $u$ by considering $f(u,1,w(u))=0$ as an implicit function 
for $w$ in terms of $u$. Expanding the curve $p=0$ first around $Q$, then around $R$ with coordinates 
\eqref{eq:points-uvw} we obtain, omitting terms of  third order and higher in $u$,
\bea
 &\hspace{-0.1cm}Q:&\!\!w =  - \frac{s_3}{s_7} u - \frac{1}{s_7} \Big( s_2 - \frac{s_3 s_6}{s_7} +\frac{s_3^2 s_9}{s_7 ^2}  \Big) u^2 
 \,,\hspace{-0.6cm} \label{eq:w@Q}\nn\\
 &\hspace{-0.1cm}R:&\!\! w = - \frac{s_7}{s_9} + \Big( \frac{s_3}{s_7} + \frac{s_7 s_8 - s_6 s_9}{s_9^2} \Big) u 
 + \Big( \frac{s_2}{s_7}-\frac{s_3 s_6}{s\,_7^2}  + \frac{s_3^2 s_9}{s_7^3} - \frac{s_7 s_8^2 - s_6 s_8 s_9 + s_5 s_9^2}{s_9^3}  \Big) u^2.\nn\hspace{-0.6cm}\label{eq:w@R}\\
\eea
Now we can pull back the section $x$ to the curve $\mathcal{E}$ by simply plugging the solution for $w(u)$ 
into the ansatz \eqref{eq:ansatzx} for $x$. Since we require an order of vanishing of two we have to
set the coefficients of $u^0$ and $u^1$ to zero. Employing the expansion \eqref{eq:w@Q} at $Q$ yields
\beq
 a = 0\,,\qquad
 b - \frac{ s_3}{s_7}d = 0\,.
\eeq
Proceeding similarly around $R$ using \eqref{eq:w@R}, we obtain two further equations
\beq
 a+  \frac{s_7^2}{s_9^2}c-  \frac{s_7}{s_9}d =0\,,\qquad 
 b + \frac{s_7^2 s_8-s_6 s_7 s_9+s_3 s_9^2}{s_7 s_9^2}d+ 2\frac{ s_6 s_7 s_9-s_7^2 s_8-s_3s_9^2}{s_9^3}c  - \frac{s_7}{s_9} e= 0\,.
\eeq
Solving these equations, setting $c=1$ by rescaling and cleaning denominators in $x$, we finally get the section
\begin{equation} \label{eq:solx}
 x = s_3 s_9 uv + (s_6 s_9-s_7 s_8) uw + s_7 s_9 vw + s_9^2 w^2  \,.
\end{equation}

Next, we need three elements of $H^0(3M)$ with triple zeros at $Q$ and $R$ on the curve $\mathcal{E}$. We know ten  sections of this bundle, only nine are 
independent and we already know two combinations that vanish appropriately, namely $u^3$ and $u x$. The other 
possible section has to be a linear combination of the remaining seven sections and we make the ansatz
\begin{equation}
 y := a u^2 v +  b v^3 + c w^3 + d v^2 w + e u w^2 + f  v w^2 + g u v w\,. 
\end{equation}
Proceeding in the same way as before,  we use the solutions \eqref{eq:w@Q} 
to pull the section $y$ to the curve $\mathcal{E}$, but now require that it vanishes at order $u^3$ both at $Q$ and $R$.
This yields three conditions for each point, i.e. a total of six conditions, by demanding that the 
terms $u^0$, $u^1$ and $u^2$ are absent. Setting $c=1$ by rescaling $y$ and cleaning some denominators, we obtain 
\begin{align} \label{eq:soly}
 y =&  \frac{s_3 (s_7^2 s_8^2 -  s_6 s_7 s_8 s_9 +  s_5 s_7 s_9^2 + s_3 s_8 s_9^2 - 
 s_2  s_9^3)}{ 
  s_6 s_9-s_7 s_8 } u^2 v   \nn \\
  &+\frac{s_7^3 s_8^2 - s_6 s_7^2 s_8 s_9 + s_5 s_7^2 s_9^2 + s_3 s_6 s_9^3 - s_2 s_7 s_9^3 }{s_6 s_9-s_7 s_8 }  u v w   + s_7 s_9^2 v w^2 \nn \\
 &+\frac{s_9(2 s_7^2 s_8^2  - 3 s_6 s_7 s_8 s_9 + s_6^2 s_9^2 + s_5 s_7 s_9^2 + 
 s_3 s_8 s_9^2 - s_2 s_9^3) }{  s_6 s_9-s_7 s_8 } u w^2 + s_9^3 w^3 .
\end{align}

Now that we know the form of the sections $x$, $y$ we  can determine the 
Tate model \eqref{eq:Tateform}
by plugging in the solutions \eqref{eq:solx}, \eqref{eq:soly}  and solving for the Tate coefficients $a_i$. We note
that in order to find a solution we have to reduce the Tate form, which is a degree six polynomial
in $u$, $v$, $w$, in the ideal generated by the original cubic constraint \eqref{eq:uvwdp2}. We present the results for the $a_i$ in
\eqref{eq:TatecoeffsdP2} of appendix \ref{app:detailsOf2U1model} from which we immediately calculate the coefficients
$b_2$, $b_4$ and $b_6$ in \eqref{eq:dP2WSF}. From the Tate model, the Weierstrass form \eqref{eq:WSF} can be obtained by the variable 
transformation \eqref{eq:varTrafoToWSF} that takes the form of \eqref{eq:xyzWSFdP2} as claimed.
We readily obtain the functions $f$ and $g$ in the Weierstrass model and the discriminant and refer to appendix \ref{app:detailsOf2U1model} 
for their explicit form.

Next, we focus on finding the rational points $Q$ and $R$ in Weierstrass form. Recalling that there is an implicit 
multiplication by a meromorphic function with poles of degree $k$ in each case, we obtain the coordinates for $Q$ ($R$) if we 
look for a section $x'$ of degree two that vanishes three times 
at  $Q$ ($R$). We have to include all elements of $H^0(2M)$:
\begin{equation}
 x' := a v^2 + b uv + c w^2 + d v w + e uw + \tilde{f} u^2,.
\end{equation}
The new parameter $\tilde{f}$ is fixed by the additional constraint from the vanishing of the coefficient of $u^2$, while 
all the other coefficients take the same values as before. Noting that $z\equiv u$, we then obtain 
\begin{equation}
 x' = x - (s_3 s_8 - s_2 s_9) z^2 \stackrel{!}{=} 0\,
\end{equation}
at $Q$ which fixes the $x$-coordinate of $Q$.
After scaling and going to the Weierstrass coordinates we obtain the generator as in \eqref{eq:MWGQ}, where the $\tilde{y}$-coordinate
of $Q$ is obtained by solving \eqref{eq:WSFdP2curve} for $\tilde{y}$ given $\tilde{x}_Q$.
Following the same procedure for the point $R$ we confirm the result \eqref{eq:MWGR}.

\subsection{Resolved Elliptic Curve in $dP_2$ and its Elliptic Fibrations}
\label{sec:blowupgeo}

In this section we discuss the key geometric properties of the
resolution of the elliptic curve \eqref{eq:uvwdp2} in $dP_2$. To this end
we have to take into account the exceptional divisors $E_i$ and their
sections $e_i$, $i=1,2$ that we have set to one in the previous discussion. 
Then, smoothness of $\mathcal{E}$ is ensured by the toric construction since the curve 
\eqref{eq:uvwdp2} lifts to the generic Calabi-Yau onefold in $dP_2$.  
The smoothness property even holds in elliptic fibrations $X$ with general fiber
given by $dP_2$, if the most generic toric Calabi-Yau hypersurface is considered. 
Indeed, as we discuss explicitly in section \ref{sec:Matter} 
the divisors $E_1$, $E_2$ resolve the singularities of the fibration at codimension two 
and higher. In particular, this allows us to determine the matter spectrum in 
F-theory from the splitting of the elliptic curve in $dP_2$ over codimension two from the 
presence of the new projective coordinates $e_i$.

We construct $dP_2$ as follows. We blow up $\mathbb{P}^2$ at 
the two points $[u:v:w]=[0:1:0]$ and $[u:v:w]=[0:0:1]$ by introducing
two new projective coordinates $e_1$, $e_2$ as
\beq \label{eq:blowdownmap}
	u=u' e_1 e_2\,,\qquad v=v' e_2\,,\qquad w=w' e_1\,.
\eeq
This is the blow-down map $\pi:\,dP_2\rightarrow \mathbb{P}^2$, where
$e_1$, $e_2$ are holomorphic sections vanishing at 
the exceptional divisors $E_1$, $E_2$. We see that the 
original codimension two loci $u=w=0$, respectively, $u=v=0$ in 
$\mathbb{P}^2$ are replaced by $e_1=0$, respectively, $e_2=0$, i.e.~by 
entire divisors $E_i$ with the geometry of $\mathbb{P}^1$. We summarize 
the toric realization of $dP_2$ encoded in its polytope $\Delta_{dP_2}$ 
and its homogeneous coordinates  as
\beq \label{eq:poldP2}
	\begin{array}{|r|rr||c|c|}
	\hline
	& \multicolumn{2}{c||}{\text{vertices}} & & \text{divisor class}\\
	\hline
		\nu_1&1& 0&  u'& H-E_1-E_2\\ 
		\nu_2&0& 1& v'& H-E_2\\ 
		\nu_3&-1&-1&  w'& H-E_1\\
		\nu_4&0& -1& e_1& E_1\\
		\nu_5&1& 1&  e_2& E_2\\ 
		\hline
	\end{array}\,.
\eeq
Here we denote the vertices of $\Delta_{dP_2}$ by $\nu_i$ in the first 
row and write them explicitly as two-dimensional row-vectors in the 
second and third column as depicted in the polytope in figure \ref{fig:dp2poly}.
The only integral internal point is the origin. In the third and fourth 
column we summarize the homogeneous coordinates on $dP_2$ and their 
divisor classes, where $H$ is the pullback of the 
hyperplane class on $\mathbb{P}^2$ and the $E_i$
are the exceptional divisors.
This in particular implies that the anti-canonical 
bundle $K^{-1}_{dP_2}$ of $dP_2$, which is computed as the negative of the sum of 
all toric divisors, has changed to $K^{-1}_{dP_2}=\mathcal{O}(3H-E_1-E_2)$, compared to 
$K_{\mathbb{P}^2}^{-1}=\mathcal{O}(3H)$ before the blow-ups.

We note the intersections in the star triangulation of 
$\Delta_{dP_2}$ with star given by the origin as
\beq \label{eq:dP2ints}
	H^2=1\,,\qquad H\cdot E_i=0\,,\qquad E_i\cdot E_j=-\delta_{ij}\,.
\eeq
Here we made use of the exceptional set, the Stanley-Reissner ideal $SR$ 
of all vertices not sharing a two dimensional cone, reading 
\beq \label{eq:SRidealdP2}
	SR=\{u'v',\,u'w',\,e_1 e_2,\,e_1 v',\,e_2 w'\}\,.
\eeq

We observe that the polytope of $\mathbb{P}^2$ is embedded by the
first three points $\nu_1$, $\nu_2$ and $\nu_3$. The addition of the 
vertices $\nu_4$, $\nu_5$ not only introduces the divisors $E_i$
but also removes the corners $(-1,2)$ and $(-1,-1)$ of the dual 
polytope of $\mathbb{P}^2$, compare figures \ref{fig:p2poly} and \ref{fig:dp2poly}. 
These corners correspond precisely to the monomials $s_4 v^3$ and 
$s_{10}w^3$ in $\mathbb{P}^2$, that are no longer admissible sections of 
the anti-canonical bundle on $dP_2$. As one can see by either looking at 
dual polytope \ref{fig:dp2poly} or by writing down all sections of the anti-canonical 
bundle $\mathcal{O}(3H-E_1-E_2)$ of $dP_2$ by hand with the charge assignments
in \eqref{eq:poldP2}, the generic Calabi-Yau hypersurface $p'$ on $dP_2$ reads
\beq \label{eq:CYindP2}
	p'\!\!=\! u' (s_1u'^2e_1^2e_2^2 + s_2 u' v' e_1 e_2^2 + s_3  v'^2e_2^2 
	+ s_5 u' w' e_1^2e_2+ s_6 v'w'e_1e_2 + s_8 w'^2 e_1^2)
	+ s_7 v'^2 w'e_2 + s_9 v' w'^2e_1\,.
\eeq

Now we investigate the pullback under $\pi$ of the points $P$, $Q$ and 
$R$ in \eqref{eq:points-uvw} that we by abuse of notation denote by the 
same symbols. We readily infer from the map 
\eqref{eq:blowdownmap} that the section $u\in\mathcal{O}(P+Q+R)$ splits 
into three components as expected and we identify $P$ as $e_2=0$, $Q$ 
as $e_1=0$ and the image of $R$ as $D_{u'}:=\{u'=0\}$. The value of all coordinates 
in the notation $[u':v':w':e_1:e_2]$ is obtained by inserting these 
components into the hypersurface equation \eqref{eq:CYindP2} and setting 
all coordinates to $1$ that do not intersect the component under 
consideration due to the exceptional set \eqref{eq:SRidealdP2}. This yields
\bea \label{eq:coordsPQRdP2}
	&P:\,\,E_2\cap p'=[-s_9:s_8:1:1:0]\,,\quad Q:\,\,E_1\cap p'=[-s_7:1:s_3:0:1]\,,\nn&\\
	&R:\,\,D_{u'}\cap p'=[0:1:1:-s_7:s_9]\,&
\eea
for the coordinates of the points $P$, $Q$ and $R$ on the resolved 
curve in $dP_2$. 

We emphasize the dependence of the points $P$, $Q$, $R$ on the $s_i$, in particular 
the situation when certain $s_i$ vanish. 
Note that it is manifest in the form \eqref{eq:coordsPQRdP2}, that the points $P$, $Q$ and $R$ are 
ill-defined if $s_i=s_j=0$ simultaneously for the combinations $(i,j)$ in \eqref{eq:coordsPQRdP2} due to the Stanley-Reissner
ideal \eqref{eq:SRidealdP2}.
In particular, when considering elliptic fibrations $\hat{X}$ with elliptic
fiber given by \eqref{eq:CYindP2}  this translates to a special behavior 
of the corresponding sections $\hat{s}_P$, $\hat{s}_Q$, $\hat{s}_R$ at codimension two loci $s_i=s_j=0$.
Note that the $s_i$ lift to sections on the base $B$ of the fibration $\hat{X}$, see the next paragraph, and can 
generically vanish. As we discuss in detail in the next section \ref{sec:Matter} this behavior is the key to understand the 
matter spectrum of an F-theory compactification on $\hat{X}$.

We would like to conclude by discussing the general construction of an elliptic fibration $\hat{X}$ over
a base $B$. For concreteness we focus on fibrations over $B=\mathbb{P}^2$ noting that the 
following can readily be generalized to different bases. First, we construct all
fibrations of $dP_2$ over $\mathbb{P}^2$. The total space of this fibration of toric varieties
is classified only by two integers, that we denote by $n_2$ and $n_{12}$ for reasons that will 
become clear below. We denote the total space as
\beq
	 dP_2\rightarrow dP_2(n_2,n_{12})\rightarrow \mathbb{P}^2\,.
\eeq
It is clear that the fibration $dP_2(n_2,n_{12})$ is only specified by two integers because we can
associate each projective coordinates $u'$, $v'$, $w'$, $e_1$, $e_2$
to a different line bundle $\mathcal{O}(k_iH_B)$, $i=1,\ldots, 5$ on $\mathbb{P}^2$, where $H_B$
is the hyperplane of the $\mathbb{P}^2$-base. 
However, by means of the three $\mathbb{C}^*$-actions on $dP_2$ we can always eliminate three $k_i$
and denote remaining two degrees as $n_2$, $n_{12}$. Then, a possible assignment is
\beq
	u'\in \mathcal{O}((n_2-3)H_B)\,,\qquad v'\in \mathcal{O}((n_2-n_{12})H_B)\,,
\eeq
with all other coordinates taking values in the trivial bundle on $\mathbb{P}^2$.
Next, we note that the anti-canonical bundle
of $dP_2(n_2,n_{12})$ is calculated by adjunction as
\beq
	K^{-1}_{dP_2(n_2,n_{12})}=\mathcal{O}(3H-E_1-E_2+(2n_2-n_{12})H_B)\,.
\eeq
From this and the Calabi-Yau constraint \eqref{eq:CYindP2} we can finally read off the 
degrees of the coefficients $s_i$, that now lift to sections of $\mathbb{P}^2$ as well:
\beq \label{eq:degree_si}
	\text{
\begin{tabular}{|c|c|c|c|c|c|c|c|}
\hline
	$s_1$&$s_2$&$s_3$&$s_5$&$s_6$&$s_7$&$s_8$&$s_9$\rule{0pt}{12pt}\\\hline
	$9-n_2-n_{12}$&$6-n_2$&$3+n_{12}-n_{2}$&$6-n_{12}$&$3$&$n_{12}$&$3+n_2-n_{12}$&$n_2$\rule{0pt}{10pt}\\ \hline
\end{tabular}
}
\eeq
Here the second line denotes the first Chern class or in other words the degree of the section $s_i$ 
with respect to the hyperplane class $H_B$ on the base. 

Finally, \eqref{eq:degree_si} illuminates the meaning of the integers $n_2$ and $n_{12}$. The sections
$s_7$ and $s_9$ vanish at curves of degree $n_2$ respectively $n_{12}$ in $\mathbb{P}^2$.
But as we see from \eqref{eq:coordsPQRdP2} these are precisely the loci where the points $Q$ and
$R$, now lifted to rational sections $\hat{s}_Q$, $\hat{s}_R$, respectively the points $P$,
also lifted to a section $\hat{s}_P$, and $R$ coincide. In other words, these integers calculate
the following intersections, cf.~\eqref{eq:height_pairing},
\beq \label{eq:defn2n12}
	n_2=\pi(S_P\cdot S_R)=S_P\cdot S_R\cdot H_B\,,\qquad n_{12}=\pi(S_Q\cdot S_R)=S_Q\cdot S_R\cdot H_B\,,
\eeq
where we denoted the divisor classes of the sections by capital letters following 
the conventions of section \ref{sec:CYwithMW}.

\section{Matter Spectrum: Codimension Two Singularities}
\label{sec:Matter}

In this section we determine the generic matter spectrum of a six-dimensional F-theory compactification on the  elliptic threefold $\hat{X}$ 
over $B$ with the general elliptic fiber $\mathcal{E}$ in $dP_2$ found in section \ref{sec:construct3points}. We mostly consider the
case $B=\mathbb{P}^2$, although our discussion is readily generalizable to other geometries. 

Determining the matter spectrum
requires an analysis of the  codimension
two singularities of this three-section elliptic fibration\footnote{We thank Antonella Grassi
for helpful discussions on the general study of codimension two singularities of elliptic fibrations and related properties of the discriminant.}. We 
see that the singularity structure that is generic to these fibrations
is completely governed by the behavior of the three sections. This study of the behavior of the sections can be conveniently performed
in the Weierstrass model \eqref{eq:WSFdP2curve}  of the elliptic curve $\mathcal{E}$.  
We determine the charges in three steps in subsections \ref{sec:charge1}, \ref{sec:charge11} and
\ref{sec:charge2}. The multiplicities of matter fields are found in \ref{sec:matter_multiplicities}. 
This analysis requires a careful counting of the codimension two points since the loci in the base $B$
supporting matter fields of different charges intersect. The right counting requires
subtraction of multiplicities of intersecting loci according to their mutual order of vanishing, which is
determined by the resultant of the polynomial systems determining the intersecting codimension two loci.
Since the matter is a codimension two effect, the matter charges we find immediately carry over to 
four-dimensional F-theory compactifications on Calabi-Yau fourfolds. Our analysis extends the methods 
of \cite{Morrison:2012ei} to elliptic fibrations with two-section models.

The following discussion is organized by the three qualitatively different codimension two singularities that can occur. The codimension
two loci of the first two types are readily found in the Weierstrass model, whereas the third type has to be addressed in the birationally related
$dP_2$-model of section \ref{sec:blowupgeo}.  The three types we distinguish correspond to the following behavior of the sections:
\begin{itemize}
	\item  Away from the collision of a section $\hat{s}$ with the zero section $z=0$ in the Weierstrass form, the existence of the section
	implies the factorization \eqref{eq:factorz=1} of the Weierstrass form. This factorization makes manifest the presence of conifold
	singularities in $X$. Their resolution
	leads to  matter of charge $1$ under the U(1) associated to $\sigma(\hat{s})$. We discuss the factorized Weierstrass form for the two sections 
	$\hat{s}_Q$, $\hat{s}_R$ in section \ref{sec:charge1}.
	\item  Two rational sections $\hat{s}$, $\hat{s}'$  both away from the zero section $z=0$  can collide and contribute charge 
	$(1,1)$-matter 
	under the two	associated U$(1)\times$ U(1)-symmetry.  This behavior is discussed in section \ref{sec:charge11}.
	\item  A section $\hat{s}$ can be ill-defined at codimension two. This is easily seen in the $dP_2$-model of section \ref{sec:blowupgeo}.
    The section $\hat{s}$ is ill-defined if its corresponding point \eqref{eq:coordsPQRdP2}  in the fiber passes through the 
    Stanley-Reissner ideal \eqref{eq:SRidealdP2} of $dP_2$. This happens at codimension two loci $s_i=s_j=0$ and we have to blow-up the base 
    $B$ at these loci. Our discussion in section 
    \ref{sec:charge2} is organized according to which of these combinations vanish.	 
    Then the section wraps a $\mathbb{P}^1$ fiber component, which is precisely the toric divisor in $dP_2$
    corresponding to the ray that subdivides the 2d-cone corresponding to the two coordinates in the Stanley-Reissner ideal \eqref{eq:SRidealdP2} . 
\end{itemize}

All these singularities have in common that their resolved fiber is of $I_2$-type, i.e.~an SU$(2)$-fiber with the original singular curve and 
one isolated $\mathbb{P}^1$. The matter hypermultiplets in six dimensions arise from the isolated rational curve $\mathbb{P}^1$ 
at the resolved  codimension two singularities of the Calabi-Yau manifold. The intersection structure of the resolved fibers with the sections 
$\hat{s}_P$, $\hat{s}_Q$, $\hat{s}_R$ determines the pattern of U(1)-charges of the matter.  The multiplicities of matter fields
is computed by counting the number of solutions to the codimension two constraints as demonstrated
in section \ref{sec:matter_multiplicities}. We summarize our findings in table
\ref{tab:MatterSpec}.

\begin{table}[ht!]
\begin{center}
	\begin{tabular}{|c||c|}
\hline
	$(q_1,q_2)$&Multiplicity\rule{0pt}{13pt} \\\hline  \hline
	 $(1,0)$& $54-15 n_2+n_2^2+\left(12+n_2\right) n_{12}-2 n_{12}^2$\rule{0pt}{11pt} \\ \hline
	 $(0,1)$& $54+2 \left(6 n_2-n_2^2+6 n_{12}-n_{12}^2\right)$\rule{0pt}{11pt}\\\hline
	 $(1,1)$&$54+12 n_2-2 n_2^2+\left(n_2-15\right) n_{12}+n_{12}^2$\rule{0pt}{11pt}\\\hline
	 $(-1,1)$&$n_{12} \left(3-n_2+n_{12}\right)$\rule{0pt}{11pt}\\\hline
	 $(0,2)$&$ n_2 n_{12}$\rule{0pt}{11pt}\\\hline
	 $(-1,-2)$&$ n_2 \left(3+n_2-n_{12}\right)$ \rule{0pt}{11pt}\\\hline
\end{tabular}
\caption{Matter spectrum with U$(1)\times$U(1)-charges $(q_1,q_2)$ in the first and multiplicities in the second column. The integers 
$n_2$, $n_{12}$ are defined in \eqref{eq:defn2n12}.}
\label{tab:MatterSpec}
\end{center}
\end{table}

We begin our analysis by summarizing the basics of the construction of a (singular) elliptic fibration $X$ over $B=\mathbb{P}^2$ in Weierstrass
form \eqref{eq:WSF}. We follow the general discussion of section \ref{sec:CYwithMW}. The Weierstrass form
of $X$ is considered as the Calabi-Yau hypersurface in the total space of projective bundle 
$\mathbb{P}^{2}(\mathcal{O}_B\oplus \mathcal{L}^2\oplus \mathcal{L}^3)$ with coordinates $[z:x:y]$\footnote{In contrast to 
section \ref{sec:FTheoryWithMW}, we denote here the coordinates in Weierstrass form by $(x,y)\equiv (\tilde{x},\tilde{y})$.}. 
By the Calabi-Yau condition we 
identify $\mathcal{L}=K^{-1}_{B}=\mathcal{O}_{\mathbb{P}^2}(3)$ as the anti-canonical bundle and 
$\mathcal{O}_B=\mathcal{O}_{\mathbb{P}^2}(0)$ as the trivial line bundle on $\mathbb{P}^2$. The parameters $f$ and $g$ in \eqref{eq:WSF} lift 
to sections of $\mathcal{O}_{\mathbb{P}^2}(12)$ and $\mathcal{O}_{\mathbb{P}^2}(18)$. The zero section is located at $[z:x:y]=[0:1:1]$.
The two Mordell-Weil generators \eqref{eq:MWGQ} and \eqref{eq:MWGR} are of the form
\begin{equation} \label{eq:secWSFP2}
 \hat{s} \,:\quad  [x,y,z]=[g_{2n+6},g_{3n+9},g_n]\,,
\end{equation}
where the subscript indicates the degree of the polynomials on $\mathbb{P}^2$. The intersection of the section 
$\hat{s}$ with the zero section $z=0$ in the Weierstrass model is the integer $n$, and we denote it by $n_1$, 
$n_2$ for the sections $\hat{s}_Q$ and $\hat{s}_R$ respectively\footnote{$n_1=0$ in our class of models.}. 
The intersection of $\hat{s}_Q$, $\hat{s}_R$ is denoted as $n_{12}$, cf.~\eqref{eq:defn2n12}.

\subsection{Factorized Weierstrass Form: charges $(1,0)$ and $(0,1)$}
\label{sec:charge1}

\begin{figure}[ht!]
\centering
\includegraphics[scale=0.37]{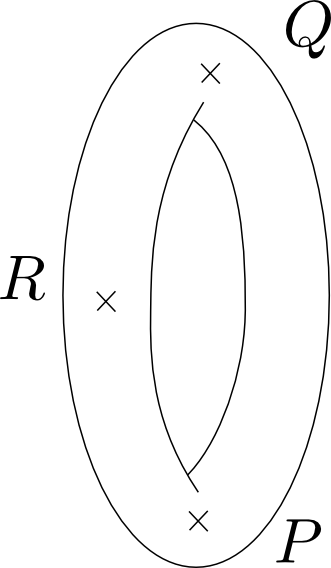} \qquad \qquad 
 \includegraphics[scale=0.55]{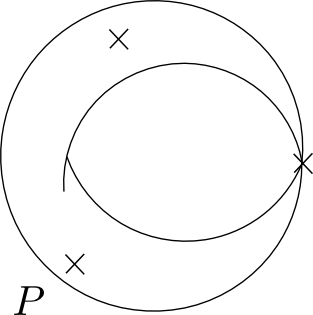} \qquad \qquad 
 \includegraphics[scale=0.5]{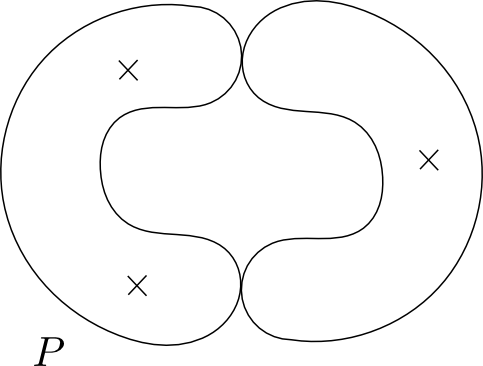}
 \caption{On the left it is shown a regular fiber with rational sections at generic points. In the center, a singular fiber is displayed with a section crossing the singularity where charged matter sits. On the right, the curve is shown after resolution. The isolated curve gives rise to an hypermultiplet charged under the corresponding Abelian gauge field.}
 \label{fig:secAtSing}
\end{figure}

Charge one loci occur when the singular point of the degenerated fiber coincides with the point marked by a section $\hat{s}$, see figure \ref{fig:secAtSing}. 
This ensures, that on the resolution the exceptional $\mathbb{P}^1$ in the fiber intersects the section 
$\hat{s}$ and render the hypermultiplet charged under the U(1) corresponding to $\hat{s}$. In our case, each of the
sections $\hat{s}_Q$, $\hat{s}_R$ can  pass separately through the singular point and its resolving $\mathbb{P}^1$, 
thus, giving rise to matter of charge $(1,0)$ respectively $(0,1)$ under  U$(1)\times$U(1). See section 
\ref{sec:charge11} for the case when both sections go through the singular point simultaneously.

We begin by spelling out the general strategy for the identification of these loci and specialize to 
$\hat{s}_Q$, $\hat{s}_R$ below. We emphasize that the structure of the  Weierstrass model of the fibration $X$ with fiber $\mathcal{E}$ with two sections  
implies a particular factorization. This facilitates the search for codimension two singularities in the fiber in $X$ that coincide with 
the point marked by the section $\hat{s}$. As we have seen in section \ref{sec:ellipticCurvesWithRP} the presence of a rational point on 
$\mathcal{E}$ with $z=1$ implies the factorization \eqref{eq:factorz=1}. Noting the form \eqref{eq:secWSFP2} of a section in Weierstrass
form, we expect this factorization where $g_n \ne 0$. Indeed, the $\mathbb{C}^*$-action on $[x:y:z]$ then allows us to cast the Mordell-Weil
(MW) generator \eqref{eq:secWSFP2} as
\begin{equation}
 \hat{s} : [x,y,z]=[g_{6},g_{9},1]\,,
\end{equation}
where we set $g_6=\frac{g_{6+2n}}{g^2_n}$, $g_9=\frac{g_{9+3n}}{g^3_n}$. Then, the Weierstrass model factors in the form
\begin{equation} \label{eq:factorizedp1}
 p_1:= -(y^2-g_9^2)+(x-g_6)(x^2+g_6x+g_{12})=0\,,
\end{equation}
with appropriate $g_{12}$, cf.~\eqref{eq:factorz=1_fg}. 

This is a singular Weierstrass model with singularities  located at the points satisfying $p_1=dp_1=0$, where $d$ 
denotes the exterior derivative. The solutions to these equations are
\begin{equation} \label{eq:conifolds}
 y=0\,, \qquad x=g_6\,, \qquad g_9=0\,, \qquad \hat{g}_{12}=g_{12}+2g_6^2=0\,.
\end{equation}
We readily confirm that the discriminant \eqref{eq:factorz=1_fg} vanishes quadratically at $g_9=\hat{g}_{12}=0$. The 
singularities are conifold singularities\footnote{In the case that $g_n=1$ these are the $108$ points in the 
$\mathcal{T}_0$ theory found in \cite{Morrison:2012ei}.} 
that are resolved by the birational transformation \eqref{eq:xyzWSFdP2}.  The original singularity at 
$y=(x-g_6)=0$ is replaced by a rational curve $\mathbb{P}^1$ and the whole fiber 
factors into two rational curves intersecting at two points, i.e.~an $I_2$-curve. The original singular fiber is 
identified as the rational curve intersected by the zero section $\hat{s}_P$. Then the other curve has to be 
the isolated exceptional $\mathbb{P}^1$. We can confirm this identification by using the birational transformation 
to the Weierstrass form \eqref{eq:factorizedp1} and by checking explicitly which curve is mapped to the singular 
point \eqref{eq:conifolds} at $y=(x-g_6)=0$. Let us study this explicitly for the two sections $\hat{s}_Q$, 
$\hat{s}_R$ of $X$.

\subsubsection*{Factorization at the $\hat{s}_Q$ section}

The MW-generator of this section has been worked out in equation \eqref{eq:MWGQ}. 
In this case, the section is located at $z=1$ and therefore $z\neq 0$ globally. The Weierstrass equation thus
factorizes with
\bea
 g_{12}&=& \tfrac{1}{72} \left[-s_6^4 + 4 s_6^2 (2 s_5 s_7 + 5 s_3 s_8 + 2 s_2 s_9) - 
 36 s_6 (s_2 s_7 s_8 + s_3 s_5 s_9 + s_1 s_7 s_9)  \right. \\ 
 &-& \left. 8 (2 s_5^2 s_7^2 - s_3^2 s_8^2 + s_2 s_3 s_8 s_9 + 2 s_2^2 s_9^2 + 
    s_5 s_7 (s_3 s_8 - 5 s_2 s_9) - 9 s_1 (s_7^2 s_8 + s_3 s_9^2)) \right] \nn
\eea
and the singular loci \eqref{eq:conifolds} at codimension two are given by
\bea \label{eq:charge10Loci}
g^Q_9&=&\frac{1}{2} (s_3 s_6 s_8 - s_2 s_7 s_8 - s_3 s_5 s_9 + s_1 s_7 s_9)\stackrel{!}{=}0\,,  \\
 \hat g^Q_{12}&=&\frac{1}{2} (2 s_3^2 s_8^2 + s_7 (-s_2 s_6 s_8 + 2 s_1 s_7 s_8 + 2 s_2 s_5 s_9 - s_1 s_6 s_9)  \nn \\
 &+& s_3 (s_6^2 s_8 - 2 s_5 s_7 s_8 - s_5 s_6 s_9 - 2 s_2 s_8 s_9 + 2 s_1 s_9^2) \stackrel{!}{=}0\,.
\eea
We can use these two constraints to solve for any two coefficients $s_i$, $s_j$, as long as the solutions are rational. 
Plugging these solutions into the resolved elliptic fiber \eqref{eq:uvwdp2} 
reveals its $I_2$ nature. In fact, solving and replacing $s_1$ and $s_2$ yields a split of the fiber class as $\mathcal{E}=c_1 + c_2$ for two rational 
curves $c_1$, $c_2$. The equations of the rational curves read
\begin{align}
 c_1:& \quad s_3 u + s_7 w=0\,, \\
 c_2:& \quad s_5 s_7 u^2 - s_3 s_8 u^2 + s_6 s_7 u v - s_3 s_9 u v + s_7^2 v^2 + s_7 s_8 u w + s_7 s_9 v w=0\,,
\end{align}
from which it is clear that the curve $c_2$ intersects the zero section, $\hat{s}_P=[u=0:v=0:w=1]$ and the rational section 
$\hat{s}_R=[u=0:v=s_9:w=-s_7]$. Thus $c_2$ is identified with the original singular fiber, which is also confirmed
by the fact that it maps to the original singular curve \eqref{eq:factorizedp1} by the birational map 
\eqref{eq:xyzWSFdP2}. 
The curve $c_1$ does not intersect $\hat{s}_P$ nor $\hat{s}_R$ and is the isolated rational curve, see figure
\ref{fig:secAtSing} for a depiction of this situation. 
This is also confirmed by noting that $c_1$ is blown-down to the singular point \eqref{eq:conifolds} by the map 
\eqref{eq:xyzWSFdP2}. It thus gives rise to a massless hypermultiplet. Since $c_1$ clearly intersects the rational 
section $\hat{s}_Q=[u=0:v=1:w=0]$ once, the hypermultiplet, according to \eqref{eq:U1charge1} in the absence of Cartan divisors $D_{i_I}$, 
has U$(1)\times$U(1)-charges $(q_1,q_2)=(1,0)$ as claimed. We note that we could have performed the same
analysis on the resolved geometry \eqref{eq:CYindP2} with the same result.

\subsubsection*{Factorization at the  $\hat{s}_R$ section}

The coordinates of this section in Weierstrass form are given in \eqref{eq:MWGR}. We have to take into account that the $z$-coordinate of
the section reads $z=s_9$ and can vanish. When it is non-zero, the Weierstrass equation factorizes as \eqref{eq:factorizedp1} along the loci
\bea \label{eq:gRs}
g^R_{9+3n_2}\!\!\!\!\!\!&\!\!\!\!=\!\!\!\! &\!\!\!\! \frac{1}{2} [2 s_7^3 s_8^3 + s_3 s_9^3 (-s_6 s_8 + s_5 s_9) + 
 s_7^2 s_8 s_9 (-3 s_6 s_8 + 2 s_5 s_9)  \nn\\
 &+& s_7 s_9^2 (s_6^2 s_8 + 2 s_3 s_8^2 - s_5 s_6 s_9 - s_2 s_8 s_9 + s_1 s_9^2)]\stackrel{!}{=}0\,, \\
 \hat g^R_{12+4n_2}\!\!\!\!\!\!&\!\!\!\!=\!\!\!\! &\!\!\!\! \frac{1}{2} \left[ 6 s_7^4 s_8^4 
 + 4 s_7^3 s_8^2 s_9 (2 s_5 s_9-3 s_6 s_8 ) + 
 s_3 s_9^4 (s_6^2 s_8 + 2 s_3 s_8^2 - s_5 s_6 s_9 - 2 s_2 s_8 s_9 + 2 s_1 s_9^2)  \right. \nn \\
 &-& s_7 s_9^3 (s_6^3 s_8 - s_5 s_6^2 s_9 + 2 s_5 s_9 (-3 s_3 s_8 + s_2 s_9) + 
     s_6 (8 s_3 s_8^2 + s_9 (-3 s_2 s_8 + s_1 s_9)))  \nn \\
&+ & s_7^2 s_9^2 (7 s_6^2 s_8^2 - 8 s_5 s_6 s_8 s_9 + 
    2 (4 s_3 s_8^3 + s_9 (-2 s_2 s_8^2 + s_5^2 s_9 + s_1 s_8 s_9))) \left .\right]\stackrel{!}{=}0\,.  
\eea
We find the rational solutions of these equations for $s_1$ and $s_2$. By plugging this into \eqref{eq:uvwdp2}, we
see that the fiber again splits into $\mathcal{E}=c_1 +c_2$ with
\begin{align}
 c_1:& \quad -s_3 s_9^2 u + s_7^2 (-s_8 u + s_9 v) + s_7 s_9 (s_6 u + s_9 w)=0\,, \\
 c_2:& \quad s_7^2 s_8^2 u^2 + s_3 s_9^2 u (s_8 u + s_9 v) + 
 s_7 s_9 (-s_6 s_8 u^2 + s_9 (s_5 u^2 + s_8 u w + s_9 v w))=0\,.
\end{align}
The curve $c_2$ intersects the zero section, $\hat{s}_P$ and also $\hat{s}_Q$, but not $\hat{s}_R$. It is the original singular fiber
as can also be checked using the birational map \eqref{eq:xyzWSFdP2}.
The curve $c_1$ intersects the rational section $\hat{s}_R$ at one point but does not intersect $\hat{s}_P$ and $\hat{s}_Q$. It is
blown-down to the singular point \eqref{eq:conifolds} in the birational map \eqref{eq:xyzWSFdP2}. Thus it is an isolated rational
curve and contributes a hypermultiplet of charges $(q_1,q_2)=(0,1)$ according to \eqref{eq:U1charge1}.

\subsection{Doubly-Factorized Weierstrass Form: charge $(1,1)$}
\label{sec:charge11}

Hypermultiplets charged under both sections can and should exist, both from a geometric point of view and for anomaly cancellation, see
section \ref{sec:anomalies}. Their location in the base can be found by determining singular points of the elliptic fiber $\mathcal{E}$
due to a factorization \eqref{eq:factorizedp1} of the Weierstrass form, where in addition both sections $\hat{s}_Q$, $\hat{s}_R$ coincide. 
As we will demonstrate next, this indeed is a codimension two phenomenon.

From the perspective of the $dP_2$ fiber, the collision of the two sections happens when $s_7=0$. In the Weierstrass 
form there are even more loci, where the two sections collide and at which the Weierstrass model becomes
automatically singular. We first summarize the structure that we find at these loci, before going into 
specifics. In the appropriate patch,  $z=s_9$, we first require the coincidence 
condition $\hat{s}_Q=\hat{s}_R$, that reads
\begin{equation} \label{eq:charge11cond1}
 s_9^2 g_6^Q - g_{6+2n_2}^R=0\,.
\end{equation}
If this is true, then also $s_9^3g_9^Q=g_{9+3n_2}^R$ because the $y$-coordinates of the sections $\hat{s}_Q$, 
$\hat{s}_R$ have been determined in section \ref{sec:construct3points} by plugging in the $x$- and $z$-coordinates 
into the Weierstrass form and solving for $y$. But since $s_9^2x_Q:=s_9^2g^Q_6=g_{6+2n_2}^R=:x_R$, the equality 
$s_9^3g_9^Q=g_{9+3n_2}^R$ is automatic.
For the loci under consideration, it requires only a single further constraint to make the Weierstrass 
model singular, which reads
\beq \label{eq:charge11cond2}
	g_9:= s_9^3g_9^Q=g_{9+3n_2}^R\stackrel{!}{=}0\,.
\eeq 
These two conditions \eqref{eq:charge11cond1} and \eqref{eq:charge11cond2} bring the Weierstrass equation 
\eqref{eq:factorizedp1}, using the notation $g_6\equiv s_9^2 g_6^Q=g_{6+2n_2}^R$, into the form
\begin{equation}
 y^2 = (x-g_6)^2(x-f_6)\,,
\end{equation}
for $f_6=-2g_6$, where the double zero at $x=g_6$, $y=0$ is manifest. We note that the polynomial 
$\hat{g}_{12+4n_2}^Q=s_9^4\hat{g}_{12}^R=0$ in \eqref{eq:charge10Loci} respectively \eqref{eq:gRs} and 
we see that fiber is automatically singular if the sections coincide and $g_9=0$. Here we denote the third root 
in $x$ by $f_6$, however the fiber is smooth $x=f_6$.  

In our concrete situation, the difference between the polynomials can be read off from \eqref{eq:MWGQ} and \eqref{eq:MWGR}  and
takes the form
\begin{equation} \label{eq:deltag6}
 \delta g_6:= s_9^2 g_6^Q - g_{6+2n_2}^R=0= s_7 (s_7 s_8^2 + s_9 (-s_6 s_8 + s_5 s_9))\,.
\end{equation}
We see that $s_7=0$ is a solution as expected, but also the vanishing of the polynomial in parenthesis 
makes the sections collide. In the first case, $s_7=0$, the fiber becomes singular iff
in addition either $s_3=0$ or $s_9=0$. However, these cases have to be treated differently, as discussed
in section \ref{sec:charge2}, because the rational sections are ill-defined at these loci. Thus,
we assume $s_7\neq 0$ in the following and explore the second possibility, the vanishing of the expression in the 
parenthesis in \eqref{eq:deltag6}.
By solving $\delta g_6=0$ we see explicitly that $g_9\equiv s_9^3g_9^Q=g_{9+3n_2}^R$ now take the same form. 
Solving $\delta g_6=0$ e.g.~for $s_6$ we determine
\beq \label{eq:g9}
	g_9 = \tfrac{1}{2} s_7 s_9^2 (s_3 s_8^2 - s_2 s_8 s_9 + s_1 s_9^2)\,.
\eeq
Noting that $s_9\ne0$ by assumption, we drop now the coefficients of $s_7$ and $s_9$ in front of $\delta g_6$ and 
$g_9$. Then the codimension two loci for matter read
\beq \label{eq:loci11}
 \delta g'_6:= s_7 s_8^2 + s_9 (-s_6 s_8 + s_5 s_9)=0 , \qquad g'_9 := s_3 s_8^2 - s_2 s_8 s_9 + s_1 s_9^2=0 \,.
\eeq
To check the order of vanishing of the discriminant, cf.~appendix \ref{app:detailsOf2U1model} and section 
\ref{sec:construct3points}, along the codimension 
two locus we consider a small neighborhood around \eqref{eq:charge11cond1} and \eqref{eq:charge11cond2}  defined as
\beq   \label{eq:locimat11}
\delta g_6'= \epsilon \,, \qquad \ g_9' = \epsilon\,, \qquad \ s_9 \ne 0\,, \qquad s_7\ne0 \,. 
\eeq
for a small $\epsilon>0$. Again by solving for two of the $s_i$ coefficients we see that the discriminant factors 
as
\beq
\Delta=\epsilon^2 \Delta^\prime\,,
\eeq
with $\Delta'\neq 0$ in the limit $\epsilon\rightarrow 0$, which is equivalent to the observation
$\hat{g}_{12}^Q\sim\hat{g}_{12+4n_2}^R=0$ if \eqref{eq:locimat11} is obeyed.
This is an indication for an $I_2$ singularity at $\epsilon=0$. 
Indeed, by solving the conditions \eqref{eq:locimat11}  for $s_3$ and $s_6$ and by plugging back 
into \eqref{eq:uvwdp2} on the resolution,  we obtain a split of the elliptic curve as $\mathcal{E}=c_1+c_2$ with
\bea
c_1:& \quad & s_8 u + s_9 v=0\,. \\
c_2:& \quad & s_1 s_8 s_9 u^2 + s_2 s_8 s_9 u v - s_1 s_9^2 u v + s_5 s_8 s_9 u w + s_7 s_8^2 v w + s_8^2 s_9 w^2=0\,.
\eea
From this we see that the zero section $\hat{s}_P=[0:0:1]$ intersects $c_1$ at a point and does not intersect $c_2$,
while the other two sections, $\hat{s}_Q=[0:1:0]$ and $\hat{s}_R=[0:s_9,-s_7]$, intersect $c_2$ at a point but not $c_1$. This implies
as before that the curve $c_1$ is the proper transform of the original singular curve under the birational map
\eqref{eq:xyzWSFdP2} while the curve $c_2$ is blown down to a point. It is the isolated rational curve and contributes
a hypermultiplet of charges $(q_1,q_2)=(1,1)$ according to \eqref{eq:U1charge1}.

\subsection{Singular Rational Sections: charges $(\text{-}1,1)$, $(\text{-}1,\text{-}2)$, $(0,2)$}
\label{sec:charge2}

\begin{figure}[ht!]
\centering
 \includegraphics[scale=0.45]{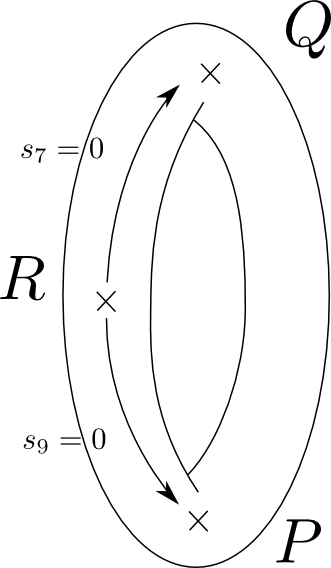}
 \caption{Fibers where two out of the three sections collide. The intersections occur when 
 $s_7$ or/and $s_9$ vanish on the base $B$.}
 \label{fig:RMoving}
\end{figure}

There is another way of looking for singularities of the elliptic fibration, that contribute
charged matter in F-theory, which is the third type
mentioned at the beginning of this section. Qualitatively, the strategy is as follows.  
For non-trivial sections $s_9$ and $s_7$ over $B$, the section $\hat{s}_R=[0:-s_9:s_7]$ can 'move' on the fiber 
$\mathcal{E}$ when varying the point on the base $B$. This  implies that 
there are loci in $B$ where two or more sections intersect in the fiber $\mathcal{E}$. This is achieved when $s_7=0$ 
or $s_9=0$ as can be seen from \eqref{eq:points-uvw} in the $\mathbb{P}^2$-model for $\mathcal{E}$, 
see figure \ref{fig:RMoving}. 
Plugging this ansatz into the discriminant $\Delta$ admits a factorization. The results of this analysis, that
are precisely the loci not discussed in section \ref{sec:charge11}, are summarized in table 
\ref{tab:LocusSingDisc}. 

\begin{table}[ht!]
 \centering
 \begin{tabular}{|c|c|c|c|c|c|} \hline 
 Locus 		& Isolated Curve 		& Charge ($U(1)_Q,U(1)_R$) \rule{0pt}{13pt}\\ \hline
 $s_3=s_7=0$	& $e_1=0$			&$(-1,1)$	  	\\
 $s_7=s_9=0$	& $u=0$			& $(0,2)$	  	\\
 $s_8=s_9=0$	& $c_2$			&$(-1,-2)$	  	\\ \hline
 \end{tabular}
\caption{Codimension two loci with singular behavior of the sections. The identifications of
the isolated curves are explained in \eqref{eq:curves-11}, \eqref{eq:curves02} and \eqref{eq:curves-1-2}.}
\label{tab:LocusSingDisc}
\end{table}

We emphasize that at each of these loci one section $\hat{s}_P$, $\hat{s}_Q$, $\hat{s}_R$
is ill-defined. This can be seen most easily on the resolved model 
of the elliptic curve $\mathcal{E}$ in $dP_2$ introduced in section \ref{sec:blowupgeo}.
For the convenience of the reader we recall the relevant properties of the resolved geometry.
After resolutions along $u=w=0$ and $u=v=0$ the sections take the values
\beq	 \label{eq:secsdP2}
	\hat{s}_P=[-s_9:s_8:1:1:0]\,,\qquad \hat{s}_Q=[-s_7:1:s_3:0:1]\,,\qquad 
	\hat{s}_R=[0:1:1:-s_7:s_9]\,,
\eeq
cf.~\eqref{eq:coordsPQRdP2}. 
We also recall the  Stanley-Reissner ideal, of the resolution\footnote{In the following we denote the coordinates 
on $dP_2$ as $[u:v:w:e_1:e_2]$ in contrast to section \ref{sec:blowupgeo}.}
\beq \label{eq:SRdP2}
	SR=\{u v,\,u w,\,e_1 e_2,\,e_1 v,\,e_2 w\}\,.
\eeq
Indeed, at each locus in table \ref{tab:LocusSingDisc} one of the sections \eqref{eq:secsdP2} is 
ill-defined since it passes through the Stanley-Reissner ideal \eqref{eq:SRdP2}. 
This requires additional blow-ups in the base $\pi_B:\,\hat{B}\rightarrow B$ as advertised in 
section \ref{sec:CYwithMW} precisely along the loci where the sections are not well-defined. 
We emphasize that this implies that the birational map \eqref{eq:xyzWSFdP2} is not enough 
to completely understand and resolve the fibration $X$, in contrast to the codimension two loci
discussed in sections \ref{sec:charge1} and \ref{sec:charge11}. 

To understand what is happening when one of the sections $\hat{s}_P$, $\hat{s}_Q$, $\hat{s}_R$, collectively 
denoted as $\hat{s}$ for the following paragraph, is singular, we first look at the sections at a 
location in the base where the fiber is not singular. In this case, after performing the blowups in the 
ambient space to $dP_2$, the exceptional $\mathbb{P}^1$'s intersect the Calabi-Yau polynomial at exactly 
one point in the fiber with coordinates shown in \eqref{eq:secsdP2}.
At any of the loci of table \ref{tab:LocusSingDisc}, when two coefficients, $s_i$ and $s_j$ vanish simultaneously, 
the two coordinates of the exceptional $\mathbb{P}^1$ vanish simultaneously and the coordinates stop making sense. 
This is another way of saying that the section passes through the Stanley-Reissner ideal \eqref{eq:SRdP2}.
To resolve this problem we blow up the base at this locus by introducing coordinates $[\ell_1:\ell_2]\in\mathbb{P}^1$ and writing
\beq \label{eq:blowupBase}
 \ell_1 s_i - \ell_2 s_j = 0\,.
\eeq
The coordinates \eqref{eq:secsdP2} of the section $\hat{s}$ will be now given by the point 
$[s_j:s_i]=[\ell_1:\ell_2]$ in $\mathbb{P}^1$. When the two coordinates vanish simultaneously, $s_i=s_j=0$, 
the values of $\ell_1$ and $\ell_2$ in \eqref{eq:blowupBase} become unrestricted and we obtain the entire 
$\mathbb{P}^1$ in the fiber. 

Another way of seeing this is to notice that the section $\hat{s}$ takes different values at the singular locus 
$s_i=s_j=0$ depending on the direction (parametrized by $[\ell_1:\ell_2]$) at which we approach the locus in the base. Coming in all possible directions 
makes the section take all possible values at the same locus: the section becomes the entire $\mathbb{P}^1$ of slopes.

\begin{figure}[ht!]
\centering
\includegraphics[scale=0.4]{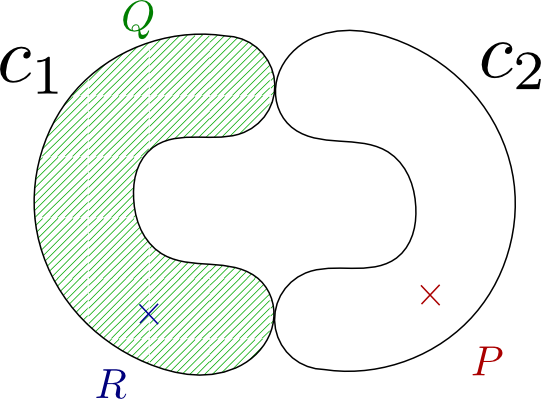} \qquad
\includegraphics[scale=0.4]{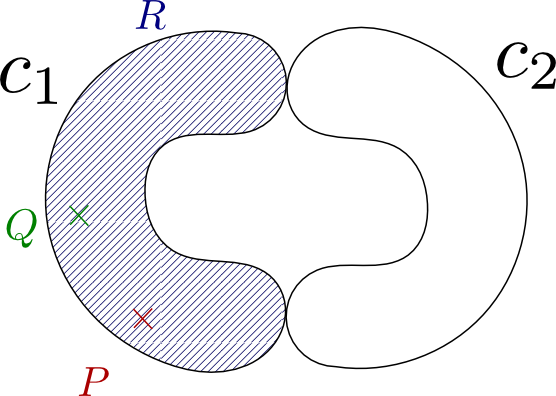} \qquad
\includegraphics[scale=0.4]{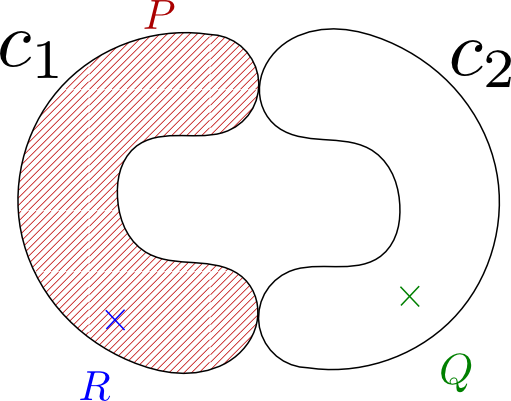}
\caption{How the fiber degenerates at the three loci (-1,1), (0,2) and (-1,-2) respectively.}
\label{fig:resolvedFibers}
\end{figure}

Now that we understand that the sections can become entire $\mathbb{P}^1$'s, we demonstrate the factorization 
of the equation \eqref{eq:CYindP2}  into two rational curves along all codimension two loci
in table \ref{tab:LocusSingDisc}. We discuss these loci successively and refer to figure \ref{fig:resolvedFibers}
to keep track of which of the three sections wraps a $\mathbb{P}^1$ in the resolved fiber.

\subsubsection*{Singularity of $\hat{s}_Q$ \& blow-up at $s_3=s_7=0$: charge $(-1,1)$}

As can be seen from \eqref{eq:secsdP2} at the loci $s_3=s_7=0$ the section $\hat{s}_Q$ passes through the
ideal \eqref{eq:SRdP2} and is ill-defined. It has to be blown-up and wraps an entire $\mathbb{P}^1$-component
of the fiber, which is given by the divisor $E_1:=\{e_1=0\}$ in $dP_2$. 
In addition, the polynomial \eqref{eq:CYindP2} factorizes into two rational curves, i.e.~$\mathcal{E}=c_1+c_2$, 
reading
\bea \label{eq:curves-11}
  c_1 :&\quad&  e_1 =0\,, \\
 c_2 :&\quad&   s_1 e_1 e_2^2 u^3 + s_2 e_2^2  u^2 v + s_5 e_1 e_2  u^2 w + s_6 e_2  u v w +s_8 e_1  u w^2 + s_9 v w^2=0\,.
\eea
We notice using \eqref{eq:secsdP2}, \eqref{eq:SRdP2} and the toric intersections \eqref{eq:dP2ints} on $dP_2$ the 
intersections of the rational sections with the two curves as
\beq \label{eq:intscharge1-1}
	S_P\cdotp c_1 =0\,,\quad S_P\cdotp c_2 =1\,, \quad S_Q\cdotp c_1 =-1\,,\quad S_Q\cdotp c_2 =0\,, \quad
	S_R \cdotp c_1=1\,,\quad S_R\cdotp c_2 =0\,,
\eeq  
where we denoted the homology classes of the sections by capital letters $S_\bullet$ as before.
Here we note that the last two intersections follow, besides from the toric intersections \eqref{eq:dP2ints}, also from the defining 
property of a section, $1=S_R\cdot \mathcal{E} =
S_R\cdotp (c_1+c_2)$, which with $S_R\cdot c_2=0$ immediately yields $S_R \cdotp c_1=1$. 

We note that the intersections \eqref{eq:intscharge1-1} imply that $c_2$ is the original singular curve since it intersects the
zero section $\hat{s}_P$ and $c_1$ is the exceptional $\mathbb{P}^1$ contributing a hypermultiplet.
Thus it follows from \eqref{eq:U1charge1} that the  U$(1)_Q\times$U$(1)_R$-charges are $(q_1,q_2)=(-1,1)$.

\subsubsection*{Singularity of $\hat{s}_R$ \&  blow-up at $s_7=s_9=0$: charge $(0,2)$}

Proceeding in a similar way as before  we see that the section $\hat{s}_R$ is blown up in a $\mathbb{P}^1$, which 
we identify with the toric divisor $D_u:=\{u=0\}$ in $dP_2$. Furthermore, the polynomial \eqref{eq:CYindP2} factorizes at 
$s_7=s_9=0$ into two rational curves given by
\bea \label{eq:curves02}
c_1 :&\quad & u=0\,, \\ 
c_2 :&\quad & s_1e_1^2 e_2^2  u^2 + s_2e_1 e_2^2  u v + s_3e_2^2  v^2 + s_5e_1^2 e_2  u w +s_6  
   e_1 e_2 v w +  s_8 e_1^2w^2=0\,.
\eea
The intersections of the sections are evaluated as before employing \eqref{eq:secsdP2}, \eqref{eq:SRdP2} and the toric intersections \eqref{eq:dP2ints} on $dP_2$ as
\beq 
\label{eq:instcharge02}
 	S_P\cdotp c_1=1\,,\quad S_P\cdotp c_2=0\,,\quad  S_Q\cdotp c_1=1\,,\quad S_Q\cdotp c_2=0\,,\quad S_R\cdotp c_1=-1\,,\quad S_R\cdotp c_2=2\,. 
\eeq
Here we have determined the intersections of $S_R$ by noting that due to the blow-up in the base at $s_7=s_9=0$, cf.~\eqref{eq:secsdP2}, 
it wraps the entire $\mathbb{P}^1$ corresponding to the ray subdividing the cone formed $e_1$, $e_2$. As we see
by recalling figure \ref{fig:dp2poly} this is nothing else but the toric divisor with $D_u$ section $u$. 
This then implies that $S_P\cdot c_2=2$ because the equation for $c_2$ has two solutions with $u=0$. From the defining property 
$1=S_P\cdot \mathcal{E}=S_P\cdot (c_1+c_2)=S_P\cdot c_1+2$ we infer $S_P\cdot c_1=-1$ as claimed in \eqref{eq:instcharge02}.

From the location of the zero section we infer that $c_1$ is the original singular fiber and $c_2$ the isolated rational curve contributing 
a hypermultiplet. Its charges are calculated employing \eqref{eq:U1charge1} as $(q_1,q_2)=(0,2)$.

\subsubsection*{Singularity of $\hat{s}_P$ \& blow-up at $s_8=s_9=0$:  charge $(-1,-2)$}

At this locus, the zero section $\hat{s}_P$ is ill-defined and is blown up into a rational fiber component.
It is identified with the toric divisor $E_2:=\{e_2=0\}$ in $dP_2$.
The constraint \eqref{eq:CYindP2} of the elliptic curve implies a split $\mathcal{E}=c_1+c_2$  at these loci 
with 
\bea 
 c_1 :&\quad &   e_2 =0\,, \\
 c_2 :&\quad &    s_1 e_1^2 e_2  u^3 + s_2 e_1 e_2 u^2 v + s_3 e_2 u v^2 +s_5  e_1^2 u^2 w + 
  s_6e_1 u v w + s_7 v^2 w=0\,.\label{eq:curves-1-2}
\eea
As before the intersections of the sections are computed from \eqref{eq:secsdP2}, \eqref{eq:SRdP2} and the toric intersections \eqref{eq:dP2ints} 
on $dP_2$ as
\beq
\label{eq:instcharge-1-2}
	S_P \cdotp c_1=-1\,,\quad S_P \cdotp c_2=2\,,\quad S_Q \cdotp c_1=0 \,,\quad S_Q \cdotp c_2=1\,,\quad S_R \cdotp c_1=1\,,\quad 
	S_R \cdotp c_2=0\,,
\eeq
where the intersection $S_P \cdotp c_2=2$ can be obtained by setting $e_2=0$ in \eqref{eq:curves-1-2}. Alternatively, it can also be calculated 
using $1=S_P\cdot \mathcal{E}=S_P\cdot (c_1+c_2)=-1+S_P\cdot c_2$.

Here we have used that the blow-up in the base along $s_8=s_9=0$ has blown-up the zero section with coordinates \eqref{eq:secsdP2} 
into $E_2$, which is the toric divisor of the ray subdividing the cone spanned by $u$ and $v$. The curve $c_1$
is part of the zero section $c_1$, thus it is the original singular fiber, whereas $c_2$ maps to the singular point of the original fiber.
Thus $c_2$ is the isolated curve and it contributes according to \eqref{eq:U1charge1} 
a hypermultiplet of total charge $(q_1,q_2)=(-1,-2)$. We note that this locus is special in the sense that it 
supports the only matter multiplet where the second term in \eqref{eq:U1charge1} contributes to the charge 
due to the ill-defined zero section $\hat{s}_P$.

\subsection{Calculating Matter Multiplicities}
\label{sec:matter_multiplicities}

Now that we have determined the full matter representations and their codimension two loci in the base $B$, 
we can count their multiplicities. The main complication that arises in these calculations is due to the intersection 
of different codimension two loci. In order to avoid double-counting we thus have to appropriately subtract 
the number of intersection points of a codimension two loci under consideration with other codimension two
loci. 

For this analysis, we first have to recall the counting of the multiplicity of 
a root in a system of two polynomials with two independent variables. Given two polynomials
\bea \label{eq:fg=0}
 f(x,y) =0 , \qquad g(x,y)=0,
\eea
we define
\bea
  h(y) := \text{Res}_x (f,g)
\eea
as the resultant of $f,g$ with respect to $x$. The degree of a root $(x_0,y_0)$ of \eqref{eq:fg=0} is given by the 
multiplicity of $y_0$ as a zero of $h(y)$\footnote{The degree can also be defined with respect to the
variable $x$, however, yielding the same multiplicity.}. 

Now that we know how to count roots of a polynomial system, the strategy is simple: take the degrees of the   
polynomials for the codimension two loci we are interested, multiply them and then subtract
all intersecting loci we are not interested in with the appropriate multiplicity. As a warm up and a 
demonstration of this technique we rederive the multiplicity of the charge one loci of the elliptic 
fibration over $\mathbb{P}^{2}$ with elliptic fiber $Bl_{(0,1,0)}\mathbb{P}^{1,1,2}$ considered in  \cite{Morrison:2012ei}.

\subsubsection*{Example: Charge one multiplicity of the $Bl_{(0,1,0)}\mathbb{P}^2(1,1,2)$ model}
The loci of the charge one hypermultiplets were given in equations (5.64) of \cite{Morrison:2012ei}
\bea \label{eq:parkLoci1}
f^3_{3+n}-3 f_6 f_{3+n} b_n^2 + b_n^4 f_{9-n} &=& 0\,, \nn \\
f^4_{3+n}-6 f_6 f^2_{3+n} b_n^2 + 9 f_6^2 b_n^4 - f_{12-2n} b_n^6 &=& 0\,, 
\eea
with $b_n \ne 0$ and $f_{3+n} \ne 0$ because those were identified as the charge two loci. We 
see that $f_{3+n}=b_n=0$ obviously  solves the system. 

Viewing now the equations \eqref{eq:parkLoci1} as polynomials in the variables $f_{3+n}$ and $b_n$, the resultant 
with respect to the variable $f_{3+n}$ is given by
\beq
 h(b_n)=b_n^{16} (b_n^2 f_{12-2n}^3 - 9 f_{12-2n}^2 f_6^2 + 6 f_{12-2n} f_6 f_{9-n}^2 - f_{9-n}^4),
\eeq
from where it can be seen that $f_{3+n}=b_n=0$ is a root of degree $16$ of the system \eqref{eq:parkLoci1}. Now we can calculate 
the multiplicity of the charge one hypermultiplets. The degrees of \eqref{eq:parkLoci1} are $(9+3n)$ and $(12+4n)$ respectively and 
there are $n(3+n)$ roots we are not interested in and that have to be subtracted, each of which of degree $16$. Thus  we obtain
\beq
 (9+3n)(12+4n) - 16 n(3+n) = 4(9-n)(3+n)
\eeq
charge one hypermultiplets.

\subsubsection*{Matter multiplicities of the $dP_2$ model}

In this subsection we will finally apply the resultant technique to determine the multiplicity
of all six codimension two matter loci found in sections \ref{sec:charge1}, \ref{sec:charge11} and
\ref{sec:charge2}. We denote the multiplicities by the same variables $x_i$, $i=1,\ldots,6$, as in
\eqref{eq:multNames} to facilitate a straightforward comparison with results from anomaly cancellation.

The basis of our analysis are the degrees of the coefficients $s_i$ calculated in \eqref{eq:degree_si} 
for the base $B=\mathbb{P}^2$. Let us recall that the degree of $s_9$ is equal to the intersection number $n_2$
of the zero section $\hat{s}_P$ and the section $\hat{s}_R$ and the degree $n_{12}$ of $s_7$ equals the intersection 
number of $\hat{s}_Q$ and $\hat{s}_R$, cf.~\eqref{eq:defn2n12}. 
With these degrees, the multiplicities of the hypermultiplets of section \ref{sec:charge2} are straightforward: they 
are simply the multiplication of the degrees of the two coefficient that vanish, see table \ref{tab:multCharge2}. For the other hypermultiplets we have to work a little more.

\begin{table}[ht!]
 \centering
 \begin{tabular}{|c|c|c|} \hline 
 Locus 		& Charge ($U(1)_Q,U(1)_R$) 	& Multiplicity \rule{0pt}{13pt}\\ \hline
 $s_3=s_7=0$	&$(-1,1)$	  		& $x_4=(3-n_2+n_{12})n_{12}$\rule{0pt}{12pt}	\\
 $s_7=s_9=0$	& $(0,2)$	  		& $x_5=n_{12} n_{2}$	\\
 $s_8=s_9=0$	&$(-1,-2)$	  		& $x_6=(3+n_2-n_{12})n_2$\\ \hline
 \end{tabular}
\caption{Matter multiplicities.}
\label{tab:multCharge2}
\end{table}

\subsubsection*{Charge (1,1)}

The charge $(1,1)$ hypermultiplets are located at the roots of equations \eqref{eq:loci11}, that we recall for 
convenience of the reader as
\beq 
 \delta g'_6= s_7 s_8^2 + s_9 (-s_6 s_8 + s_5 s_9) \stackrel{!}{=}0\,, \qquad g'_9 = s_3 s_8^2 - s_2 s_8 s_9 + s_1 s_9^2 \stackrel{!}{=}0\,. \nn
\eeq
They have degrees 
\bea \label{eq:11loc}
 \text{deg}(\delta g'_6) & = & n_{12}+2(3+n_2-n_{12}) =6+2n_2-n_{12}, \nn \\
 \text{deg}(g'_9 ) &= & (3-n_2+n_{12})+2(3+n_2-n_{12}) = 9+ n_{2} -n_{12}\,.
\eea
The loci $s_8=s_9=0$ are obviously roots of these equations. The degree of this root is four as we see by working 
out the resultant of these constraints with respect to $s_8$ and $s_9$. No other codimension two loci satisfy
the equations \eqref{eq:11loc}. Thus, the multiplicity of the hypermultiplets with charges $(1,1)$ is 
\beq
 x_3=(6+2n_2-n_{12})(9+ n_{2} -n_{12})-4n_2(3+n_2-n_{12}) = 54 + 12 n_2 - 2 n_2^2 +( n_{2} - 15 ) n_{12} + n_{12}^2
\eeq

\subsubsection*{Charge (1,0)}

We proceed with the calculation of the multiplicity of charge $(1,0)$ hypermultiplets. The constraints of the
corresponding codimension two loci is given by equation \eqref{eq:charge10Loci}, that takes the schematic form
\beq
g^Q_9 = 0 \,, \qquad \hat g^Q_{12}=0\,.
\eeq
Recalling table \ref{tab:LocusSingDisc} we notice that all hypermultiplets with charges $(1,1)$, $(-1,1)$ and 
$(-1,-2)$ satisfy these equations. The multiplicity is calculated from the resultant as one in each case. 
Thus, the multiplicity of the loci supporting charge $(1,0)$ hypermultiplets is then
\bea
 x_1 \!\!&\!\!=\!\!&\!\! 108 - (54 + 12 n_2 - 2 n_2^2 + ( n_2-15) n_{12} + n_{12}^2) - n_{12} (3 - n_2 + n_{12})-n_2 (3 + n_2 - n_{12}) \nn \\
 \!\!&\!\!=\!\!&\!\! 54 - 15 n_2 + n_2^2 + 12 n_{12} + n_2 n_{12} - 2 n_{12}^2\,.
\eea

\subsubsection*{Charge (0,1)}
The defining equations for the loci supporting these hypermultiplets are given by the equations in 
\eqref{eq:gRs}, that schematically read as
\beq \label{eq:gRs1}
g^R_{9+3n_2} = 0 \,, \qquad \hat g^R_{12+4n_2}=0\,.
\eeq
Again by looking at table \ref{tab:LocusSingDisc} we see that all hypermultiplets with
charge $(1,1)$, $(-1,1)$, $(0,2)$ and  $(-1,-2)$ satisfy the equations with multiplicities. Their
multiplicities are $1$, $1$, $16$ and $16$, respectively, and by subtracting from the 
degree of \eqref{eq:gRs1} we obtain the multiplicity as
\bea
 x_2 &=& (9+3n_2)(12+4n_2) - x_3 - x_4 - 16 x_5 - 16 x_6 \nn \\
  &=& 54 + 12 n_2 - 2 n_2^2 + 12 n_{12} - 2 n_{12}^2\,.
\eea

\section{Anomaly Cancellation: a Consistency Check}
\label{sec:anomalies}

In this section we check the consistency of the matter spectrum determined in the 
previous section via constraints arising from anomaly cancellation. We demonstrate that anomaly cancellation
allows us to classify all F-theory compactifications with base $B=\mathbb{P}^2$ and the two section
elliptic fiber determined in section \ref{sec:construct3points}. This
confirms the geometric analysis of the previous section that has led to determination of the charges of matter fields.
We also see that the computed multiplicities agree with the ones predicted by anomaly cancellation. This 
analysis ensure a consistent low-energy theory of F-theory in six dimensions with two Abelian gauge fields. 

Anomaly cancellation in six-dimensional F-theory vacua, based on general results on 6d anomaly
cancellation \cite{Erler:1993zy,Honecker:2006dt}, has been discussed in \cite{Park:2011ji,Morrison:2012ei},
whose notations and conventions we follow here. See also the analysis in \cite{Cvetic:2012xn} of anomaly
cancellation in four dimensional F-theory compactifications.
We note that the following discussion holds more generally for any Abelian gauge theory.

We are interested in anomaly cancellation in an Abelian gauge theory with matter arising from 
F-theory compactifications.
In this case, there are only the purely gravitational anomaly, the mixed Abelian-gravitational
and the pure Abelian gauge anomalies. We assume that there are $H$ hypermultiplets, $T$
tensor multiplets and $V$ vector multiplets. The gauge and mixed anomalies induced by the six-dimensional
charged matter in hypermultiplets have to be canceled by a classical effect, the generalized
Green-Schwarz mechanism \cite{Sadov:1996zm}. These cancellation conditions are the anomaly equations.
The mixed Abelian-gravitational, purely Abelian and purely gravitational anomaly equations read 
\bea \label{eq:6dAnomalies}
 	K_B\cdotp b_{mn} &= &-\frac{1}{6}\sum_{\underline{q}} x_{q_m, q_n} q_m q_n\,,\nn\\
 	b_{mn} \cdotp b_{kl} + b_{mk} \cdotp b_{nl} +b_{ml} \cdotp b_{nk}  &=& \sum_{\underline{q}} x_{q_m,q_n,q_k,q_l} q_m q_n q_k q_l\,,\nn\\
 	273=H-V+ 29T\,,&&\,\, K_B\cdot K_B=9-T\,.
\eea
Here we have labeled the $r$ U(1)-fields in our theory by $A^m$, where $r$ denotes the rank of the
Mordell-Weil group of the Calabi-Yau threefold $X$.
On the right hand side of \eqref{eq:6dAnomalies}, the sum runs over all hypermultiplets with charge vector 
$\underline{q}=(q_1,\ldots,q_r)$. The integers
$x_{q_m,q_n}$ respectively $x_{q_m,q_n,q_k,q_l}$ denote the number of hypermultiplets with charges
$(q_m,q_n)$ under U$(1)_m\times$U$(1)_n$ respectively charges $(q_m,q_n,q_k,q_l)$ under U$(1)_m\times$U$(1)_n\times$U$(1)_k\times$U$(1)_l$. The left hand side of \eqref{eq:6dAnomalies}
is to be evaluated in the cohomology of the base $B$ of the fibration. $K_B$ denotes
the canonical class of $B$, $\cdot$ is the intersection pairing on $B$ and
the expressions $b_{mn}$ are defined as the curves
\bea \label{eq:b_mnAbelian}
 b_{mn} &=& -\pi( \sigma(\hat{s}_m) \cdotp \sigma(\hat{s}_n) ) \nn\\
  &=& -\pi( S_m \cdotp S_n )-[K_B] + \pi(S_m \cdotp S_P) + \pi(S_n \cdotp S_P)\,. 
\eea
We recall that $S_m$ denotes the divisors class of a section $\hat{s}_m$ and the homology
class $S_P$ of the zero section $\hat{s}_P$.
Here the first line has been evaluated employing \eqref{eq:anomalycoeff} in the presence
of only Abelian gauge fields.
We note that the pure gravitational anomaly even without further specification of the spectrum 
puts an upper bound on a theory with $T=0$, which is the case for $B=\mathbb{P}^2$, as 
\begin{equation}
 \sum_f N_f \le 275\,.
\end{equation}

We now analyze the anomaly constraints \eqref{eq:6dAnomalies} for the F-theory compactification
on the elliptic fibration over $\mathbb{P}^2$ with two U(1) gauge fields and the matter content
determined in section \ref{sec:Matter}. We see that anomaly cancellation fixes the possible
matter multiplicities in terms of two integers. We first evaluate the anomaly coefficients $b_{mn}$ in 
\eqref{eq:6dAnomalies}.
In the case of $\mathbb{P}^2$ as the base, there is only one element generating $H_2(B)$, the hyperplane class $H_B$. 
The anti-canonical line bundle is $K_{\mathbb{P}^2}=-3H_B$. Then the coefficients are just numbers, namely the 
coefficients of $H_B$, and evaluated to
\begin{equation}
 b_{kl}=  \left\lbrace \begin{array}{cc}
                    2(n_k+3), & k=l \\
                    3+n_k+n_l-n_{kl}, & k \ne l,
                   \end{array} \right.
\end{equation}
where we defined $n_{kl}=\pi(  S_k \cdotp  S_l )$ and $n_k=\pi( S_k \cdotp B)$. 
In addition, we have employed the second relation in \eqref{eq:intsSections} to replace 
$\pi(S_k\cdot S_k)=(S_k\cdot S_k\cdot H_B)H_B=K=3H_B$.
Then, we evaluate the right hand side of \eqref{eq:6dAnomalies} using the spectrum found in section \ref{sec:Matter} with arbitrary multiplicities $x_i$ assigned as
\beq \label{eq:multNames}
\text{
\begin{tabular}{|c||c|c|c|c|c|c|}
\hline
	$(q_1,q_2)$&$(1,0)$&$(0,1)$&$(1,1)$&$(-1,1)$&$(0,2)$&$(-1,-2)$ \rule{0pt}{12pt}\\\hline
	multiplicity &$x_1$&$x_2$&$x_3$&$x_4$&$x_5$&$x_6$\rule{0pt}{11pt}\\ \hline
\end{tabular}}
\eeq
Upon inserting this into the anomaly constraints \eqref{eq:6dAnomalies} for two sections $S_1\equiv S_Q$, $S_2\equiv S_R$ 
and the zero section $S_P$ we obtain a set of linear equations for the $x_i$ reading
\bea \label{eq:linearSystem}
&36 (3+n_1)=12 (3+n_1)^2=x_1+x_3+x_4+x_6\,,&\nn\\ 
&\,\,\,\,\,18 (3+n_1-n_{12}+n_2)=6 (3+n_1) (3+n_1-n_{12}+n_2)=x_3-x_4+2 x_6\,, & \nn\\
   &12 (3+n_2)^2=x_2+x_3+x_4+16 x_5+16 x_6\,,&\nn\\
   &6 (3+n_2) (3+n_1-n_{12}+n_2)=x_3-x_4+8 x_6\,,&\nn\\
   &36 (3+n_2)=x_2+x_3+x_4+4 x_5+4 x_6\,,&\nn\\
   & 4 (3+n_1) (3+n_2)+2 (3+n_1-n_{12}+n_2)^2=x_3+x_4+4 x_6\,.&
\eea

We see that the first equation in this system immediately requires $n_1=\pi(S_Q\cdot S_P)=0$.
This nicely agrees with the finding of \eqref{eq:MWGQ} respectively \eqref{eq:coordsPQRdP2}
that the section $S_Q$ does not intersect the zero section $S_P$. The solutions of \eqref{eq:linearSystem}
then take the form
\bea \label{eq:generalmatter}
	x_1\!\!\!&=\!\!\!& 54-15 n_2+n_2^2+\left(12+n_2\right) n_{12}-2 n_{12}^2\,,\,\,\,\,\,\,\,\qquad\quad x_4= n_{12} \left(3-n_2+n_{12}\right)\,,\nn\\
	x_2\!\!\!&=\!\!\!& 54+2 \left(6 n_2-n_2^2+6 n_{12}-n_{12}^2\right)\,,\,\,\,\quad\,\,\,\quad \qquad\quad\quad x_5= n_2 n_{12}\,,\nn\\
	 x_3\!\!\!&=\!\!\!& 54+12 n_2-2 n_2^2+\left(n_2-15\right) n_{12}+n_{12}^2\,, 
	 \,\,\,\,\,\,\,\qquad\quad x_6= n_2 \left(3+n_2-n_{12}\right)\,.
\eea
This provides a full classification of the possible matter multiplicities with two U(1)-gauge fields
in terms of the intersections 
\beq \label{eq:defn2n121}
	n_2=\pi(S_Q\cdot S_P)\,,\quad n_{12}=\pi(S_Q\cdot S_R)
\eeq
of the two sections $S_Q$, $S_R$ and of $S_P$ introduced in \eqref{eq:height_pairing}. It is satisfying
that the matter multiplicities found in section \ref{sec:matter_multiplicities} are reproduced by 
\eqref{eq:generalmatter}. 

We conclude by evaluating two special cases. First we consider $n_2=n_{12}=0$ where we obtain
\beq \label{tab:example1Matter}
\text{
\begin{tabular}{|c||c|c|c|c|c|c|}
\hline
	$(q_1,q_2)$&$(1,0)$&$(0,1)$&$(1,1)$&$(-1,1)$&$(0,2)$&$(-1,-2)$\rule{0pt}{12pt}\\\hline
	multiplicity &$54$&$54$&$54$&$0$&$0$&$0$\rule{0pt}{11pt}\\ \hline
\end{tabular}
}
\eeq 
In this case, charge two loci are completely absent. A less trivial example can be constructed by setting $n_2=n_{12}=1$ in which case we obtain
\beq \label{tab:example2Matter}
\text{\begin{tabular}{|c||c|c|c|c|c|c|}
\hline
	$(q_1,q_2)$&$(1,0)$&$(0,1)$&$(1,1)$&$(-1,1)$&$(0,2)$&$(-1,-2)$\rule{0pt}{12pt}\\\hline
	multiplicity &$51$&$74$&$51$&$3$&$1$&$3$\rule{0pt}{11pt}\\ \hline
\end{tabular}}
\eeq
The global geometries of both examples are constructed in section \ref{sec:examples}. As we see there, we 
will find perfect agreement with results from anomaly cancellation by using the general geometric techniques of 
section \ref{sec:Matter}.

\section{Toric Elliptic Calabi-Yau Manifolds with Two\\ Rational Sections}
\label{sec:examples}

In this section we construct explicitly toric elliptically fibered Calabi-Yau threefolds realizing concrete elliptic fibrations
with two rational sections.  In all examples the base of the fibration is $B=\mathbb{P}^2$ and the generic fiber is the elliptic curve in
$dP^2$. The first two examples both have U$(1)\times$U(1)-gauge symmetry, but with rational sections of different complexity and 
consequently different matter sectors. In a third example we add an SU$(5)$-GUT sector and explicitly present the corresponding 
four-dimensional toric polytope.

In all examples considered below, we construct the toric ambient variety $V$ as a toric fibration of $dP_2$ 
over a toric base $B$,
\beq
 dP_2\rightarrow V\rightarrow B\,.
\eeq 
The presence of a toric fibration can be detected by analyzing the toric polytope $\Delta_{V}$ defining 
$V$. The polytope $\Delta_{dP_2}$ of the fiber has to be cut out by a 
two-dimensional plane $E$ through the origin ${0}$ as
\beq \label{eq:ToricFib}
	\Delta_{dP_2}=E\cap \Delta_V\,.
\eeq 
Then a toric morphism $\varphi$ to the base $B$ is defined by first finding
the two normal vectors $n_i$, $i=1,2$, to $E$ and then by 
defining the $2\times 4$-matrix
\beq \label{eq:ToricMor}
	\varphi=\begin{pmatrix}
		 n_1^T\\
		\vdots\\
		n_{k-1}^T
		\end{pmatrix}\,.
\eeq  
This matrix acts as $\varphi^T\cdot p$ on the integral points $p$ of the 
polytope $\Delta_V$ yielding all the integral points of the polytope 
$\Delta_B$ of the base $B$. We denote this in the suggestive form
\beq
	\Delta_B=\varphi^T\cdot \Delta\,.
\eeq

\subsection{Example 1}
\label{sec:Example1}

The first and simplest example we consider has  three non-intersecting sections. As we have seen in section \ref{sec:Matter} 
even in this simple case, matter with charges $(1,0)$, $(0,1)$ and $(1,1)$ can exist and we check explicitly that this is the case. 
The vertices of the polytope along with their divisor classes and charge vectors $\ell^{(i)}$ read:
\beq \label{eq:ExampleTD1}
\text{
 \begin{tabular}{|c||c|c|c|c||c|| c|c|c|c|} \hline
 variable & \multicolumn{4}{c||}{vertices} &  divisor class  & $\ell^{(0)}$ & $\ell^{(1)}$ & $\ell^{(2)}$ & $\ell^{(3)}$  \rule{0pt}{13pt}\\  \hline
  $z_0$ & 1 & 1 & 1 & 0	& $H_B$   	& 1 	& 0	& 0 	& 0 	\\
  $z_1$ & -1 & 0 & 1 & 0 	& $H_B $  	& 1 	& 0 	& 0 	& 0 	\\
  $z_2$ & 0 & -1 & 1 & 0	& $H_B $ 		& 1 	& 0 	& 0 	& 0 	\\ \hline
  $u$  & 0 & 0 & 1 &  0	& $-3H_B+H-E_1-E_2$  	& -3  	& 1 	& 1 	& -1  	\\
  $v$ &0 & 0 & 0 &  1& 	$H-E_2$  		& 0 	& 1	& 0 	& 0\rule{0pt}{11pt} 	\\
  $w$ &0 & 0 &  -1 & -1	& $H-E_1$  		& 0 	& 0 	& 1 	& 0 	\\
  $e_1$ & 0 & 0 &  0 & -1	& $E_1$  	& 0 	& 0  	&-1  	& 1  	\\ 
  $e_2$ & 0 & 0 &  1 & 1	& $E_2$  		& 0  	& -1 	& 0 	& 1 	\\ \hline
 \end{tabular}
 }
\eeq

We readily check by the condition \eqref{eq:ToricFib} that the polytope \eqref{eq:ExampleTD1} describes indeed a
toric fibration with the hyperplane $E=\{(0,0,x,y)\vert x,y\in \mathbb{R}\}$. In other words, all points of the form
$(0,0,x,y)$ describe the polytope of $dP_2$, which are the fourth to eigth points in \eqref{eq:ExampleTD1}.
The toric morphism $\varphi$ in \eqref{eq:ToricMor} is then constructed as the projection onto the first two coordinates
of the vertices and reveals the polytope of the base $\mathbb{P}^2$
as the first three points  in \eqref{eq:ExampleTD1}.
The toric Calabi-Yau hypersurface
$\hat{X}$ in the toric variety $V$ specified by \eqref{eq:ExampleTD1} has Euler number and Hodge numbers
\beq
	\chi(\hat{X})=-216\,,\qquad h^{(1,1)}(\hat{X})=4\,,\qquad  h^{(2,1)}(\hat{X})=112\,.
\eeq
It is satisfying to note that there are neither non-toric divisors nor non-toric complex structure deformations in $\hat{X}$ as expected
by construction.

This model is characterized by $n_2=n_{12}=0$ as can be seen for example by calculating the toric intersections \eqref{eq:defn2n12}.
Using the general formula \eqref{eq:degree_si} or by constructing explicitly the sections of the anti-canonical bundle
of \eqref{eq:ExampleTD1}  we can readily determine the coefficients $s_i$. They are the most generic polynomials in the projective 
coordinates $[z_0:z_1:z_2]$ on $\mathbb{P}^2$ of the following degrees or line bundles (LB): 
\beq \label{eq:OrderCoeffEx1}
\text{
 \begin{tabular}{|c|c|}  \hline
 coefficients 		& section of LB \rule{0pt}{12pt}\\ \hline
 $s_9$, $s_7$		& $\mathcal{O}(0)$\rule{0pt}{11pt} \\
 $s_8$, $s_6$ , $s_3$	& $\mathcal{O}(3H_B)$ \\
 $s_5$, $s_2$		& $\mathcal{O}(6H_B)$ \\
 $s_1$			& $\mathcal{O}(9H_B)$ \\ \hline
 \end{tabular}
}
\eeq

The three sections are given by the intersection of the Calabi-Yau hypersurface with
the three toric divisors $D_u:=\{u=0\}$, $E_1$ and $E_2$. We summarize this schematically as 
\begin{equation}
 \hat{s}_P : \,\,\{e_2=0\} \cap p\, \qquad  \hat{s}_Q :\,\, \{e_1=0\} \cap p\,, \qquad \hat{s}_R : \,\,\{u=0\} \cap p\,,
\end{equation}
where $p$ is the polynomial of the Calabi-Yau hypersurface. All three sections intersect the fiber once and are holomorphic 
everywhere on the base $B=\mathbb{P}^2$. 
To be consistent with the notation in the rest of this work, we choose $\hat{s}_P$ as the zero section. The other two sections 
$\hat{s}_Q$ and $\hat{s}_R$ generate the rank two Mordell-Weill group, responsible for the Abelian $U(1) \times U(1)$-sector
of the theory.

The matter content of this example can be determined explicitly in this toric example or by using the results  of sections \ref{sec:Matter}. 
We summarized our findings in equation \eqref{tab:example1Matter} and emphasize that anomaly cancellation is automatic.

\subsection{Example 2}
\label{sec:Example2}

In order to create a concrete geometry with more complicated  rational sections that in particular realize all the matter content
found in section \ref{sec:Matter} it is necessary to fiber the $dP_2$-polytope slightly differently over the base $\mathbb{P}^2$. 
This requires a change of the line bundle on the base $B$ under which the coordinates $[u:v:w;e_1:e_2]$ transform. 
A simple way to realize this torically is  by changing the location of the base polytope $\Delta_B$ in the polytope of the toric fibration as:
\beq \label{eq:ExampleTD2}
\text{
 \begin{tabular}{|c||c|c|c|c|} \hline
 variable & \multicolumn{4}{c|}{vertices\rule{0pt}{13pt} }  \\  \hline
  $z_0$ & 1 & 1 & 1 &  0  	 		\rule{0pt}{11pt}\\
  $z_1$ & -1 & 0 & 0 &  0 	 		\\
  $z_2$ &  0 & -1 & 1 & 0		 	\\ \hline
 \end{tabular}
}
\eeq
The remaining points take the same form as in \eqref{eq:ExampleTD1} of the previous example. Thus, the only change of the charge vectors 
occurs for  $\ell^{(0)}$, whose fourth entry $-3$ is to be replaced by$-2$. 
This means that the coordinate $u$ now transforms in the bundle $\mathcal{O}_{\mathbb{P}^2}(-2)$ over $\mathbb{P}^2$.

As before we  check the existence of an elliptic fibration by the condition \eqref{eq:ToricFib} with the hyperplane 
$E=\{(0,0,x,y)\vert x,y\in \mathbb{R}\}$.
The toric morphism $\varphi$ in \eqref{eq:ToricMor} is constructed as before as the projection onto 
the first two coordinates of the vertices. The toric Calabi-Yau hypersurface
$\hat{X}$ in the toric variety $V$ specified by \eqref{eq:ExampleTD2} has Euler number and Hodge numbers
\beq
	\chi(\hat{X})=-174\,,\qquad h^{(1,1)}(\hat{X})=4\,,\qquad  h^{(2,1)}(\hat{X})=91\,,
\eeq
and again there are neither non-toric divisors nor non-toric complex structure deformations in $\hat{X}$ as expected.

This model is characterized by $n_2=n_{12}=1$ as can be seen for example by calculating the toric intersections \eqref{eq:defn2n12}.
By means of the general formula \eqref{eq:degree_si} or by constructing explicitly the sections of its anti-canonical bundle
 we  determine the coefficients $s_i$ that are the most generic polynomials in the projective 
coordinates $[z_0:z_1:z_2]$ on $\mathbb{P}^2$ of the following degrees: 
\beq \label{eq:OrderCoeff}
\text{
 \begin{tabular}{|c|c|}  \hline
 coefficients 		& section of LB \rule{0pt}{13pt}\\ \hline
 $s_9$, $s_7$		& $\mathcal{O}(H_B)$ \rule{0pt}{11pt}\\
 $s_8$, $s_6$ , $s_3$	& $\mathcal{O}(3H_B)$ \\
 $s_5$, $s_2$		& $\mathcal{O}(5H_B)$ \\
 $s_1$			& $\mathcal{O}(7H_B)$ \\ \hline
 \end{tabular}
 }
\eeq
The three rational sections are given by the intersection of the toric divisors with the Calabi-Yau hypersurface $p=0$ as
\begin{equation}
 \hat{s}_P : \,\,\{e_2=0\} \cap p\, \qquad  \hat{s}_Q :\,\, \{e_1=0\} \cap p\,, \qquad \hat{s}_R : \,\,\{u=0\} \cap p\,,
\end{equation}

Since $n_2=n_{12}=1$  there are loci where each of the sections $\hat{s}_P$, $\hat{s}_Q$ and $\hat{s}_R$ are ill-defined,
cf.~table \ref{tab:LocusSingDisc}. Thus, the full spectrum analyzed in section \ref{sec:Matter} is  realized.  The matter
multiplicities follow readily from the classification in section \ref{sec:matter_multiplicities} and were calculated explicitly in
\eqref{tab:example2Matter}. Again, it is seen that anomalies are canceled.

\subsection{Example with SU(5)}

In this concluding example we construct  a concrete resolved elliptically fibered 
Calabi-Yau threefold over $\mathbb{P}^2$ with an $SU(5)$-GUT sector. The geometry constructed below
is an extension of the geometry constructed in section \ref{sec:Example1} by a resolved SU$(5)$-singularity on 
the divisor $z_2=0$. The full gauge symmetry of the F-theory compactification is thus SU$(5)\times$U$(1)\times$U(1).

The variables, vertices of the polytope describing the toric variety $V$ and the independent toric divisor classes  read:
\beq \label{eq:ExampleTD3}
\text{
 \begin{tabular}{|c||c|c|c|c||c|} \hline
  variable  & \multicolumn{4}{c||}{vertices}	& divisor class   \rule{0pt}{13pt}\\  \hline
  $z_0$ & 1  & 1  & 1  &  0 	& $H_B$ 		 	\rule{0pt}{11pt}\\
  $z_1$ & -1 & 0  & 1  &  0	& $H_B$ 		 	\\
  $z_2$ & 0  & -1 & 1  &  0	& $H_B-D_1-D_2-D_3-D_4$ 			\\ \hline
  $d_1$ & 0  & -1 & 0  &  0	& $D_1$		 	\rule{0pt}{11pt}\\
  $d_2$ & 0  & -1 & 0  &  -1	& $D_2$		 	\\
  $d_3$ & 0  & -1 & -1  &  -2 	& $D_3$			\\ 
  $d_4$ & 0  & -1 & -1  &  -1	& $D_4$		 	 \\ \hline
  $u$  & 0  & 0  & 1  &  0 	& $-3H_B+H-D_2-D_3-E_1-E_2$ 		\rule{0pt}{11pt}\\
  $v$ & 0  & 0  & 0  &  1 	& $H-D_1-D_2-D_3-D_4-E_2$ 			\\
  $w$ & 0  & 0  & -1 & -1		& $H-D_1-2D_2-3D_3-2D_4-E_1$ 		 	\\
  $e_1$ & 0  & 0  & 0  & -1 	& $E_1$			\\ 
  $e_2$ & 0  & 0  & 1  & 1 	& $E_2$ 			\\ \hline
 \end{tabular}
 }
\eeq
We note that the Abelian sector has not changed compared to section \ref{sec:Example1}. The sections 
are still located at the following intersections of the Calabi-Yau hypersurface $p=0$, 
\begin{equation}
 \hat{s}_P : \,\,\{e_2=0\} \cap p\, \qquad  \hat{s}_Q :\,\, \{e_1=0\} \cap p\,, \qquad \hat{s}_R : \,\,\{u=0\} \cap p\,.
\end{equation}

A constructive way to understand the  form of the polytope in \eqref{eq:ExampleTD3} is to start with the singular geometry
and perform the four blow-ups necessary to resolved codimension one and two singularities of the fibration following
\cite{Esole:2011sm,Marsano:2011hv,Lawrie:2012gg}. Then all one has to do is to translate the divisor classes back into
toric charge vectors and find the vertices realizing these charge vectors as linear relations. This analysis is performed in the following
and leads to the polytope in \eqref{eq:ExampleTD3}.

As usual an SU(5)-singularity at codimension one in the base $B$ is engineered by specializing the coefficients in the Calabi-Yau 
hypersurface equation $p=0$, that is specializing the form presented in \eqref{eq:CYindP2}. In the fibration with elliptic fiber $dP_2$ this means that we 
have to consider non-generic sections $s_i$ in \eqref{eq:CYindP2}. Concretely, we engineer an SU(5)-singularity on the  divisor $z_2=0$ in 
$\mathbb{P}^2$. One way to realize an SU(5)-singularity is to consider non-generic coefficients $s_i$ of the form
\bea \label{eq:s'SU5}
s_1 = z_2^3 s_1^\prime\,, \qquad  s_2 = z_2^2 s_2^\prime\,, 
\qquad s_3 = z_2^2 s_3^\prime, \qquad s_5 =z_2 s_5^\prime\,,
\eea
with all other $s_i$ generic. The sections $s_i$ have to take values in the bundles \eqref{eq:OrderCoeffEx1}, which determines that 
$s_1^\prime$, $s_2^\prime$ $s_3^\prime$, $s_5^\prime$ have to be generic sections of the bundles $\mathcal{O}(6 H_B)$, 
$\mathcal{O}(4 H_B)$, $\mathcal{O}(H_B)$, $\mathcal{O}(5 H_B)$, respectively. 

With these definitions we can check explicitly, 
employing the formulae of section \ref{sec:ellipticCurvesWithRP} for $f$, $g$  and $\Delta$, that the choice \eqref{eq:s'SU5} realizes
an SU(5)-singularity at $z_2=0$, i.e.~$f$, $g$ do not vanish whereas $\Delta\sim z_2^5$ as  required by the Kodaira classification for an SU(5)-
singularity \cite{kodaira1963compact}. In fact, by inserting  \eqref{eq:s'SU5} into the discriminant we observe the structure
\be \label{eq:diskSU5}
 \Delta = -z_2^5 \left( \beta_5^4 P + z_2 \beta^2_5(8\beta_4 P+\beta_5 R) +z_2^2(16\beta_3^2\beta_4^3+\beta_5 Q)+ z_2^3 S + z_2^4 T + z_2^5 U + \mathcal{O}(z_2^6) \right)
\ee
with the identification
\be \label{eq:betaP}
	\beta_5=s_6\,, \quad P := P_1 P_2 P_3 P_4 P_5 =  (s_2 s_5 - s_1 s_6) s_7 (-s_3 s_6 + s_2 s_7) s_8 (-s_7 s_8 + s_6 s_9)\,.
\ee
All the other coefficients can be determined similarly but are not shown because we are only interested in the structure at codimension two. 
We readily see that the form of the discriminant  agrees with the expected one from the local model analysis in the literature, compare 
e.g.~to \cite{Esole:2011sm} whose notation we follow. However, in  a local model the discriminant \eqref{eq:diskSU5} contains only terms up
to order five in $z_2$. As expected for global models there are higher order terms in $z_2$ in \eqref{eq:diskSU5}, in this 
case terms up to $z_2^{10}$. 

Next we turn to the resolution of the singular elliptic fibration over $z_2=0$. A completely smooth elliptic fibration is obtained by four 
consecutive blow-ups in the ambient space, two of which being ordinary blow-ups and the other two being small resolutions on the
Calabi-Yau hypersurface. 
The first blow-up has to be performed at the codimension three locus $w=v=z_2=0$ in the ambient space. Upon introducing a new section 
$d_1$ corresponding to one exceptional divisor, we formulate the blow-up relation as
\beq
 w = d_1  \tilde w\,, \qquad  v = d_1 \tilde v\,,  \qquad z_2 = d_1 \tilde z_2\,. 
\eeq
The Calabi-Yau manifold is still singular at $\tilde{w}=u=d_1=0$. We have to perform another blow-up  and upon introducing a
new section $d_2$ we obtain
\beq
 \tilde w = d_2 \tilde{\tilde w}\,, \qquad  u = d_2 \tilde u\,, \qquad d_1 = d_2 \tilde d_1\,.
\eeq
These blow-ups resolve all codimension one singularities in the elliptically fibered Calabi-Yau manifold. However, there are
still codimension two singularities of the elliptic fibration that have to be resolved. This has to be done via a small
resolution in order to maintain a flat fibration. In the following we will consider one of the six possible small resolutions
in \cite{Esole:2011sm,Marsano:2011hv}. We perform the small resolutions at $\tilde{\tilde{w}}=\tilde{d}_1=0$ and at
$\tilde{\tilde{w}}=d_2=0$  by introducing two new divisor sections $d_3$, $d_4$. Then, the resolution map reads
\beq
 \tilde{\tilde w} = d_3 d_4 \tilde{ \tilde{\tilde w}}, \qquad  \tilde d_1 = d_4 \tilde{ \tilde{d_1}}, \qquad
  d_2 = d_3 \tilde{ d_2 }.
\eeq
It can be checked explicitly that these final resolutions render the geometry smooth.
Combining all the four resolution maps, we obtain the total variable transformations
\beq \label{eq:Fullresolution}
w \rightarrow  \tilde{\tilde{d_1}} \tilde{\tilde{d_2}}^2 d_3^3 d_4^2 \tilde{ \tilde{\tilde w}}, \qquad
v \rightarrow \tilde{\tilde{d_1}} \tilde{d_2} d_3 d_4 \tilde{v},\qquad
z_2 \rightarrow \tilde{\tilde{d_1}} \tilde{d_2} d_3 d_4 \tilde{z_2}, \qquad
u \rightarrow \tilde{d_2} d_3 \tilde{u}.
\eeq
In order to shorten our notation we will in the following drop all the tildes of the variables on the resolution by abuse of notation. 
A consistent assignment of divisor classes on the resolved geometry is presented in \eqref{eq:ExampleTD3}, where we note
that the coordinates displayed there are the coordinates on the resolution after dropping tildes.

Finally, we use the resolution map \eqref{eq:Fullresolution}  to obtain the proper transform of the Calabi-Yau polynomial  $p'$
in \eqref{eq:CYindP2}. The fully resolved geometry $\hat{X}$ reads, after dropping superscripts $'$, as
\begin{align}
 p =& s_6 (e_1 e_2)  u v w + s_7 (d_1 d_4) e_2 v^2 w + s_8 (d_2 d_3^2 d_4) e_1^2  u w^2 + 
 s_9 (d_1 d_2 d_3^2 d_4^2) e_1 v w^2 \nn \\ &+ z_2 s_5 (d_2 d_3) e_1^2 e_2 u^2 w + 
  z_2^2 s_2 (d_1 d_2) e_1 e_2^2 u^2 v  + z_2^2  s_3 (d_1^2 d_2 d_4) e_2^2 u v^2  \nn \\ &
 + z_2^3 s_1 (d_1 d_2^2 d_3) e_1^2 e_2^2  u^3\, .
\end{align}

We conclude with some remarks on codimension two singularities and the expected matter multiplets in the resolved geometry
It is well-known, see e.g.~\cite{Esole:2011sm} for a review, that the SU(5)-singularity in the fiber enhances further at the codimension 
two loci $z_2=0$ with either $\beta_5=0$ or $P=0$. In our case, these loci were identified in \eqref{eq:betaP}.
At the loci $\beta_5:=s_6\stackrel{!}{=}0$  the fiber enhances to a $D_5$-type singularity which are the loci of $\mathbf{10}$-matter 
representations in F-theory. As we see from \eqref{eq:betaP} the $\beta_5$ is generic. Thus  there is only one
$\mathbf{10}$-matter representation in an model with an additional U$(1)^2$-sector at $\beta_5=0$. In contrast, 
the loci $P=0$, where the fiber enhances to an $I_6$-singularity and where  $\mathbf{5}$-matter is located, 
split into five different curves described by  $P_i=0$, $i=1,\ldots,5$. Thus in an F-theory compactification with  
SU$(5)\times$U$(1)^2$ gauge symmetry there will be five different $\mathbf{5}$-matter representations with differing U(1)-charges.
It would be interesting to investigate these geometries further and check anomaly cancellation in six dimensions by working out matter
multiplicities along the lines of section \ref{sec:matter_multiplicities}.

\section{Equivalence of $Bl_{(1,0,0)}\mathbb{P}^2(1,2,3)$ and $Bl_{(0,1,0)}\mathbb{P}^2(1,1,2)$ Elliptic Fibrations}
\label{sec:birationalmodels}

In this section we illustrate the power of birational transformations to show the equivalence
of two different representations of an elliptic curve with one rational point. One representation
is the U$(1)$-restricted Tate model \cite{Grimm:2010ez} in the blow-up $Bl_{(1,0,0)}\mathbb{P}^{2}(1,2,3)$ and
the other one is the hypersurface in $Bl_{(0,1,0)}\mathbb{P}^{2}(1,1,2)$ studied in \cite{Morrison:2012ei}.
We show that elliptic fibrations constructed from these models agree via a birational map and 
are related to models with no rational section by extremal transitions. We strongly suspect that the logic presented
in the following holds more generally and can also be applied to prove the birational equivalence with $dP_1$-model 
studied in \cite{Braun:2013yti} and equivalences of models with multiple sections. Our following discussion is based on the extremal
transitions studied in \cite{Klemm:1996hh}\footnote{We also acknowledge very useful discussions with Antonella Grassi about
the toric blow-up in $\mathbb{P}^{4}(1,1,1,3,6)$ and its effect on the toric Calabi-Yau hypersurface.}. 

We start by motivating the equivalence of the U(1)-restricted Tate model, which can be realized as the toric 
hypersurface in $Bl_{(1,0,0)}\mathbb{P}^2(1,2,3)$, with the toric hypersurface in $Bl_{(0,1,0)}\mathbb{P}^{2}(1,1,2)$
by studying the two-dimensional toric ambient varieties. 

The toric variety $Bl_{(1,0,0)}\mathbb{P}^2(1,2,3)$ is represented by a polytope with vertices
\beq \label{eq:P123bu}
	\begin{array}{|rr||c|c||rr |r|}
	\hline
	\multicolumn{2}{|c||}{\text{vertices}} & & \text{divisor class}& \ell^{(1)} &\ell^{(2)}& \tilde{\ell}^{(1)}\rule{0pt}{12pt}\\
	\hline
		-2&-3&  z&H&1&0& 1\rule{0pt}{11pt}\\
		1& 0& x_1& 2H-E_1& -1&1&0\\ 
		0& 1&  y_1& 3H-E_1 & 0&1&1\\ 
		1& 1& e_1& E_1&3&-1&2\\
		\hline
	\end{array}\,,
\eeq
with only interior point the origin. We note that there are three points interior to the one-dimensional faces of the 
polytope. 
As before we presented the vertices as row-vectors in the first
two columns, introduced the projective coordinates in the third column and displayed the toric divisor classes 
in the fourth column. $H$ denotes the proper transform of the hyperplane class on $\mathbb{P}^{2}(1,2,3)$ and $E_1$ is 
the exceptional divisor. The charge vectors $\ell^{(i)}$ in the fifth 
and sixth columns are the generators of the Mori
cone, that describe the scaling relations among the coordinates. We note that the charge vector of $\mathbb{P}^{2}(1,2,3)$ 
is obtained as $\ell^{(1)}+3\ell^{(2)}=(1,2,3,0)^T$.
In terms of the coordinates in \eqref{eq:P123bu} the blow-down map
$\pi$ to $\mathbb{P}^{2}(1,2,3)$ with coordinates $[x:y:z]$ reads
\beq
	\pi:\,\,\,[z:x_1:y_1:e_1]\,\mapsto\, [x:y:z]=[x_1e_1:y_1e_1:z]\,.
\eeq
We also obtain the charge vector $\tilde{\ell}^{(1)}=\ell^{(1)}+\ell^{(2)}=(1,0,1,2)^T$,
as denoted in the last column of \eqref{eq:P123bu},
with respect to which $[z:1:y_1:e_1]$ scale as the homogeneous coordinates $[u:v:w]$ in $\mathbb{P}^{(2)}(1,1,2)$.  
In addition, we can make this manifest by forming the following $\mathbb{C}^*$-covariant combination of the 
coordinates $[z:x_1:y_1:e_1]$ as
\beq \label{eq:P123toP112}
	U\equiv\frac{z}{x_1}\,,\quad V\equiv \frac{y_1}{x_1}\,,\quad W\equiv e_1\,,\quad T\equiv x_1\,.
\eeq
We readily  check using \eqref{eq:P123bu} that these coordinates enjoy the $\mathbb{C}^*$-actions
\beq \label{eq:P112buscalings}
	\begin{array}{|l|rrrr|}
	\hline
		&U & V & W & T\rule{0pt}{13pt}\\ \hline\hline
		\tilde{\ell}^{(1)}&1 & 1 &2 &0 \rule{0pt}{13pt}\\
		\tilde{\ell}^{(2)}&-1 & 0& -1 & 1\\
		\hline
	\end{array}\,,
\eeq
which are precisely the $\mathbb{C}^*$-actions defining $Bl_{(0,1,0)}\mathbb{P}^2(1,1,2)$, cf.~\eqref{eq:P112bu}. Thus, we 
have a second blow-down map $\pi'$ to $\mathbb{P}^{(2)}(1,1,2)$ resolving $[u:v:w]=[0:1:0]$ is obvious from \eqref{eq:P123toP112} 
reading
\beq \label{eq:blowdownP112}
	\pi':\,\,\,[U:V:W:T]\,\mapsto\, [u:v:w]=[UT:V:WT]\,.
\eeq
In fact, by writing down all sections
of the anti-canonical bundle of $Bl_{(1,0,0)}\mathbb{P}^2(1,2,3)$ in the coordinates \eqref{eq:P123toP112} we precisely reproduce all
sections of the canonical bundle of $Bl_{(0,1,0)}\mathbb{P}^{(2)}(1,1,2)$. Thus, we see the birational 
equivalence of the two models. In the following 
we work out the birational maps in detail and recover the birational map, that was found in \cite{Morrison:2012ei} algebraically, 
by a straightforward toric analysis and one toric blow-up.
  
Before proceeding we present the toric data of $Bl_{(0,1,0)}\mathbb{P}^{(2)}(1,1,2)$. Its polytope reads
\beq \label{eq:P112bu}
	\begin{array}{|rr||c|c||rr|}
	\hline
	\multicolumn{2}{|c||}{\text{vertices}} & & \text{divisor class}& \ell'^{(1)} &-\tilde{\ell}^{(2)}\rule{0pt}{13pt}\\
	\hline
		1&0&  U& H-E_1&-1&1\rule{0pt}{11pt}\\
		-1& -2& V& H&1&0\\
		0& 1& W& 2H-E_1& 0&1\\ 
		1& 1&  T& E_1 & 2&-1\\ 
		\hline
	\end{array}\,,
\eeq
where again the only interior point is the origin and there is one point interior to an one-dimensional face.
The homogeneous coordinates in the third column scale under the relations displayed in the last two columns, that form
a basis of the two-dimensional Mori-cone. We recover the scaling relation of $Bl_{(0,1,0)}\mathbb{P}^{(2)}(1,1,2)$ in \eqref{eq:P112buscalings} 
as $\ell'^{(1)}=\tilde{\ell}^{(1)}+2\tilde{\ell}^{(2)}$.

We  perform our derivation of the birational equivalence of these two presentation of elliptic curves by first matching their respective 
singular Tate models and then by performing the toric resolution.   We begin with recalling the 
U(1)-restricted Tate model. The Tate form \eqref{eq:Tateform} of the elliptic curve takes the form
\beq \label{eq:U1restrTate}
	y^2+a_1 xyz+a_3y z^3=x^3+a_2x^2z^2+a_4x z^4\,
\eeq
where we work in the affine patch with the coordinate of the exceptional divisor $e_1=1$.
The coefficient $a_6$ has been removed torically, $a_6\equiv 0$,  by adding the last point in \eqref{eq:P123bu} of the 
exceptional divisor to the polytope of $\mathbb{P}^2(1,2,3)$, which removes the point in the dual polytope to \eqref{eq:P123bu} 
corresponding to the monomial $a_6z^6$. Thus, an elliptically fibered Calabi-Yau manifold over a base $B$ with the global 
Tate model of the form \eqref{eq:U1restrTate} with generic coefficients 
$a_i\in \mathcal{O}(-iK_B)$
is readily constructed as the Calabi-Yau hypersurface or complete 
intersection in the ambient space obtained by fibering 
$Bl_{(1,0,0)}\mathbb{P}^2(1,2,3)$ over the base $B$. In particular if $B$ 
admits a toric description, i.e.~is either a toric variety or a toric complete intersection, 
we can construct this ambient space torically by fibering the polytope 
\eqref{eq:P123bu} over the polytope describing $B$. We refer to 
section \ref{sec:examples} for the required structure of the 
polytope to construct such a toric fibration

Now we turn to the hypersurface in 
$Bl_{(0,1,0)}\mathbb{P}^{(2)}(1,1,2)$. It has been shown  in 
\cite{Morrison:2012ei} that its Tate form reads
\beq \label{eq:Bl112Tate}
	y^2+\tilde{a}_1 x'yz
	+\tilde{a}_3y z^3=x'^3+\tilde{a}_2x'^2z^2+\tilde{a}_4x' z^4
	+\tilde{a}_6  z^6
\eeq
with the Tate coefficients $\tilde{a}_i$ given by
\beq
	\tilde{a}_1= \frac{2c_3}{b}\,,\quad \tilde{a}_2= \frac{b^2 c_2-c_3^2}{b^2}\,,\quad \tilde{a}_3=b c_1\,,\quad \tilde{a}_4=-b^2 c_0\,,\quad \tilde{a}_6=\tilde{a}_2\tilde{a}_4=-b^2 c_0(b^2c_2-c_3^2)\,.
\eeq
We note immediately that these coefficients are not generic, namely we 
have five different coefficients parametrized by only four independent
variables $\tilde{a}_1$, $\tilde{a}_2$, $\tilde{a}_3$ and $\tilde{a}_4$. Indeed we can eliminate this 
redundancy by a simple variable transformation,
\beq \label{eq:Trafoa6=0}
	x'=x-\tilde{a}_2 z^2\,,
\eeq
which is a subgroup of the automorphisms of $\mathbb{P}^2(1,2,3)$.
The Tate coefficients $a_i'$ with respect to the new
coordinates $[x':y:z]$ read
\beq \label{eq:Tatematch}
	a'_1=\tilde{a}_1\,,\quad a_2'=-2\tilde{a}_2\,,\qquad 
	a'_3=\tilde{a}_3-\tilde{a}_1\tilde{a}_2\,,\qquad a_4'=\tilde{a}_2^2+\tilde{a}_4\,,\qquad a_6'=0\,.
\eeq
When considering an elliptic fibration, these coefficients are 
identified with generic sections of appropriate line bundles on the base 
$B$ of the fibration. We then identify $a_i'\equiv a_i$ in the 
U(1)-restricted Tate model \eqref{eq:U1restrTate}. Thus we have shown by 
matching the Tate forms \eqref{eq:U1restrTate} and 
\eqref{eq:Bl112Tate} that the elliptic curves in 
$Bl_{(1,0,0)}\mathbb{P}^{2}(1,2,3)$, i.e.~the U(1)-restricted Tate 
model, and in $Bl_{(0,1,0)}\mathbb{P}^{2}(1,1,2)$ are birationally 
equivalent. 

We emphasize that this equivalence does not hold for 
general elliptically fibered Calabi-Yau manifolds. We claim that the 
equivalence only holds if the Calabi-Yau constraint with general fiber 
in $Bl_{(0,1,0)}\mathbb{P}^{2}(1,1,2)$ takes globally the form
\beq \label{eq:P112model}
	w^2+b w v^2=u(c_0u^3+c_1 u^2 v+c_2 uv^2+c_3 v^3)\,,
\eeq
where it is crucial that the coefficient of $w^2$ is one, which puts certain 
restrictions on the construction of the elliptic fibration.
Fibrations of these type have been studied in  \cite{Morrison:2012ei} 
for the example of $B=\mathbb{P}^2$. In this case the coefficients were
identified  as
\beq \label{eq:coeffsP2base}
	b\equiv b_n\,,\qquad c_3=f_{3+n}\,,\qquad  c_2=3f_6\,,\qquad c_1=2f_{9-n}\,,\qquad c_0=f_{12-2n}.
\eeq
where the integers in the subscript denote the degree of the polynomial 
on $\mathbb{P}^2$. We see with these identifications that the Tate
coefficients \eqref{eq:Tatematch} are indeed general polynomials and the 
fibration becomes equivalent to a generic U(1)-restricted Tate model 
\eqref{eq:U1restrTate} as claimed.

After having proved this equivalence of the two elliptic fibrations with 
Tate models of the form \eqref{eq:U1restrTate} and \eqref{eq:Bl112Tate}
let us rederive the birational transformation between 
\eqref{eq:U1restrTate}
in $Bl_{(1,0,0)}\mathbb{P}^2(1,2,3)$ and the elliptic curve 
\eqref{eq:P112model} in $Bl_{(0,1,0)}\mathbb{P}^2(1,1,2)$. As we will 
demonstrate momentarily this will involve the single toric blow-up 
$Bl_{(1,0,0)}\mathbb{P}^2(1,2,3)\rightarrow \mathbb{P}^{2}(1,2,3)$ at 
$x=y=0$, the identification of coordinates of coordinates 
\eqref{eq:P123toP112} and the action of a simple automorphism bringing 
the constraint into the form \eqref{eq:P112bu}.  

First we note that the Tate model \eqref{eq:U1restrTate} is 
singular at codimension two in the base when $a_3=a_4=0$, or 
equivalently in the parametrization \eqref{eq:Tatematch} when
\beq
	\frac{1}{2} b c_1 - c_2 \frac{c_3}{b} + \frac{c_3^3}{b^3}=
	 \left(\frac{c_3^2}{b^2}-c_2\right)^2-b^2 c_0 =0
\eeq
All these singular loci in the base are contained in the intersection of 
the surface $x=y=0$ with the  elliptic curve \eqref{eq:Bl112Tate}.
Thus all these singularities are resolved by a single toric blow-up 
along at $x=y=0$. We perform the toric blow-up  in \eqref{eq:U1restrTate} by setting 
\beq \label{eq:U1restrTateToP112}
	x=x_1 e_1=W T\,,\qquad y=y_1 e_1=VW T\,,\quad z=U T
\eeq 
where we have also used the coordinates \eqref{eq:P123toP112} and have at the same time performed the coordinate transformation
\eqref{eq:Trafoa6=0} to eliminate $\tilde{a}_6$ in \eqref{eq:Bl112Tate}.
Inserting this into \eqref{eq:U1restrTate} we obtain
\bea \label{eq:resConstraint}
	&&W^2 T - V^2 W - 2 \frac{c_3}{b} UVWT+2\big( \frac{c_3^2}{b^2}- c_2 \big)U^2 W T^2\nn\\
   &=&U\left(\big(b^2 c_0- c_2^2+2 c_2 \frac{c_3^2}{b^2}-\frac{c_3^4}{b^4}\big) (UT)^3+\big(b c_1-2  c_2 \frac{c_3}{b}+2 \frac{c_3^3}{b^3}\big)  V (UT)^2\right)\,,
\eea
where we have canceled a factor $T^2 W$.  We readily observe that this 
constraint is mapped under the blow-down map \eqref{eq:blowdownP112} 
to the quartic in $\mathbb{P}^{2}(1,1,2)$ as claimed. Then we perform 
the simple variable transformation
\beq \label{eq:Varshift}
	W=W'+\frac{c_3}{b}  U V+(c_2-\frac{c_3^2}{b^2})T U^2\,,\quad T=T'/b
\eeq
in order to bring \eqref{eq:resConstraint} into the form 
\eqref{eq:P112model}, 
\beq
	W^2 T - bV^2 W=U( c_0 U^3 T^3+ c_1 V U^2T^2+c_2 V^2 U T+c_3 V^3)\,,
\eeq
where we multiplied by $b$ and, by abuse of notation, denoted the new
coordinates again by $W$ and $T$. In these coordinates \eqref{eq:U1restrTateToP112} reads
\bea \label{eq:xyzasUVWT}
	x&=&\frac{1}{b^2}T\left(b W+c_3  U V+(c_2-\frac{c_3^2}{b^2})T U^2\right) \,, \nn \\ y&=&\frac{1}{b^2}VT \left(bW+c_3  U V+( c_2-\frac{c_3^2}{b^2})T U^2\right)\,, \nn \\  z&=&\frac{1}{b}TU \,,
\eea
where we again denote the shifted coordinates \eqref{eq:Varshift}
by $W$ and $T$.

Finally, we relate the coordinates $\tilde{x}$, $\tilde{y}$
defined in \eqref{eq:varTrafoToWSF} that bring the U(1)-restricted Tate 
model \eqref{eq:U1restrTate} into 
Weierstrass form to the coordinates of $[U:V:W:T]$ of 
$Bl_{(0,1,0)}\mathbb{P}^{2}(1,1,2)$. This amounts to simply writing out 
the transformation \eqref{eq:varTrafoToWSF} in terms of $x$ and $y$ 
given by \eqref{eq:xyzasUVWT}. First, we evaluate using the Tate 
form \eqref{eq:U1restrTate} the parameters
\beq
	b_2=12 \big(  \frac{c_3^2}{b^2}-\frac{2}{3} c_2\big)\,,\quad a_1
	=2 \frac{c_3}{b}\,,\quad a_3= 2\big(\frac{1}{2}b c_1 
	- c_2\frac{c_3}{b} + \frac{c_3^3}{b^3}\big)\,.
\eeq
Then we readily plug this into \eqref{eq:varTrafoToWSF} with 
\eqref{eq:xyzasUVWT} to obtain
\bea \label{eq:blowupP112WSF}
	\tilde{x}&=& \frac{1}{b^2}T \big(c_3 U V +\frac{1}{3} c_2 T U^2 
	+b W\big) \,,\nn\\ 		\tilde{y}&=& 
	\frac{1}{b^3}T\big(b c_3U V^2+ b c_2 T U^2 V
	+\frac{1}{2} b c_1 T^2 U^3 
	+c_3 T U W + b^2 V W\big)\,,\nn\\
	z&=&\frac{1}{b} TU\,,
\eea
where, as before, $W$ and $T$ are to be understood as the shifted 
coordinates in \eqref{eq:Varshift}.
We note that we can use the $\mathbb{C}^*$-action on 
$[z:\tilde{x}:\tilde{y}]$ to cancel the prefactors $\frac{1}{b^k}$, 
$k=1,2,3$. 

We readily recognize the birational transformation\footnote{Its inverse
is readily constructed by simply solving \eqref{eq:blowupP112WSF}
for $U$, $V$, $W$.} \eqref{eq:blowupP112WSF} as
the birational map worked out in \cite{Morrison:2012ei} to
map the Weierstrass model of the elliptic fibration under consideration 
to the quartic \eqref{eq:P112model}. In fact, by using the
$\mathbb{C}^*$-action we can eliminate the 
prefactor of $T$  in \eqref{eq:blowupP112WSF} in front of $z$. Then we 
identify the coordinates of $\mathbb{P}^{2}(1,1,2)$ in the conventions of \cite{Morrison:2012ei} as
\beq
	u:=U\,,\quad v:=\frac{V}{T}\,,\quad w:=\frac{W}{T}\,.
\eeq
Thus, we have found the same birational map from a simple
toric resolution \eqref{eq:U1restrTateToP112} and identification 
of charge vectors. Although both maps \eqref{eq:U1restrTateToP112} and
\eqref{eq:blowupP112WSF} are blow-ups at $x=y=0$ respectively 
$\tilde{x}=\tilde{y}=0$, the former map is considerably more natural.

\section{Outlook}
\label{sec:conclusions}

In this work we have analyzed F-theory vacua with U$(1)\times$U(1) gauge group
from compactifications on elliptically fibered Calabi-Yau 
manifolds with rank two Mordell-Weil group. We have derived from first principles that one has to
depart from the Tate and Weierstrass form to represent an elliptic curve $\mathcal{E}$ with 
two rational points and its resolved elliptic fibrations: the  appropriate presentation  is the toric 
Calabi-Yau one-fold in $dP_2$. 
We have found the birational maps to its Tate and Weierstrass model and the coordinates of its
two rational points in Weierstrass form.
In addition we have explicitly studied its resolved elliptic fibrations $\hat{X}$ over a base $B$, that we have classified 
for $B=\mathbb{P}^2$. We have seen explicitly that one has to give up the
paradigm of a holomorphic zero section in F-theory. This has also been confirmed by analyzing the
codimension two singularities of the fibration $\hat{X}$, that determine the F-theory matter
spectrum. The spectrum that we work out in detail and in generality is only consistent with six-dimensional 
anomaly cancellation if the zero section wraps fiber components over codimension two loci.

We have also constructed explicitly resolved toric Calabi-Yau threefolds with Mordell-Weil group of rank two
for concrete six-dimensional F-theory compactifications with U$(1)\times$U(1)-sectors.
We have two examples with only U$(1)\times$U(1)-gauge group, that differ in the complexity of
their rational sections and matter sector. In both cases we determine the full anomaly-free matter spectrum.
In a third example we add the non-Abelian sector of an SU$(5)$-gauge theory, i.e.~construct
an F-theory compactification with SU$(5)\times$U$(1)\times$U(1) gauge group. We explicitly write down
the corresponding four-dimensional polytope. We observe a splitting of the $\mathbf{5}$-matter curve into five
different components, in other words matter in the $\mathbf{5}$-representation splits into five matter
representation $\mathbf{5}_{(q_1,q_2)}$ differing by their U$(1)^2$-charges. The $\mathbf{10}$ matter
curves does not split.  We conclude by
an analysis of extremal transitions and birational equivalences of elliptic fibrations with different fiber
types. We employ these techniques to relate the U$(1)$-restricted Tate model with fiber in $Bl_{(1,0,0)}\mathbb{P}^2(1,2,3)$ 
to the fibration with general elliptic fiber in $Bl_{(0,1,0)}\mathbb{P}^2(1,1,2)$.

It would be interesting to explore the `landscape' of elliptic fibrations with different
fiber types in F-theory. A possible starting point to study such different fibrations is to construct and analyze
elliptic fibrations with general elliptic fiber realized as a Calabi-Yau onefold in all the different toric 
varieties corresponding to the 16 two-dimensional reflexive polytopes. This might also be relevant
for a partial classification of Abelian sectors in F-theory. Furthermore, the systematic inclusion of 
non-Abelian sectors as well as the study of their matter spectrum and anomaly cancellation along the 
lines of sections \ref{sec:Matter} and \ref{sec:anomalies} would be desirable. In particular, the 
general check of the gravitational anomaly, done here only for examples, can be used to prove the completeness
of the study of codimension two singularities.  Mathematically this translates into the Euler 
number of the resolution $\hat{X}$ \cite{Grassi:2011hq}.  We also refer to the early study of \cite{Klemm:1996ts} 
on the connection between the rank of the Mordell-Weil group and the Euler number for elliptic fibrations
of different fiber types. Finally and of immediate importance for the full understanding of four-dimensional 
F-theory vacua is the systematic extension of our techniques to Calabi-Yau fourfolds, codimension three phenomena
and the structure of $G_4$-fluxes. We will return to the above issues in a future work \cite{Cvetic:2013uta}.

\subsubsection*{Acknowledgments}

We gratefully acknowledge discussions and correspondence with Lasha Berezhiani,Yi-Zen Chu, Thomas Grimm, Albrecht Klemm,
Daniel Park and Maximilian Poretschkin. We thank especially Antonella Grassi for many useful discussions and comments. DK thanks the
Bethe Center for Theoretical Physics Bonn for hospitality. This work is supported 
in part by DOE grant DE-SC0007901, the Fay R.~and Eugene L.~Langberg Endowed Chair
and the Slovenian Research Agency (ARRS).

\appendix

\section{More Details on the Elliptic Curve in $dP_2$}
\label{app:detailsOf2U1model}

In this appendix we present the results of the explicit calculation of
the Tate and Weierstrass model for the elliptic curve \eqref{eq:uvwdp2} in $dP_2$ that we
omitted in the main text in section \ref{sec:construct3points}. 

First we summarize the coefficients $a_i$ of the Tate form
\begin{equation}
  y^2-x^3+  a_1  xyz - a_2 x^2 z^2 + a_3 y z^3 - a_4 x z^4 -a_6 z^6 = 0\,.
\end{equation}
Using the results \eqref{eq:solx} and \eqref{eq:soly} for the sections $x$ and $y$ in Tate form  we obtain 
\bea \label{eq:TatecoeffsdP2}
 a_1 &=& \frac{ 1 }{s_7 s_8 -s_6 s_9}(s_6 s_7 s_8 - s_6^2 s_9 + 2 s_5 s_7 s_9 + 2 s_3 s_8 s_9 - 2 s_2 s_9^2)\,, \nn \\
 a_2  &=& -\frac{1}{ (s_7 s_8-s_6 s_9)^2}  \left( s_5 s_7^3 s_8^2 + s_3 s_7^2 s_8^3 - s_5 s_6 s_7^2 s_8 s_9 - s_3 s_6 s_7 s_8^2 s_9 -  2 s_2 s_7^2 s_8^2 s_9 \right. \nn \\ 
  && + s_5^2 s_7^2 s_9^2 + 2 s_3 s_5 s_7 s_8 s_9^2 + 
  3 s_2 s_6 s_7 s_8 s_9^2 + s_3^2 s_8^2 s_9^2 - s_2 s_6^2 s_9^3 - 
  2 s_2 s_5 s_7 s_9^3  \nn \\
  && \left.- 2 s_2 s_3 s_8 s_9^3 + s_2^2 s_9^4 \right) \,,\nn \\
 a_3 &=& -s_2 s_7 s_8 - s_3 s_5 s_9 + s_2 s_6 s_9 - s_1 s_7 s_9\,, \nn \\
 a_4 &=& \frac{1}{s_7 s_8 - s_6 s_9} \left( s_3 s_5 s_7^2 s_8^2 + s_1 s_7^3 s_8^2 - s_3 s_5 s_6 s_7 s_8 s_9 - 
 s_1 s_6 s_7^2 s_8 s_9 + s_3 s_5^2 s_7 s_9^2 \right. \nn \\
  && \left. + s_1 s_5 s_7^2 s_9^2 + s_3^2 s_5 s_8 s_9^2 + 2 s_1 s_3 s_7 s_8 s_9^2 - s_2 s_3 s_5 s_9^3 - 
 s_1 s_3 s_6 s_9^3 - s_1 s_2 s_7 s_9^3 \right)\,, \nn\\
 a_6 &=& s_1 s_3 (-s_7^2 s_8^2 + s_6 s_7 s_8 s_9 - s_5 s_7 s_9^2 - s_3 s_8 s_9^2 + s_2 s_9^3)\,. 
\eea

From this we readily obtain the polynomials
\bea 
 	b_2&=&s_6^2 - 4 s_5 s_7 - 4 s_3 s_8 + 8 s_2 s_9\,,\nn\\ 
b_4&=&2 s_2^2 s_9^2 + s_1 s_7 (2 s_7 s_8 - s_6 s_9) + 
 s_2 ( s_6^2 s_9-s_6 s_7 s_8 - 2 s_5 s_7 s_9)  \nn \\
& &+s_3 (2 s_5 s_7 s_8 - s_5 s_6 s_9 - 2 s_2 s_8 s_9 + 2 s_1 s_9^2)\,,\nn\\ 
b_6&=&(s_2 s_7 s_8 + s_3 s_5 s_9 - s_2 s_6 s_9 + s_1 s_7 s_9)^2\nn\\ 
&&-4 s_1 s_3 (s_7^2 s_8^2 + s_9^2 (s_3 s_8 - s_2 s_9) +  s_7 s_9 (s_5 s_9-s_6 s_8 ))\,,
\eea
in terms of which we calculate the polynomials $f$, $g$
for the Weierstrass model according to \eqref{eq:defsWSF} as 
\begin{footnotesize}
\begin{eqnarray} 
 f &=& \frac{1}{48} \left[-s_6^4+8 s_6^2 \left(s_5 s_7+s_3 s_8+s_2 s_9\right)-24 s_6 \left(s_2 s_7 s_8+s_3 s_5 s_9+s_1 s_7 s_9\right)\right.\nn\\ &+&\left.16 \left(-s_5^2 s_7^2+3 s_1 s_7^2 s_8-s_3^2 s_8^2+s_2 s_3 s_8 s_9-s_2^2 s_9^2+3 s_1 s_3 s_9^2+s_5 s_7 \left(s_3 s_8+s_2 s_9\right)\right)\right]\,,\nn\\
   g &=& \frac{1}{864} \left[s_6^6-12 s_6^4 \left(s_5 s_7+s_3 s_8+s_2 s_9\right)+36 s_6^3 \left(s_2 s_7 s_8+s_3 s_5 s_9+s_1 s_7 s_9\right)\right.\nn\\
   &+&\left.24 s_6^2 \left(2 s_5^2 s_7^2+2 s_3^2 s_8^2+s_2 s_3 s_8 s_9+2 s_2^2 s_9^2+s_5 s_7 \left(s_3 s_8+s_2 s_9\right)-3 s_1 \left(s_7^2 s_8+s_3 s_9^2\right)\right)\right.\nn\\
   &+&\left.8 \left(-8 s_5^3 s_7^3-72 s_1 s_3 s_7^2 s_8^2-8 s_3^3 s_8^3+27 s_1^2 s_7^2 s_9^2-72 s_1 s_3^2 s_8 s_9^2-8 s_2^3 s_9^3+3 s_2^2 s_8 \left(9 s_7^2 s_8+4 s_3 s_9^2\right)\right.\right.\nn\\
   &+&\left.\left.6 s_5 s_7 \left(6 s_1 s_7^2 s_8+2 s_3^2 s_8^2+s_2 s_3 s_8 s_9+2 s_2^2 s_9^2-3 s_1 s_3 s_9^2\right)+6 s_2 s_9 \left(-3 s_1 s_7^2 s_8+2 s_3^2 s_8^2+6 s_1 s_3 s_9^2\right)\right.\right.\nn\\
   &+&\left.\left.3 s_5^2 \left(4 s_3 s_7^2 s_8+4 s_2 s_7^2 s_9+9 s_3^2 s_9^2\right)\right)-144 s_6 \left(s_2^2 s_7 s_8 s_9+\left(s_1 s_5 s_7^2+s_3^2 s_5 s_8+s_3 s_7 \left(s_5^2-5 s_1 s_8\right)\right) s_9\right.\right.\nn\\
  &+& \left.\left.s_2 \left(s_5 s_7^2 s_8+s_3 s_7 s_8^2+s_3 s_5 s_9^2+s_1 s_7 s_9^2\right)\right)\right]\,.
\end{eqnarray}
\end{footnotesize}
With these results the discriminant is straightforward and can be provided upon request.

\section{Nagell's Algorithm: Cubic to Weierstrass}
\label{app:NagellsAlgorithm}

Any  Calabi-Yau hypersurfaces in a two-dimensional toric variety is an elliptic curve $\mathcal{E}$. In the 
following we will find the birational map of the cubic in $\mathbb{P}^2$ to Weierstrass form.  In contrast to the 
method used in the main
text in section \ref{sec:construct3points} we employ Nagell's algorithm\footnote{We note that the following 
results have been worked out independently by Maximilian Poretschkin.}, which is an alternative way to
obtain the discriminant of $\mathcal{E}$.

Before delving into the details of this algorithm, let us summarize the general idea on which it is based. 
Every elliptic curve $\mathcal{E}$ can be described as a double cover of an $\mathbb{P}^1$ with all information about the curve
$\mathcal{E}$ encoded in the  four branch points and the two branch cuts in-between. Such a double cover is generically described as
\beq \label{eq:doublecover}
	\xi^2=q_4(\tau)=(t-t_1)(t-t_2)(t-t_3)(t-t_4)
\eeq
where $t$ is the coordinate on $\mathbb{P}^1$ and the $\tau_i$ are the branch points. By the action of $\text{SL}(2,\mathbb{C})$ we
can always move these points to $0$, $1$ and $\infty$ with only one movable branch point that specifies the complex structure of $\mathcal{E}$.
If we move only one of the branch points to infinity, we recover upon identifying $\xi\equiv y$ and $t\equiv x$ the cubic polynomial on the 
right hand side of the Weierstrass form \eqref{eq:WSF} in the affine patch $z$. Thus, finding the Weierstrass model in all of the above cases
reduces to finding an appropriate map to a $\mathbb{P}^1$ such that the Calabi-Yau equation takes the form of a double cover over this 
$\mathbb{P}^1$. Upon mapping one branch point in $t$ to infinity one immediately recovers the Weierstrass model.

The application of this idea to the cubic curve is the foundation of Nagell's algorithm that we present in the following.
Consider a cubic curve in $\mathbb{P}^2$ with projective coordinates $[u:v:w]$ of the form
\beq \label{eq:cubic}
	p= s_1 u^3 + s_2 u^2 v + s_3 uv^2  + s_4 v^3 + s_5 u^2 w +s_6 u vw+ s_7 v^2 w  +   s_8 u w^2 +s_9 v w^2+ s_{10} w^3\,.
\eeq
For the moment we assume that the coefficients $s_i$ are in a field $K$, e.g.~$K=\mathbb{C}$. Then, for generic 
$s_i$ the cubic \eqref{eq:cubic} defines a smooth elliptic curve $\mathcal{E}$. In applications to elliptic 
fibrations $\pi:\,X\rightarrow B$ of 
$\mathcal{E}$ over a base $B$, the $s_i$ are sections of a line bundle $\mathcal{L}^i$ and are locally represented
by polynomials in local coordinates on the base $B$. 

The cubic \eqref{eq:cubic} can be brought into Weierstrass form by application of Nagell's algorithm, where we 
follow the exposition of \cite{connell1996elliptic}. 
In the case at hand, the above mentioned map to an appropriate $\mathbb{P}^1$, which turns the cubic 
\eqref{eq:cubic} into a double cover of the form \eqref{eq:doublecover}, is given by mapping to the slopes $t$ of 
lines through a given point $Q$ on $\mathcal{E}$. Concretely, we first assume that $s_{10}=0$ which can be achieved 
by a coordinate transformation\footnote{In general, however, this coordinate transformation involves the third roots 
of the coefficients $b_i$ which might not be in the field $K$ 
under consideration.} of $u$ and $v$. Then the point $P=[0,0,1]$ is on the curve \eqref{eq:cubic}. Next we assume 
that $s_9\neq 0$ without loss of generality since we can always interchange $u$ and $v$ and both $s_8=s_9=0$ implies 
that the corresponding elliptic curve $\mathcal{E}$ is singular at $P$.

\begin{figure}[ht!]
\centering
\begin{tabular}{cc}
 \includegraphics[scale=0.6]{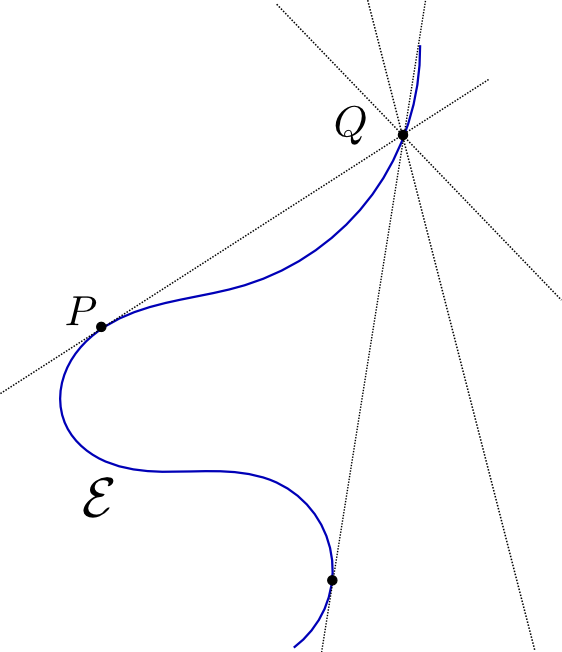} &\hspace{2cm}
 \end{tabular}
 \caption{Cubic curve $\mathcal{E}$ (in blue) with the two points $P$ and $Q$. The tangent at $P$ intersects $\mathcal{E}$
 at the point $Q$. This implies that the slope $t_0$ of the tangent is a root of the discriminant $\delta(t)$ in \eqref{eq:delta}, since $\delta(t)=0$
 for double points of $\mathcal{E}$ along lines through $Q$.}
 \label{fig:nagell}
\end{figure}

Then we define 
\beq
	p=F_3(u,v) +F_2(u,v)  w+F_1(u,v)  w^2\,,
\eeq
i.e.~$F_i(u,v)$ is the coefficient polynomial of the term $w^{3-i}$ in \eqref{eq:cubic}. Every line in $\mathbb{P}^2$ meets the curve
\eqref{eq:cubic} in three points. The tangent at $P$ is the line meeting $\mathcal{E}$ twice at $P$. The point $P$ becomes a double point of 
\eqref{eq:cubic} precisely when $F_1=0$ along the line $v=t_0u$, thus, the tangent at $P$ is described by the equation
\beq \label{eq:tangent@P}
	F_1=s_8 u+s_9 v=0\,.
\eeq
In other words, the slope  $t_0$ of the tangent at $P$ is $t_0=-\frac{s_8}{s_9}$. The tangent meets $p$ in another point $Q$, see figure
\ref{fig:nagell} for an illustration of this situation.
Its coordinates are $Q=[-e_2 s_9:e_2s_8:e_3]$ where we defined $e_i=F_i(s_9,-s_8)$\footnote{Note that the $e_i$ here
should not be confused with the coordinates $e_1$, $e_2$ of the exceptional divisors in $dP_2$ in the main text.}. If $e_2=0$, then $P=Q$ 
is a triple point or flex of $\mathcal{E}$ and if $e_3=0$, then $Q=[-s_9,s_8,0]$ is at infinity. Both $e_2=e_3=0$ implies that $p=0$ along 
the line $F_1=0$ or $p=F_1(u,v) q_2(u,v,w) $ for a quadratic polynomial in $[u:v:w]$. This means that the curve $\mathcal{E}$ is reducible 
with components $\mathbb{P}^1$ and the quadric in $\mathbb{P}^2$, thus, not elliptic. 

Next we perform a variable transformation that centers $Q$ at $u'=v'=0$. The variable change reads
\bea \label{eq:vartrafo}
	&e_3 \neq 0\,:\qquad &u=u'-s_9\frac{e_2}{e_3}w\,,\quad\, v=v'+s_8\frac{e_2}{e_3}w\,,\\
	& e_3=0\,:\qquad&u=u'-s_9w\,,\quad\quad\quad v=v'+s_8w\,,\nn
\eea
where we have to distinguish again the cases $e_3\neq0$ and $e_3=0$. 
We denote the constraint $p$ in the new coordinates as
\beq \label{eq:p'}
	p'=f_3(u',v')+f_2(u',v')w+f_1(u',v')w^2\,,
\eeq
where $f_i$ denote homogeneous polynomials of degree $i$ in $(u',v')$.
Now we consider the lines $v'=tu'$ through $Q$. Their intersections
with $\mathcal{E}$ in generically two other points are determined by
the intersection with the constraint $p'$ reading
\beq \label{eq:linesQ}
	u'(\phi_3(t) u'^2 +\phi_2(t)u' w+\phi_1(t)w^2)=0\,.
\eeq
Here we readily cancel the factor $u'=0$ which is the point $Q$ itself. 
Furthermore, we have defined $\phi_i(t)=f_i(1,t)$.
We determine the two roots of this quadratic equation as a function of 
the slope $t$ by multiplying \eqref{eq:linesQ} with $4\phi_3(t)$. Then 
we complete the square yielding
\beq \label{eq:delta}
	(2\phi_3(t)u'+\phi_2(t)w)^2=\delta(t)w^2\,,\qquad \delta(t)=\phi_2^2(t)-4\phi_1(t)\phi_3(t)
\eeq
where in the case at hand the $\phi_i$ are lengthy polynomials in $t$
that we omit.
The double roots are the zeros $t_i$ of the discriminant $\delta(t)$ of 
the quadratic equation in $u'$, of which there are generically four as 
$\delta(t)$ is a fourth order polynomial in $t$. Geometrically, the 
lines with the slopes $t_i$ are those lines through $Q$, that are a
tangent or in other words are a double point of $\mathcal{E}$ at a 
different point. By construction, we already know one such line that is 
the tangent at $P$ given by \eqref{eq:tangent@P} with slope 
$t_0=-\frac{s_8}{s_9}$, cf.~figure \ref{fig:nagell}.

In other words we know the linear factor $(t-t_0)$ of $\delta(t)$.  Thus we can perform the variable transformation $t=t_0+\frac{1}{\tau}$
to move this root to infinity $\tau=\infty$. Consequently we obtain a third order polynomial 
 \beq
 	\rho(\tau)=\tau^4 \delta(t_0+\frac{1}{\tau})=c \tau^3+d\tau^2+e \tau+ k\,,
 \eeq
for coefficients $c$, $d$, $e$ and $k$, that in the case at hand read
\begin{footnotesize}
\bea
	&c=4( s_2 s_8+\frac{ s_4 s_8^3}{s_9^2}-\frac{ s_3 s_8^2}{s_9}- s_1 s_9)\,,
	\quad d=s_6^2-4 s_5 s_7+8 s_3 s_8-\frac{12 s_4 s_8^2}{s_9}-4 s_2 s_9\,,&\nn \\
 &e=\frac{2}{{s_4 s_8^3-s_3 s_8^2 s_9+s_2 s_8 s_9^2-s_1 s_9^3}}\left(6 s_4^2 s_8^4-s_4 s_6^2 s_8^2 s_9+4 s_4 s_5 s_7 s_8^2 s_9-8 s_3 s_4 s_8^3 s_9-s_4 s_5 s_6 s_8 s_9^2\right.& \nn \\ &+s_3 s_6^2 s_8 s_9^2-2 s_3 s_5 s_7 s_8 s_9^2-s_2 s_6 s_7 s_8 s_9^2+2 s_1 s_7^2 s_8 s_9^2+2 s_3^2 s_8^2 s_9^2+6 s_2 s_4 s_8^2 s_9^2+2 s_4 s_5^2 s_9^3-s_3 s_5 s_6 s_9^3& \nn \\ &\left.+2 s_2 s_5 s_7 s_9^3-s_1 s_6 s_7 s_9^3-2 s_2 s_3 s_8 s_9^3-6 s_1 s_4 s_8 s_9^3+2 s_1 s_3 s_9^4\right)\,,&\nn\\
 &k=\frac{-s_9 }{\left(s_4 s_8^3-s_3 s_8^2 s_9+s_2 s_8 s_9^2-s_1 s_9^3\right)^2}\left(4 s_4^3 s_8^6-s_4^2 s_6^2 s_8^4 s_9+4 s_4^2 s_5 s_7 s_8^4 s_9-8 s_3 s_4^2 s_8^5 s_9-2 s_4^2 s_5 s_6 s_8^3 s_9^2\right.&\nn\\
 &+2 s_3 s_4 s_6^2 s_8^3 s_9^2-4 s_3 s_4 s_5 s_7 s_8^3 s_9^2-2 s_2 s_4 s_6 s_7 s_8^3 s_9^2+4 s_1 s_4 s_7^2 s_8^3 s_9^2+4 s_3^2 s_4 s_8^4 s_9^2+8 s_2 s_4^2 s_8^4 s_9^2&\nn\\
 &+3 s_4^2 s_5^2 s_8^2 s_9^3-s_3^2 s_6^2 s_8^2 s_9^3+6 s_2 s_4 s_5 s_7 s_8^2 s_9^3+2 s_2 s_3 s_6 s_7 s_8^2 s_9^3-6 s_1 s_4 s_6 s_7 s_8^2 s_9^3-s_2^2 s_7^2 s_8^2 s_9^3&\nn\\
 &-8 s_2 s_3 s_4 s_8^3 s_9^3-8 s_1 s_4^2 s_8^3 s_9^3-2 s_3 s_4 s_5^2 s_8 s_9^4+2 s_3^2 s_5 s_6 s_8 s_9^4-4 s_2 s_4 s_5 s_6 s_8 s_9^4+4 s_1 s_4 s_6^2 s_8 s_9^4&\nn\\
 &-2 s_2 s_3 s_5 s_7 s_8 s_9^4+2 s_1 s_4 s_5 s_7 s_8 s_9^4-2 s_1 s_3 s_6 s_7 s_8 s_9^4+2 s_1 s_2 s_7^2 s_8 s_9^4+4 s_2^2 s_4 s_8^2 s_9^4+8 s_1 s_3 s_4 s_8^2 s_9^4&\nn\\
 &\left.-s_3^2 s_5^2 s_9^5+4 s_2 s_4 s_5^2 s_9^5-4 s_1 s_4 s_5 s_6 s_9^5+2 s_1 s_3 s_5 s_7 s_9^5-s_1^2 s_7^2 s_9^5-8 s_1 s_2 s_4 s_8 s_9^5+4 s_1^2 s_4 s_9^6\right)\,.&
\eea
\end{footnotesize}
This yields the sought for cubic on the right hand side of 
\eqref{eq:Tateform}. Finally upon the variable transformation 
$\tau=\frac{x}{c}$ the quadric \eqref{eq:delta} reads, after multiplying 
by $c^2\tau^4$
\beq \label{eq:Tatecubic}
	y^2\equiv[\tfrac{1}{c}x^2 (2\phi_3(t)u'+\phi_2(t)w)]^2=x^3+dx^2+ecx+kc^2\,.
\eeq
This describes the double cover over $\mathbb{P}^1$ with coordinate $y$, 
that finally brings \eqref{eq:Tatecubic} in the Tate form 
\eqref{eq:Tateform}.
In the concrete calculation we obtain the coefficients $c$, $d$, $e$ as
well as the Tate coefficients $a_i$ in \eqref{eq:Tateform} as
\bea
	a_2 &=&-s_6^2+4 s_5 s_7-8 s_3 s_8+\frac{12 s_4 s_8^2}{s_9}+4 s_2 s_9\,,\nn \\
	a_4&=&-\frac{1}{s_9^2}8 \left(6 s_4^2 s_8^4-s_4 s_9 \left(s_6^2 s_8^2-4 s_5 s_7 s_8^2+8 s_3 s_8^3+s_5 s_6 s_8 s_9-6 s_2 s_8^2 s_9-2 s_5^2 s_9^2+6 s_1 s_8 s_9^2\right)\right. \nn \\ && +s_9^2 \left(2 s_3^2 s_8^2+s_7 \left(-s_2 s_6 s_8+2 s_1 s_7 s_8+2 s_2 s_5 s_9-s_1 s_6 s_9\right) \right. \nn \\ &&\left.\left. +s_3 \left(s_6^2 s_8-2 s_5 s_7 s_8-s_5 s_6 s_9-2 s_2 s_8 s_9+2 s_1 s_9^2\right)\right)\right)\,,\nn \\
	a_6&=&\frac{1}{s_9^3}16 \left(4 s_4^3 s_8^6-s_9^3 \left(s_7 \left(s_2 s_8-s_1 s_9\right)
	+s_3 \left(-s_6 s_8+s_5 s_9\right)\right)^2\right. \nn \\ 
	&&+s_4^2 s_8^2 s_9 \left(s_6^2 s_8^2-4 s_5 s_7 s_8^2+2 s_5 s_6 s_8 s_9
	-3 s_5^2 s_9^2+8 s_8 \left(s_3 s_8^2-s_2 s_8 s_9
	+s_1 s_9^2\right)\right) \nn \\ 
	&&+2 s_4 s_9^2 \left(2 s_3^2 s_8^4+2 s_2^2 s_8^2 s_9^2
	+s_3 s_8 \left(s_6^2 s_8^2-2 s_5 s_7 s_8^2-s_5^2 s_9^2
	+4 s_8 s_9 \left(-s_2 s_8+s_1 s_9\right)\right)\right. \nn \\ 
	&&+s_1 \left(2 s_7^2 s_8^3+s_7 s_8 s_9 \left(-3 s_6 s_8+s_5 s_9\right)+2 s_9^2 \left(s_6^2 s_8-s_5 s_6 s_9+s_1 s_9^2\right)\right) \nn \\ &&\left.\left.+s_2 \left(3 s_5 s_7 s_8^2 s_9+2 s_5^2 s_9^3-4 s_1 s_8 s_9^3-s_6 \left(s_7 s_8^3+2 s_5 s_8 s_9^2\right)\right)\right)\right)
\eea
with all $a_i=0$ for odd $i=1,3$.

The birational coordinate transformation from $(u,v)$ to $(x,y)$ can be 
determined by going backwards through the relations 
\eqref{eq:Tatecubic} and \eqref{eq:vartrafo} using
\beq \label{eq:relvarTrafo}
	t=-\frac{s_8}{s_9}+\frac{c}{x}=\frac{v'}{u'}\,.
\eeq
We readily obtain from this and the identification of $y$ in 
\eqref{eq:Tatecubic}
\bea
	 x&=&c\frac{u+s_9\frac{e_2}{e_3}w}{v+\frac{s_8}{s_9}u}\,,\nn \\  
	 y&=&c\frac{u'^2(2\phi_3(t)u'+\phi_2(t)w)}{(v+\frac{s_8}{s_9}u)^2}
	 =c\left.\frac{-f_2(u',v')w-2f_1(u',v')w^2}{(v+\frac{s_8}{s_9}u)^2}\right\vert_{u'=u+s_9 \frac{e_2}{e_3},\,v'=v-s_8\frac{e_2}{e_3}} 
\eea
with $f_i$ defined in \eqref{eq:p'}. Here we have used the elementary
relation $2f_3+f_2=-f_2-2f_1$ that directly follows from \eqref{eq:p'}
to rewrite the numerator of $y$.
In the case at hand we have
\beq
	e_2=s_7 s_8^2-s_6 s_8 s_9+s_5 s_9^2\,,\qquad e_3=-s_4 s_8^3+s_3 s_8^2 s_9-s_2 s_8 s_9^2+s_1 s_9^3\,.
\eeq
This implies
\begin{footnotesize}
\bea \label{eq:xycubic}
x&=&-\frac{4 \left( s_9 \left(s_7 s_8^2+s_9 \left(-s_6 s_8+s_5 s_9\right)\right)w+ \left(-s_4 s_8^3+s_9 \left(s_3 s_8^2-s_2 s_8 s_9+s_1 s_9^2\right)\right)u\right)}{s_9 \left(u s_8+v s_9\right)}\,.\\
y&=&-4w\frac{m_1 u^2+m_2 uv+ m_3 v^2+ m_4 u w+m_5 vw}{\left(u s_8+v s_9\right)^2}
\eea
\end{footnotesize}
where the coefficients $m_i$ are given by
\begin{footnotesize}
\bea
	m_1&=&s_4 s_5 s_8^3+s_2 s_8^2 \left(-s_7 s_8+s_6 s_9\right)+s_9 \left(-s_3 s_5 s_8^2+s_1 \left(3 s_7 s_8^2-3 s_6 s_8 s_9+2 s_5 s_9^2\right)\right)\,,\nn\\
	m_2&=&s_4 s_6 s_8^3+s_3 s_8 \left(-2 s_7 s_8^2+s_9 \left(s_6 s_8-2 s_5 s_9\right)\right)+s_9 \left(-s_1 s_6 s_9^2+s_2 \left(2 s_7 s_8^2-s_6 s_8 s_9+2 s_5 s_9^2\right)\right)\,,\nn\\
	m_3&=&-(s_9^2 \left(s_7 \left(-s_2 s_8+s_1 s_9\right)+s_3 \left(s_6 s_8-s_5 s_9\right)\right)+s_4 s_8 \left(2 s_7 s_8^2+3 s_9 \left(-s_6 s_8+s_5 s_9\right)\right))\,, \\
	m_4&=&s_6^2 s_8^2 s_9-s_6 \left(s_7 s_8^3+3 s_5 s_8 s_9^2\right)+2 \left(s_4 s_8^4+s_9 \left(s_5 s_7 s_8^2-s_3 s_8^3+s_2 s_8^2 s_9+s_5^2 s_9^2-s_1 s_8 s_9^2\right)\right)\,,\nn\\
	m_5&=&-2 s_7^2 s_8^3+s_7 s_8 s_9 \left(3 s_6 s_8-2 s_5 s_9\right)+s_9 \left(2 s_4 s_8^3+s_9 \left(-s_6^2 s_8-2 s_3 s_8^2+s_5 s_6 s_9+2 s_2 s_8 s_9-2 s_1 s_9^2\right)\right)\,.\nn 
\eea
\end{footnotesize}
We readily obtain the inverse variable transformation determining $u$ and $v$ in terms of $x$, $y$ by inverting \eqref{eq:xycubic}. Equivalently, 
one can directly invert the relation \eqref{eq:Tatecubic} to obtain
\beq \label{eq:uvPrime}
	u'_{1/2}=\left.\frac{-x^3\phi_2(t)\pm cxy}{2x^3\phi_3(t)}\right\vert_{t=-\frac{s_8}{s_9}+\frac{c}{x}}=-\left.\frac{2\phi_1(t) x^2}{\phi_2(t) x^2\pm cy}\right\vert_{t=-\frac{s_8}{s_9}+\frac{c}{x}}\,,\qquad v'=u'\Big(-\frac{s_8}{s_9}+\frac{c}{x}\Big)\,.
\eeq
Here we have made us of the equation \eqref{eq:p'} to arrive at the second equality for $u'$ and inverted \eqref{eq:relvarTrafo} to obtain $v'$.
In order to simply the expressions on the right side we have to use the
Weierstrass equation \eqref{eq:Tatecubic} and we obtain for $u$ and $v$, 
employing \eqref{eq:vartrafo},
\beq
	u=2s_9\frac{q_1 x+q_2 y+q_3}{r_1x^2+r_2x+r_3 y+r_4 }\,,\qquad v=2s_9\frac{n_1x+n_2y+n_3}{p_1 x^2+p_2x+p_3y+p_4}\,,
\eeq
where the explicit form of $q_i$, $r_j$, $n_k$ and $p_l$ follow from \eqref{eq:uvPrime} and are provided upon
request.

\bibliographystyle{utphys}	
\bibliography{ref}

\providecommand{\href}[2]{#2}\begingroup\raggedright\begin{thebibliography}{10}

\bibitem{Donagi:2008ca}
R.~Donagi and M.~Wijnholt, ``{Model Building with F-Theory},''
  \href{http://arxiv.org/abs/0802.2969}{{\ttfamily arXiv:0802.2969 [hep-th]}}.

\bibitem{Beasley:2008dc}
C.~Beasley, J.~J. Heckman, and C.~Vafa, ``{GUTs and Exceptional Branes in
  F-theory - I},'' \href{http://dx.doi.org/10.1088/1126-6708/2009/01/058}{{\em
  JHEP} {\bfseries 01} (2009) 058},
\href{http://arxiv.org/abs/0802.3391}{{\ttfamily arXiv:0802.3391 [hep-th]}}.

\bibitem{Beasley:2008kw}
C.~Beasley, J.~J. Heckman, and C.~Vafa, ``{GUTs and Exceptional Branes in
  F-theory - II: Experimental Predictions},''
  \href{http://dx.doi.org/10.1088/1126-6708/2009/01/059}{{\em JHEP} {\bfseries
  01} (2009) 059},
\href{http://arxiv.org/abs/0806.0102}{{\ttfamily arXiv:0806.0102 [hep-th]}}.

\bibitem{Donagi:2008kj}
R.~Donagi and M.~Wijnholt, ``{Breaking GUT Groups in F-Theory},'' {\em
  Adv.Theor.Math.Phys.} {\bfseries 15} (2011) 1523--1604,
\href{http://arxiv.org/abs/0808.2223}{{\ttfamily arXiv:0808.2223 [hep-th]}}.

\bibitem{Blumenhagen:2009yv}
R.~Blumenhagen, T.~W. Grimm, B.~Jurke, and T.~Weigand, ``{Global F-theory
  GUTs},'' \href{http://dx.doi.org/10.1016/j.nuclphysb.2009.12.013}{{\em
  Nucl.Phys.} {\bfseries B829} (2010) 325--369},
\href{http://arxiv.org/abs/0908.1784}{{\ttfamily arXiv:0908.1784 [hep-th]}}.

\bibitem{Marsano:2009wr}
J.~Marsano, N.~Saulina, and S.~Schafer-Nameki, ``{Compact F-theory GUTs with
  U(1) (PQ)},'' \href{http://dx.doi.org/10.1007/JHEP04(2010)095}{{\em JHEP}
  {\bfseries 1004} (2010) 095},
\href{http://arxiv.org/abs/0912.0272}{{\ttfamily arXiv:0912.0272 [hep-th]}}.

\bibitem{Chen:2010ts}
C.-M. Chen, J.~Knapp, M.~Kreuzer, and C.~Mayrhofer, ``{Global SO(10) F-theory
  GUTs},'' \href{http://dx.doi.org/10.1007/JHEP10(2010)057}{{\em JHEP}
  {\bfseries 1010} (2010) 057},
\href{http://arxiv.org/abs/1005.5735}{{\ttfamily arXiv:1005.5735 [hep-th]}}.

\bibitem{Grimm:2009yu}
T.~W. Grimm, S.~Krause, and T.~Weigand, ``{F-Theory GUT Vacua on Compact
  Calabi-Yau Fourfolds},''
  \href{http://dx.doi.org/10.1007/JHEP07(2010)037}{{\em JHEP} {\bfseries 1007}
  (2010) 037},
\href{http://arxiv.org/abs/0912.3524}{{\ttfamily arXiv:0912.3524 [hep-th]}}.

\bibitem{Knapp:2011ip}
J.~Knapp and M.~Kreuzer, ``{Toric Methods in F-theory Model Building},''
  \href{http://dx.doi.org/10.1155/2011/513436}{{\em Adv.High Energy Phys.}
  {\bfseries 2011} (2011) 513436},
\href{http://arxiv.org/abs/1103.3358}{{\ttfamily arXiv:1103.3358 [hep-th]}}.

\bibitem{Heckman:2010bq}
J.~J. Heckman, ``{Particle Physics Implications of F-theory},''
  \href{http://dx.doi.org/10.1146/annurev.nucl.012809.104532}{{\em
  Ann.Rev.Nucl.Part.Sci.} {\bfseries 60} (2010) 237--265},
\href{http://arxiv.org/abs/1001.0577}{{\ttfamily arXiv:1001.0577 [hep-th]}}.

\bibitem{Weigand:2010wm}
T.~Weigand, ``{Lectures on F-theory compactifications and model building},''
  \href{http://dx.doi.org/10.1088/0264-9381/27/21/214004}{{\em
  Class.Quant.Grav.} {\bfseries 27} (2010) 214004},
\href{http://arxiv.org/abs/1009.3497}{{\ttfamily arXiv:1009.3497 [hep-th]}}.

\bibitem{Maharana:2012tu}
A.~Maharana and E.~Palti, ``{Models of Particle Physics from Type IIB String
  Theory and F-theory: A Review},''
\href{http://arxiv.org/abs/1212.0555}{{\ttfamily arXiv:1212.0555 [hep-th]}}.

\bibitem{Vafa:1996xn}
C.~Vafa, ``{Evidence for F theory},''
  \href{http://dx.doi.org/10.1016/0550-3213(96)00172-1}{{\em Nucl.Phys.}
  {\bfseries B469} (1996) 403--418},
\href{http://arxiv.org/abs/hep-th/9602022}{{\ttfamily arXiv:hep-th/9602022
  [hep-th]}}.

\bibitem{Morrison:1996na}
D.~R. Morrison and C.~Vafa, ``{Compactifications of F theory on Calabi-Yau
  threefolds. 1},'' \href{http://dx.doi.org/10.1016/0550-3213(96)00242-8}{{\em
  Nucl.Phys.} {\bfseries B473} (1996) 74--92},
\href{http://arxiv.org/abs/hep-th/9602114}{{\ttfamily arXiv:hep-th/9602114
  [hep-th]}}.

\bibitem{Morrison:1996pp}
D.~R. Morrison and C.~Vafa, ``{Compactifications of F theory on Calabi-Yau
  threefolds. 2.},'' \href{http://dx.doi.org/10.1016/0550-3213(96)00369-0}{{\em
  Nucl.Phys.} {\bfseries B476} (1996) 437--469},
\href{http://arxiv.org/abs/hep-th/9603161}{{\ttfamily arXiv:hep-th/9603161
  [hep-th]}}.

\bibitem{kodaira1963compact}
K.~Kodaira, ``On compact analytic surfaces: Ii,'' {\em The Annals of
  Mathematics} {\bfseries 77} no.~3, (1963) 563--626.

\bibitem{tate1975algorithm}
J.~Tate, ``Algorithm for determining the type of a singular fiber in an
  elliptic pencil,'' {\em Modular functions of one variable IV} (1975) 33--52.

\bibitem{Bershadsky:1996nh}
M.~Bershadsky, K.~A. Intriligator, S.~Kachru, D.~R. Morrison, V.~Sadov, {\em
  et~al.}, ``{Geometric singularities and enhanced gauge symmetries},''
  \href{http://dx.doi.org/10.1016/S0550-3213(96)90131-5}{{\em Nucl.Phys.}
  {\bfseries B481} (1996) 215--252},
\href{http://arxiv.org/abs/hep-th/9605200}{{\ttfamily arXiv:hep-th/9605200
  [hep-th]}}.

\bibitem{Esole:2011sm}
M.~Esole and S.-T. Yau, ``{Small resolutions of SU(5)-models in F-theory},''
\href{http://arxiv.org/abs/1107.0733}{{\ttfamily arXiv:1107.0733 [hep-th]}}.

\bibitem{Marsano:2011hv}
J.~Marsano and S.~Schafer-Nameki, ``{Yukawas, G-flux, and Spectral Covers from
  Resolved Calabi-Yau's},''
  \href{http://dx.doi.org/10.1007/JHEP11(2011)098}{{\em JHEP} {\bfseries 1111}
  (2011) 098},
\href{http://arxiv.org/abs/1108.1794}{{\ttfamily arXiv:1108.1794 [hep-th]}}.

\bibitem{Lawrie:2012gg}
C.~Lawrie and S.~Schafer-Nameki, ``{The Tate Form on Steroids: Resolution and
  Higher Codimension Fibers},''
\href{http://arxiv.org/abs/1212.2949}{{\ttfamily arXiv:1212.2949 [hep-th]}}.

\bibitem{Candelas:1996su}
P.~Candelas and A.~Font, ``{Duality between the webs of heterotic and type II
  vacua},'' \href{http://dx.doi.org/10.1016/S0550-3213(96)00410-5}{{\em
  Nucl.Phys.} {\bfseries B511} (1998) 295--325},
\href{http://arxiv.org/abs/hep-th/9603170}{{\ttfamily arXiv:hep-th/9603170
  [hep-th]}}.

\bibitem{Candelas:1997eh}
P.~Candelas, E.~Perevalov, and G.~Rajesh, ``{Toric geometry and enhanced gauge
  symmetry of F theory / heterotic vacua},''
  \href{http://dx.doi.org/10.1016/S0550-3213(97)00563-4}{{\em Nucl.Phys.}
  {\bfseries B507} (1997) 445--474},
\href{http://arxiv.org/abs/hep-th/9704097}{{\ttfamily arXiv:hep-th/9704097
  [hep-th]}}.

\bibitem{Braun:2011ux}
V.~Braun, ``{Toric Elliptic Fibrations and F-Theory Compactifications},''
  \href{http://dx.doi.org/10.1007/JHEP01(2013)016}{{\em JHEP} {\bfseries 1301}
  (2013) 016},
\href{http://arxiv.org/abs/1110.4883}{{\ttfamily arXiv:1110.4883 [hep-th]}}.

\bibitem{Donagi:2009ra}
R.~Donagi and M.~Wijnholt, ``{Higgs Bundles and UV Completion in F-Theory},''
\href{http://arxiv.org/abs/0904.1218}{{\ttfamily arXiv:0904.1218 [hep-th]}}.

\bibitem{Marsano:2009gv}
J.~Marsano, N.~Saulina, and S.~Schafer-Nameki, ``{Monodromies, Fluxes, and
  Compact Three-Generation F-theory GUTs},''
  \href{http://dx.doi.org/10.1088/1126-6708/2009/08/046}{{\em JHEP} {\bfseries
  0908} (2009) 046},
\href{http://arxiv.org/abs/0906.4672}{{\ttfamily arXiv:0906.4672 [hep-th]}}.

\bibitem{Dudas:2009hu}
E.~Dudas and E.~Palti, ``{Froggatt-Nielsen models from E(8) in F-theory
  GUTs},'' \href{http://dx.doi.org/10.1007/JHEP01(2010)127}{{\em JHEP}
  {\bfseries 1001} (2010) 127},
\href{http://arxiv.org/abs/0912.0853}{{\ttfamily arXiv:0912.0853 [hep-th]}}.

\bibitem{Cvetic:2010rq}
M.~Cvetic, I.~Garcia-Etxebarria, and J.~Halverson, ``{Global F-theory Models:
  Instantons and Gauge Dynamics},''
  \href{http://dx.doi.org/10.1007/JHEP01(2011)073}{{\em JHEP} {\bfseries 1101}
  (2011) 073},
\href{http://arxiv.org/abs/1003.5337}{{\ttfamily arXiv:1003.5337 [hep-th]}}.

\bibitem{Dudas:2010zb}
E.~Dudas and E.~Palti, ``{On hypercharge flux and exotics in F-theory GUTs},''
  \href{http://dx.doi.org/10.1007/JHEP09(2010)013}{{\em JHEP} {\bfseries 1009}
  (2010) 013},
\href{http://arxiv.org/abs/1007.1297}{{\ttfamily arXiv:1007.1297 [hep-ph]}}.

\bibitem{Dolan:2011iu}
M.~J. Dolan, J.~Marsano, N.~Saulina, and S.~Schafer-Nameki, ``{F-theory GUTs
  with U(1) Symmetries: Generalities and Survey},''
  \href{http://dx.doi.org/10.1103/PhysRevD.84.066008}{{\em Phys.Rev.}
  {\bfseries D84} (2011) 066008},
\href{http://arxiv.org/abs/1102.0290}{{\ttfamily arXiv:1102.0290 [hep-th]}}.

\bibitem{Marsano:2012yc}
J.~Marsano, H.~Clemens, T.~Pantev, S.~Raby, and H.-H. Tseng, ``{A Global SU(5)
  F-theory model with Wilson line breaking},''
  \href{http://dx.doi.org/10.1007/JHEP01(2013)150}{{\em JHEP} {\bfseries 1301}
  (2013) 150},
\href{http://arxiv.org/abs/1206.6132}{{\ttfamily arXiv:1206.6132 [hep-th]}}.

\bibitem{neron1964modeles}
A.~N{\'e}ron, ``Modeles minimaux des vari{\'e}t{\'e}s ab{\'e}liennes sur les
  corps locaux et globaux,'' {\em Publications Math{\'e}matiques de L'IH{\'E}S}
  {\bfseries 21} no.~1, (1964) 5--125.

\bibitem{shioda1989}
T.~Shioda, ``{Mordell-Weil lattices and Galois representation. I},'' {\em Proc.
  Japan Acad.} {\bfseries 65A} (1989) 268--271.

\bibitem{shioda1990mordell}
T.~Shioda, ``{On the Mordell-Weil lattices},'' {\em Comment. Math. Univ. St.
  Paul} {\bfseries 39} no.~2, (1990) 211--240.

\bibitem{Wazir:2001}
R.~Wazir, ``{Arithmetic on Elliptic Threefolds},''
  \href{http://arxiv.org/abs/math/0112259}{{\ttfamily arXiv:math/0112259
  [math.NT]}}.

\bibitem{silverman2009arithmetic}
J.~H. Silverman, {\em The arithmetic of elliptic curves}, vol.~106.
\newblock Springer, 2009.

\bibitem{Aspinwall:1998xj}
P.~S. Aspinwall and D.~R. Morrison, ``{Nonsimply connected gauge groups and
  rational points on elliptic curves},'' {\em JHEP} {\bfseries 9807} (1998)
  012,
\href{http://arxiv.org/abs/hep-th/9805206}{{\ttfamily arXiv:hep-th/9805206
  [hep-th]}}.

\bibitem{Aspinwall:2000kf}
P.~S. Aspinwall, S.~H. Katz, and D.~R. Morrison, ``{Lie groups, Calabi-Yau
  threefolds, and F theory},'' {\em Adv.Theor.Math.Phys.} {\bfseries 4} (2000)
  95--126,
\href{http://arxiv.org/abs/hep-th/0002012}{{\ttfamily arXiv:hep-th/0002012
  [hep-th]}}.

\bibitem{Grimm:2010ez}
T.~W. Grimm and T.~Weigand, ``{On Abelian Gauge Symmetries and Proton Decay in
  Global F-theory GUTs},''
  \href{http://dx.doi.org/10.1103/PhysRevD.82.086009}{{\em Phys.Rev.}
  {\bfseries D82} (2010) 086009},
  \href{http://arxiv.org/abs/1006.0226}{{\ttfamily arXiv:1006.0226 [hep-th]}}.

\bibitem{Braun:2011zm}
A.~P. Braun, A.~Collinucci, and R.~Valandro, ``{G-flux in F-theory and
  algebraic cycles},''
  \href{http://dx.doi.org/10.1016/j.nuclphysb.2011.10.034}{{\em Nucl.Phys.}
  {\bfseries B856} (2012) 129--179},
\href{http://arxiv.org/abs/1107.5337}{{\ttfamily arXiv:1107.5337 [hep-th]}}.

\bibitem{Krause:2011xj}
S.~Krause, C.~Mayrhofer, and T.~Weigand, ``{$G_4$ flux, chiral matter and
  singularity resolution in F-theory compactifications},''
  \href{http://dx.doi.org/10.1016/j.nuclphysb.2011.12.013}{{\em Nucl.Phys.}
  {\bfseries B858} (2012) 1--47},
\href{http://arxiv.org/abs/1109.3454}{{\ttfamily arXiv:1109.3454 [hep-th]}}.

\bibitem{Grimm:2011fx}
T.~W. Grimm and H.~Hayashi, ``{F-theory fluxes, Chirality and Chern-Simons
  theories},'' \href{http://dx.doi.org/10.1007/JHEP03(2012)027}{{\em JHEP}
  {\bfseries 1203} (2012) 027},
\href{http://arxiv.org/abs/1111.1232}{{\ttfamily arXiv:1111.1232 [hep-th]}}.

\bibitem{Morrison:2012ei}
D.~R. Morrison and D.~S. Park, ``{F-Theory and the Mordell-Weil Group of
  Elliptically-Fibered Calabi-Yau Threefolds},''
  \href{http://dx.doi.org/10.1007/JHEP10(2012)128}{{\em JHEP} {\bfseries 1210}
  (2012) 128},
\href{http://arxiv.org/abs/1208.2695}{{\ttfamily arXiv:1208.2695 [hep-th]}}.

\bibitem{Cvetic:2012xn}
M.~Cvetic, T.~W. Grimm, and D.~Klevers, ``{Anomaly Cancellation And Abelian
  Gauge Symmetries In F-theory},''
  \href{http://dx.doi.org/10.1007/JHEP02(2013)101}{{\em JHEP} {\bfseries 1302}
  (2013) 101},
\href{http://arxiv.org/abs/1210.6034}{{\ttfamily arXiv:1210.6034 [hep-th]}}.

\bibitem{Mayrhofer:2012zy}
C.~Mayrhofer, E.~Palti, and T.~Weigand, ``{U(1) symmetries in F-theory GUTs
  with multiple sections},''
\href{http://arxiv.org/abs/1211.6742}{{\ttfamily arXiv:1211.6742 [hep-th]}}.

\bibitem{Braun:2013yti}
V.~Braun, T.~W. Grimm, and J.~Keitel, ``{New Global F-theory GUTs with U(1)
  symmetries},''
\href{http://arxiv.org/abs/1302.1854}{{\ttfamily arXiv:1302.1854 [hep-th]}}.

\bibitem{Hayashi:2010zp}
H.~Hayashi, T.~Kawano, Y.~Tsuchiya, and T.~Watari, ``{More on Dimension-4
  Proton Decay Problem in F-theory -- Spectral Surface, Discriminant Locus and
  Monodromy},'' \href{http://dx.doi.org/10.1016/j.nuclphysb.2010.07.011}{{\em
  Nucl.Phys.} {\bfseries B840} (2010) 304--348},
\href{http://arxiv.org/abs/1004.3870}{{\ttfamily arXiv:1004.3870 [hep-th]}}.

\bibitem{Marsano:2010sq}
J.~Marsano, ``{Hypercharge Flux, Exotics, and Anomaly Cancellation in F-theory
  GUTs},'' \href{http://dx.doi.org/10.1103/PhysRevLett.106.081601}{{\em
  Phys.Rev.Lett.} {\bfseries 106} (2011) 081601},
\href{http://arxiv.org/abs/1011.2212}{{\ttfamily arXiv:1011.2212 [hep-th]}}.

\bibitem{Grassi:2012qw}
A.~Grassi and V.~Perduca, ``{Weierstrass models of elliptic toric K3
  hypersurfaces and symplectic cuts},''
\href{http://arxiv.org/abs/1201.0930}{{\ttfamily arXiv:1201.0930 [math.AG]}}.

\bibitem{Batyrev:1994hm}
V.~V. Batyrev, ``{Dual polyhedra and mirror symmetry for Calabi-Yau
  hypersurfaces in toric varieties},'' {\em J.Alg.Geom.} {\bfseries 3} (1994)
  493--545,
\href{http://arxiv.org/abs/alg-geom/9310003}{{\ttfamily arXiv:alg-geom/9310003
  [alg-geom]}}.

\bibitem{Klemm:1996hh}
A.~Klemm, P.~Mayr, and C.~Vafa, ``{BPS states of exceptional noncritical
  strings},''
\href{http://arxiv.org/abs/hep-th/9607139}{{\ttfamily arXiv:hep-th/9607139
  [hep-th]}}.

\bibitem{Borchmann:2013jwa}
E.~P. J.~Borchmann, C.~Mayrhofer and T.~Weigand, ``{Elliptic fibrations for
  SU(5) x U(1) x U(1) F-theory vacua},''
\href{http://arxiv.org/abs/1303.5054}{{\ttfamily arXiv:1303.5054 [hep-th]}}.

\bibitem{Katz:2011qp}
S.~Katz, D.~R. Morrison, S.~Schafer-Nameki, and J.~Sully, ``{Tate's algorithm
  and F-theory},'' \href{http://dx.doi.org/10.1007/JHEP08(2011)094}{{\em JHEP}
  {\bfseries 1108} (2011) 094},
\href{http://arxiv.org/abs/1106.3854}{{\ttfamily arXiv:1106.3854 [hep-th]}}.

\bibitem{Grimm:2010ks}
T.~W. Grimm, ``{The N=1 effective action of F-theory compactifications},''
  \href{http://dx.doi.org/10.1016/j.nuclphysb.2010.11.018}{{\em Nucl.Phys.}
  {\bfseries B845} (2011) 48--92},
\href{http://arxiv.org/abs/1008.4133}{{\ttfamily arXiv:1008.4133 [hep-th]}}.

\bibitem{Park:2011ji}
D.~S. Park, ``{Anomaly Equations and Intersection Theory},'' {\em JHEP}
  {\bfseries 1201} (2012) 093,
\href{http://arxiv.org/abs/1111.2351}{{\ttfamily arXiv:1111.2351 [hep-th]}}.

\bibitem{Grimm:2011sk}
T.~W. Grimm and R.~Savelli, ``{Gravitational Instantons and Fluxes from
  M/F-theory on Calabi-Yau fourfolds},''
  \href{http://dx.doi.org/10.1103/PhysRevD.85.026003}{{\em Phys.Rev.}
  {\bfseries D85} (2012) 026003},
\href{http://arxiv.org/abs/1109.3191}{{\ttfamily arXiv:1109.3191 [hep-th]}}.

\bibitem{Bonetti:2011mw}
F.~Bonetti and T.~W. Grimm, ``{Six-dimensional (1,0) effective action of
  F-theory via M-theory on Calabi-Yau threefolds},''
  \href{http://dx.doi.org/10.1007/JHEP05(2012)019}{{\em JHEP} {\bfseries 1205}
  (2012) 019},
\href{http://arxiv.org/abs/1112.1082}{{\ttfamily arXiv:1112.1082 [hep-th]}}.

\bibitem{griffiths2011principles}
P.~Griffiths and J.~Harris, {\em Principles of algebraic geometry}, vol.~52.
\newblock Wiley-interscience, 2011.

\bibitem{Erler:1993zy}
J.~Erler, ``{Anomaly cancellation in six-dimensions},''
  \href{http://dx.doi.org/10.1063/1.530885}{{\em J.Math.Phys.} {\bfseries 35}
  (1994) 1819--1833},
\href{http://arxiv.org/abs/hep-th/9304104}{{\ttfamily arXiv:hep-th/9304104
  [hep-th]}}.

\bibitem{Honecker:2006dt}
G.~Honecker, ``{Massive U(1)s and heterotic five-branes on K3},''
  \href{http://dx.doi.org/10.1016/j.nuclphysb.2006.04.027}{{\em Nucl.Phys.}
  {\bfseries B748} (2006) 126--148},
\href{http://arxiv.org/abs/hep-th/0602101}{{\ttfamily arXiv:hep-th/0602101
  [hep-th]}}.

\bibitem{Sadov:1996zm}
V.~Sadov, ``{Generalized Green-Schwarz mechanism in F theory},''
  \href{http://dx.doi.org/10.1016/0370-2693(96)01134-3}{{\em Phys.Lett.}
  {\bfseries B388} (1996) 45--50},
\href{http://arxiv.org/abs/hep-th/9606008}{{\ttfamily arXiv:hep-th/9606008
  [hep-th]}}.

\bibitem{Grassi:2011hq}
A.~Grassi and D.~R. Morrison, ``{Anomalies and the Euler characteristic of
  elliptic Calabi-Yau threefolds},''
\href{http://arxiv.org/abs/1109.0042}{{\ttfamily arXiv:1109.0042 [hep-th]}}.

\bibitem{Klemm:1996ts}
A.~Klemm, B.~Lian, S.~Roan, and S.-T. Yau, ``{Calabi-Yau fourfolds for M theory
  and F theory compactifications},''
  \href{http://dx.doi.org/10.1016/S0550-3213(97)00798-0}{{\em Nucl.Phys.}
  {\bfseries B518} (1998) 515--574},
\href{http://arxiv.org/abs/hep-th/9701023}{{\ttfamily arXiv:hep-th/9701023
  [hep-th]}}.

\bibitem{Cvetic:2013uta}
M.~Cveti{\v c}, A.~Grassi, D.~Klevers, and H.~Piragua, ``{Chiral
  Four-Dimensional F-Theory Compactifications With SU(5) and Multiple
  U(1)-Factors},''
\href{http://arxiv.org/abs/1306.3987}{{\ttfamily arXiv:1306.3987 [hep-th]}}.

\bibitem{connell1996elliptic}
I.~Connell, ``Elliptic curve handbook,'' {\em Preprint} (1996) .

\end{thebibliography}\endgroup

\end{document}